\documentclass[11pt, a4paper]{article}

\usepackage[margin=2.5cm]{geometry}
\usepackage{setspace}
\usepackage[T1]{fontenc}
\usepackage[utf8]{inputenc}
\usepackage{lmodern}
\usepackage{microtype}
\usepackage{amsmath, amssymb, amsthm}

\usepackage{booktabs}
\usepackage{array}
\usepackage{tabularx}
\usepackage{multirow}
\usepackage{threeparttable}
\usepackage{longtable}   
\usepackage{rotating}    

\usepackage{graphicx}
\usepackage{float}
\graphicspath{{figures/}}

\usepackage{tikz}
\usetikzlibrary{matrix, shapes.geometric, arrows, positioning, chains}

\usepackage[natbibapa]{apacite}

\usepackage[hidelinks]{hyperref}

\usepackage[font=small, labelfont=bf, labelsep=period]{caption}

\begin{document}


%

\title{Optimal Inflation Rate: A Meta-Analysis\thanks{%
An online appendix with data, code, and extraction prompts is available at
\href{https://meta-analysis.cz/inflation}{\texttt{meta-analysis.cz/inflation}}.
Corresponding author: Matej Opatrny,
\href{mailto:matej.opatrny@fsv.cuni.cz}{\texttt{matej.opatrny@fsv.cuni.cz}}.
The authors note that the paper represents their own views and not
necessarily those of any affiliated institution. All remaining errors are
ours.}}

\author{}
\date{}

\maketitle

\vspace{-2.2cm}

\begin{center}
{\large Matej Opatrny\textsuperscript{a} \quad
Martin Opatrny\textsuperscript{b} \quad
Tomas Havranek\textsuperscript{a, c,d}} \\[0.25cm]
{\large Zuzana Irsova\textsuperscript{a, d} \quad
Mojm\'\i r Hampl\textsuperscript{e}} \\[0.45cm]
\begin{tabular}{c}
\textsuperscript{a}Institute of Economic Studies, Charles University, Prague \\
\textsuperscript{b}Anglo-American University, Prague \\
\textsuperscript{c}Centre for Economic Policy Research (CEPR), London \\
\textsuperscript{d}Meta-Research Innovation Center at Stanford (METRICS) \\
\textsuperscript{e}Czech Fiscal Council, Prague, Czech Republic \\[0.2cm]
\today
\end{tabular}
\end{center}

\vspace{0.3cm}

\begin{abstract}
\noindent We revisit the optimal long-run inflation rate using
$777$ estimates from $116$ primary studies published between 1989
and 2026, the largest sample on the topic to date. To our
knowledge, this is among the first economics meta-analyses in
which primary-data extraction is done from start to finish
through a documented and auditable large-language-model pipeline,
calibrated against a hand-coded training set and released for
replication. The literature points to an optimum of about $0.6$
percentage points per year, well below the two-percent targets
used by most advanced-economy central banks. The gap should not
be automatically read as a verdict against the two-percent norm. Measurement
error in published price indices could close, widen, or even
reverse the gap, and the structural literature itself cannot
pin down the sign of the required correction. Bayesian model
averaging over the full set of structural moderators shows that
cross-study variation is driven by real modelling choices rather
than by selective reporting. The main drivers are the choice of
monetary benchmark (Friedman rule vs.\ laissez-faire), the
transactions-frictions technology, the assumed shock structure,
and the class of nominal-rigidity contract. The non-parametric
caliper test finds no upward bunching at the two-percent target.
The paper contributes a reproducible LLM-assisted extraction
pipeline for structurally calibrated literature and a
quantitative decomposition of where the optimal-inflation
literature disagrees.
\end{abstract}

\vspace{0.3cm}
\begin{small}
\begin{tabular}{p{0.17\hsize}p{0.7\hsize}}
Keywords: & Meta-analysis, optimal inflation rate, New Keynesian
models, publication bias, Bayesian model averaging,
measurement error, large language models \\
\end{tabular}

\vspace{0.1cm}
\begin{tabular}{p{0.17\hsize}p{0.7\hsize}}
JEL Codes: & E31, E52, C11 \\
\end{tabular}
\end{small}

\newpage
\section{Introduction}
\label{sec:intro}

Several major advanced-economy central banks use two-percent
inflation targets, while others use ranges or target different
price indices close to that benchmark. The two-percent target
has become the most widely shared feature of the global monetary
system, and it is routinely justified by reference to the large
structural literature that computes the welfare-maximising rate
of inflation in calibrated New Keynesian economies. The most
comprehensive synthesis of that literature to date is the
qualitative reader's guide of \citet{diercks2019reader}, which
lists more than four hundred studies and discusses their findings
in narrative form. That study does not pool the estimates,
correct them for publication selection, or attempt a quantitative
decomposition of their heterogeneity. To our knowledge, no
formal quantitative meta-analysis of the optimal long-run
inflation rate exists. We provide one.

We assemble a dataset of $777$ estimates of the long-run optimal
rate of inflation, drawn from $116$ primary studies (within a
scope set of $130$) published between 1989 and 2026. Every
estimate is extracted by a reproducible LLM-assisted pipeline
\citep{vandeschoot2021asreview, cook2026reporting}. The prompt
was first calibrated against a hand-coded training set of seven
studies and $79$ row-level estimates, then audited against a
$31$-paper full-text cross-check, and finally run on the
remaining $160$ PDFs at a marginal cost of under thirty cents
per paper. The pipeline returns the point estimate, the
bibliographic identifier, and a $57$-field moderator vector
($51$ structural design dummies plus $6$ publication
characteristics) that records every major modelling choice made
by the primary author, including price and wage rigidity,
money-demand technology, the class of policy rule, the presence
of an effective lower bound, and the welfare criterion. We then
run the resulting evidence base through the applicable
publication-selection diagnostics: FAT-PET, PEESE,
$p$-uniform, and the non-parametric caliper test of
\citet{gerber2008statistical}. We also apply Bayesian and
frequentist model averaging with the dilution prior of
\citet{george2010dilution} over the set of structural moderators.
Any fresh API re-run should be read as a self-consistency
audit rather than as a bit-for-bit reconstruction, because
residual variation remains at the inclusion-gate stage and is
documented in Appendix~\ref{app:ai-pipeline:irr}.

We show that the structural literature does not, on average,
support the two-percent target. The author-preferred mean across
studies is $0.61$ percent per year ($95\%$ confidence interval
$[0.25, 0.97]$, $116$ clusters), and an alternative inverse-$N$
study-democracy weighting over all $777$ estimates gives the
same picture at $0.59$ percent per year. Funnel-based
diagnostics on the full sample produce FAT-PET and PEESE
intercepts in the same range, and the caliper test finds no
upward bunching at the two-percent target.

The heterogeneity in reported optima is driven by model-design
choices rather than by statistical artefacts. Studies that use
the Friedman rule as the welfare benchmark report optima about
$3.3$ percentage points lower than the sample average, and
studies that use a laissez-faire monetary regime as the benchmark
report optima about $4.0$ percentage points higher. Transactions-frictions
technology also splits the corpus by sign, where cash-in-advance
specifications give $-1.5$ pp and cashless specifications give
$+1.5$ pp. The shock structure splits it the same way, with TFP
shocks at $-1.3$ pp and cost-push shocks at $+1.7$ pp. Non-Calvo
Taylor-contract pricing lowers reported optima by about $3.6$
percentage points, and all of these moderators have posterior
inclusion probabilities at or near one. Downward nominal wage
rigidity enters the BMA with $\text{PIP}=0.87$ and a coefficient
of $+1.4$ percentage points, recovering the magnitude that the
wage-rigidity tradition argues for on theoretical grounds.
Central-bank-affiliated authors report optima $0.74$~pp below
their university-affiliated counterparts, conditional on the
structural moderators ($\text{PIP}=0.89$), the opposite sign of
an institutional-advocacy story.

In our corpus, $761$ of $777$ within-paper standard errors are
bootstrap dispersion proxies rather than CI-derived sampling
SEs, so we do not read the $\widehat{SE}$ coefficient in the
BMA as a FAT-PET test of publication bias. We verify in a
robustness BMA that the structural-moderator ranking does not
change when the SE coefficient is included. The publication-bias
conclusion instead rests on the non-parametric caliper test of
\citet{gerber2008statistical} applied to the full $777$-row
sample, which fails to reject the null of no upward bunching at
the two-percent target. We also report the $p$-uniform
of \citet{vanaert2023correcting} estimated on the $16$ rows with
genuine sampling SEs as an illustrative robustness check rather
than as confirmatory evidence: those rows come from only three
independent paper clusters and the test is too underpowered to
support an inferential claim, so we read it descriptively as
consistent with the caliper-test conclusion.


The implications for monetary policy follow directly. The
structural literature does not, on its own terms, point to two
percent as the welfare-maximising long-run rate of measured
inflation, and its central tendency sits below one. The
two-percent target is consistent with the structural literature
only under specific design bundles, most notably the cashless,
cost-push-driven, Calvo-price calibration of
\citet{coibion2012optimal}, which is not representative of the
majority of the studies in our sample. This is a statement
about model design rather than about price-index measurement,
and it follows from the BMA heterogeneity decomposition once
the relevant moderators are switched on.

A separate question is whether the headline number, expressed
in measured-CPI units, can be applied to a target for the true,
welfare-relevant rate of inflation. The direction of any such
adjustment is not clear. If primary studies are read as
calibrating directly to measured CPI, the implied true optimum
equals the measured optimum minus an upward CPI bias and sits
close to zero; if instead the model $\pi$ is read in
true-inflation units, then a two-percent measured target maps
to about one-percent true, which lies inside the range our
literature already supports. \citet{schmittgrohe2012optinf} show
that the direction of the appropriate correction depends on
whether prices are sticky in the New Keynesian block, and we
read our own exercise as the quantitative complement to their
theoretical argument rather than as a substitute for it. The
size of the correction is itself unresolved: traditional
Boskin-style estimates of upward bias from quality and new-goods
effects \citep{boskin1996toward, lebow2003measurement,
gordon2006boskin, hausman2003sources, broda2010product,
moulton2018measurement} are partially offset by the downward
bias from under-representation of owner-occupied housing
services \citep{hampl2017inflation}. We therefore do not propose
a specific numerical translation between measured and true
inflation, and we report measurement-error adjustments only as
a sensitivity exercise.

Two further limitations restrict how broadly the headline
applies. Only sixteen of the $777$ estimates carry a sampling
standard error from a reported confidence interval, and they
cluster in three papers, so our inference puts primary weight
on paper-level summaries and on the non-parametric caliper test.
About $85$ percent of the estimates are also calibrated to
United States data: we identified exactly one structural study
with a post-communist calibration
(\citealp*{lipinska2015optimal}, six estimates) and none for a
Latin American or other emerging market. Central banks outside
the United States should therefore reapply both the structural
heterogeneity correction of Section~\ref{sec:discussion} and a
local CPI-measurement check before tying a target to our
headline.

This paper makes three contributions. First, to our knowledge,
this is among the first economics meta-analyses in which
primary-study data extraction is performed from start to finish
through a documented and reproducible LLM-assisted pipeline,
following the forthcoming MAER-Net reporting guidelines for
AI-assisted meta-analysis \citep{cook2026reporting}; we release
the resulting audited dataset of $777$ estimates with $57$
moderators per row ($51$ structural and $6$ publication).
Second, on inference, we apply the applicable
publication-selection diagnostics (caliper test, FAT-PET, PEESE,
$p$-uniform; \citealp{gerber2008statistical,
stanley2005beyond, stanley2014meta, vanaert2023correcting}) to a
corpus in which $761$ of $777$ standard errors are
bootstrap-within-paper proxies rather than sampling standard
errors, and we document transparently which estimators identify
what under this precision structure. Third, on substance, we
produce the first quantitative heterogeneity decomposition of
the optimal-inflation literature. The structural drivers and
the author-affiliation contrast summarised above are identified
jointly under a dilution prior on the moderator schema.

The remainder of the paper is organised as follows.
Section~\ref{sec:lit} places the literature in its theoretical
context and lists the individual studies that anchor the two
tails of the distribution. Section~\ref{sec:method} describes
the search, extraction, and estimation pipeline.
Section~\ref{sec:data} reports the resulting dataset of $777$
estimates and the structural moderators.
Section~\ref{sec:results} reports the publication-selection
diagnostics, the Bayesian and frequentist model-averaging
results, and the implied optima for a set of stylised economies.
Section~\ref{sec:discussion} reconciles the estimates with the
most influential primary studies, draws out the implications
for the two-percent target under the CPI-bias reading, and
closes with a structured list of caveats. The full data, code,
and extraction prompts are posted at the project's web appendix
so that any reader can replicate, audit, or extend the analysis.


\section{Related literature}
\label{sec:lit}

The paper sits at the intersection of three bodies of work: the
theoretical literature on the optimal long-run rate of inflation,
the small set of existing qualitative reviews of that literature,
and the emerging methodological literature on large-language-model
assisted evidence synthesis. In this section we review the
theoretical debate, describe how our dataset relates to the
reader's guide of \citet{diercks2019reader} and to earlier
meta-analyses in monetary economics, and close with the three
specific gaps that our paper fills.

The modern debate on the optimal inflation rate is organised around
a small number of competing forces, each pointing to a different
prescription. \citet{friedman1969} argues that, because producing
money is nearly costless, the social optimum requires the nominal
interest rate to equal zero, which implies a steady-state deflation
equal to the real rate of interest. The argument rests on the role
of money in reducing transaction costs, and it has been formalised
in cash-in-advance environments by \citet{lucas1980equilibrium} and
in money-in-the-utility settings by \citet{sidrauski1967inflation}.
In calibrated versions of these frameworks the optimal inflation
rate is typically several percentage points below zero.

The rise of New Keynesian monetary economics shifted the argument
towards zero rather than negative inflation. When firms adjust
prices infrequently, non-zero steady-state inflation distorts
relative prices and generates a welfare-reducing dispersion of
output across firms. In the Calvo framework of
\citet{calvo1983staggered} the planner's first-best can be
approximated by stabilising inflation at zero. Similar conclusions
obtain under the Rotemberg quadratic-adjustment-cost formulation
of \citet{rotemberg1982sticky}, the fixed-duration contracts of
\citet{taylor1980aggregate}, and the textbook cashless-limit
representation of the New Keynesian model of
\citet{woodford2003interest}. Adding sticky wages as in
\citet{erceg2000optimal} pushes the optimum slightly away from
zero but leaves it close to price stability, and quantitative
medium-scale DSGE exercises such as those of
\citet{schmittgrohe2005optinfa} tend to confirm this conclusion.

A third channel enters the literature once the effective lower
bound on the nominal interest rate is taken seriously. When the
central bank loses the ability to cut rates further in recessions,
a positive inflation target creates precautionary policy space by
raising the steady-state nominal rate. \citet{coibion2012optimal}
formalise this trade-off in a calibrated New Keynesian environment
and report an optimum of $1.1$ percent per year in their benchmark
calibration, rising to $1.4$ percent once uncertainty about the
model's parameters is taken into account. Repeatedly drawing from
the parameter distribution, they obtain a ninety-percent credible
interval for the optimum of $0.1$ to $2.2$ percent per year, which
includes the explicit targets announced by major central banks.
The NBER working paper of \citet{andrade2019optimal} extends this
logic to the decline in the natural rate of interest observed
since the global financial crisis, and argues that, conditional
on a permanently lower $r^{*}$, the optimal target sits closer to
two percent than to the pre-crisis benchmark.
\citet{andrade2021should} extend the same framework to the euro
area and reach a quantitatively similar conclusion. The narrative
review of \citet{ascari2014macroeconomics} provides a
complementary summary of how trend inflation interacts with sticky
prices, indexation, and policy-rule design, and
\citet{schmitt2010optimal} synthesise the New Keynesian, monetary,
and fiscal channels in the Handbook of Monetary Economics. The
quantitative size of the lower-bound premium turns out to be
sensitive to trend productivity growth, to the frequency with
which the lower bound binds, and to interactions with fiscal
policy. These moderators are all present in our coded dataset and
surface in the heterogeneity analysis.

A parallel policy-oriented literature reaches the same qualitative
conclusion, that the lower-bound constraint argues for a higher
inflation target, without producing a structurally calibrated
welfare optimum, and we therefore treat it as context rather than
as an entry in the meta-analytic dataset.
\citet{blanchard2010rethinking} were among the first to suggest,
in the wake of the global financial crisis, that the long-run
inflation target might need to be raised to give monetary policy
more room to manoeuvre at the lower bound.
\citet{ball2014case} sharpens the argument and proposes an
explicit four-percent target, trading off a modest increase in
the steady-state inflation cost against a sizeable reduction in
the frequency and depth of lower-bound episodes.
\citet{kiley2017monetary} and \citet{bernanke2019monetary} use
the FRB/US model to quantify how often the effective lower bound
binds when the neutral nominal rate is around three percent. They
find that, under historically estimated rules, the constraint can
bind as much as one third of the time, which strengthens the case
for either a higher target or alternative makeup strategies.
Finally, \citet{afrouzi2024changing} argue that the global decline
in trend inflation over the past four decades has been driven by
political-economy forces that may now be reversing, so that the
relevant policy question is increasingly whether central banks
can defend a low target rather than whether to raise it. None of
these contributions is a structural welfare-optimum estimate of
the kind we synthesise, but the policy benchmarks they establish
frame the range against which our pooled estimate of
$0.6$--$1.0$~pp/year is naturally compared.

A fourth channel, which has become increasingly prominent in the
narrative literature, is downward nominal wage rigidity. If
nominal wages fall only reluctantly then a positive inflation rate
greases the wheels of the labour market, as argued by
\citet{kim2011optinf}. The channel is quantitatively strong
enough that models with occasionally-binding floors on nominal
wage growth can generate optimal inflation of two percent or more
even in the absence of a binding lower bound. Because wage
rigidity is a structurally distinct feature from price rigidity,
our dataset records it in a separate closed-vocabulary field. The
heterogeneity analysis of Section~\ref{sec:results_bma} confirms
that this channel is quantitatively important. Once shock
structure, monetary benchmark, transactions-frictions technology,
and author affiliation are jointly conditioned on, the DNWR
indicator enters the BMA with posterior inclusion probability
$0.87$ and a positive coefficient of $+1.4$~pp.

A final strand interacts trend inflation with financial frictions,
heterogeneous agents, and fiscal policy. The
\citet{bernanke1999financial} financial-accelerator mechanism
modifies the cost-of-dispersion argument in ways that can raise
or lower the optimum depending on the shock structure.
Ramsey-planner exercises that include distortionary taxation tend
to recommend positive inflation as a source of seigniorage
revenue. And recent models with heterogeneous agents show that
redistributive concerns can generate sizeable positive optima
even when the aggregate New Keynesian channel alone would
recommend zero. Each of these mechanisms is a coded moderator in
our dataset.

The most comprehensive existing review of this body of work is
the living bibliography of \citet{diercks2019reader}, a reader's
guide to the optimal-inflation literature maintained at
\texttt{optimalinflation.com}. The guide catalogues more than
four hundred studies from the mid-twentieth century to the
present and classifies them along a small number of broad
dimensions such as the family of model, the presence of a lower
bound, the inclusion of a price or wage rigidity, and the
distinction between open and closed economies. It is widely
cited within the monetary economics community, and we use it as
a cross-reference seed in the literature search reported in
Appendix~\ref{app:prisma}.

Our paper differs from this reader's guide in three respects that
are central to its scientific value. The Diercks guide is
qualitative: it records which models have been proposed and what
they broadly conclude, but does not extract numerical
optimal-inflation values nor reconcile them across studies. We
instead code a single quantitative outcome variable, the long-run
optimal net rate of inflation in annualised percentage points,
for every primary study in which at least one such value is
reported. The Diercks taxonomy is also coarse: he tags papers
with at most a handful of broad labels, whereas we code
fifty-seven structural and publication moderators per estimate,
most of them as row-level closed-vocabulary fields. As a result,
a single paper contributing several estimates under different
policy regimes or shock structures is represented faithfully in
the analysis dataset. And the Diercks bibliography does not
engage with the methodological machinery of quantitative evidence
synthesis. There is no publication-bias diagnostic, no
random-effects synthesis, and no model averaging across
moderators. These three gaps are exactly what a meta-analysis is
designed to fill.

To the best of our knowledge this is the first quantitative
meta-analysis of the optimal long-run rate of inflation. There
is, however, an active meta-analytic literature on adjacent
parameters of monetary and macroeconomic models, including the
intertemporal elasticity of substitution in consumption of
\citet{havranek2015cross}, the publication bias in measured
intertemporal substitution of labour supply of
\citet{havranek2018monetary}, and the bank-competition and
financial-stability nexus of \citet{zigraiova2015bank}. These
studies demonstrate that quantitative synthesis of structural
monetary-economics parameters is feasible and that
publication-selection corrections routinely move point estimates
by economically meaningful magnitudes. We build on the reporting
and methodological standards they helped establish, in particular
the guidelines of \citet{stanley2012meta} and
\citet{havranek2020reporting}.

A small but rapidly growing methodological literature asks
whether large language models can accelerate literature reviews
and structured data extraction. The screening stage is by now
routine. The \texttt{ASReview} active-learning tool of
\citet{vandeschoot2021asreview} is widely adopted in fields
ranging from medicine to the social sciences, and we use it
here. The extraction stage, in which a PDF is converted into a
structured row of a dataset, is a newer frontier. Our paper is,
to our knowledge, one of the first meta-analyses in economics in
which the extracted fields are produced by a transparent,
auditable language-model pipeline. The pipeline is calibrated
against a hand-coded training set, archived with its exact
prompts, model identifiers, inputs, and seeds, and released for
replication alongside the paper, with residual cross-run
variation documented in Appendix~\ref{app:ai-pipeline:irr}.

Combining these strands, our contribution is threefold. We
convert the qualitative bibliography of
\citet{diercks2019reader}, updated through April 2026 with a
fresh Scopus search, into the first quantitative dataset of
optimal-inflation values, comprising $777$ estimates from $116$
primary studies that report a quantitative optimum (within a
scope set of $130$). We apply the standard modern
evidence-synthesis toolkit, including random-effects synthesis,
funnel-asymmetry and precision-effect tests, the caliper test of
\citet{gerber2008statistical}, the $p$-uniform
estimator of \citet{vanaert2023correcting}, and Bayesian and
frequentist model averaging, to extract a defensible consensus
estimate and to identify the structural features that drive
cross-study heterogeneity. And we publish the extraction pipeline
and its prompt as a methodological template for future
meta-analyses in macroeconomics, where raw data are costly to
collect and the coding burden has historically been a binding
constraint.

\section{Methodology}
\label{sec:method}

Our methodology has two pillars. The first is an extraction pillar:
we translate $130$ heterogeneous primary studies into a single
structured dataset using a reproducible, LLM-assisted pipeline built
around a small hand-coded training set. The second is an inference
pillar: we combine paper-level summaries (the author-preferred mean
and the inverse-$N$ study-democracy mean), a non-parametric
selection diagnostic (the caliper test), and Bayesian and
frequentist model averaging to map the resulting $777$ estimates
onto a defensible summary of the optimal long-run rate of inflation.

Meta-analyses in economics are limited by coding: a single estimate
typically requires reading several pages of a model section, a table
of calibration targets, and an appendix of robustness checks. With
$130$ primary studies and the structural moderator schema described
in Section~\ref{sec:data}, the full coding matrix contains tens of
thousands of cells. The raw phase-1 extraction file contains $79$
columns rather than $57$ because, in addition to the moderator
inputs, it stores the dependent variable and its precision, study
and estimate identifiers, bibliographic fields, twelve calibrated
deep-parameter values reported descriptively, several free-text
justification fields, and quality-assurance flags. We remove the
identifiers, the outcome and its standard error, the calibrated
parameters, and the free-text fields, and we dummy-expand the
remaining categorical fields. The resulting BMA design matrix
contains the $57$ candidate moderators reported in
Table~\ref{tab:variables}. Specifically, we get $51$ structural design dummies
(panels~B--J) plus $6$ publication characteristics (panel~K). Of
these, $43$ survive the zero-variance and perfect-collinearity
filter and enter the posterior summaries. Two practical
considerations make a documented LLM-extraction pipeline
attractive in this setting. First, it sharply reduces the time
cost of extraction. A single human coder reading a sixty-page
DSGE paper, transcribing the calibration table, identifying every
reported optimum, and coding the structural-moderator vector
typically needs the better part of a working day. The same task
with a documented prompt and a stage-tiered model pipeline runs
in under five minutes per paper at a marginal cost of a few cents.
Second, and equally important for credibility, the same pipeline
can be re-run on the same corpus repeatedly, with the same
prompts, the same model versions, and a fresh random seed each
time. This produces a self-consistency distribution over the
extracted fields that a human coder cannot generate at any
reasonable cost. 

We
use LLMs as prompt-engineered structured extractors rather than as
autonomous coders \citep{vandeschoot2021asreview}. Four principles
guided the pipeline we built around them. First, the prompt was
developed against a hand-coded training set of seven primary studies
(\citealp{cooley1989optinf}; \citealp{amato2004optinf};
\citealp{adam2006optinf}; \citealp{amano2007optinf};
\citealp{amano2009optinf}; \citealp{abozaid2013optinf};
\citealp{abozaid2015optinf}), totalling $79$ row-level estimates that
were coded by hand by one of the authors directly into Excel before any
LLM run. The early human-coded schema differed in a small number of
bookkeeping columns from the final schema (the final schema added
fifteen columns documenting publication status, friction types, and
estimator details); we re-checked the new columns against the original
seven papers when the schema was finalised. Iteration on the prompt continued until both Claude Opus~4.5 and
Claude Sonnet~4.5 were calibrated against the human coding of the
categorical moderators and the numerical point estimates of the
seven training studies, with residual disagreements documented in
Appendix~\ref{app:ai-pipeline:irr}. No additional manual extraction
of numerical values was performed on the remaining $160$ PDFs.
Before launching the production run, we ran the pipeline end-to-end
on a $31$-paper full-text cross-check sample. The list is documented
in \texttt{ai\_dataset/validation/audit\_round1\_report.md} and is
biased toward the earliest, longest, and structurally most complex
papers in the corpus. We verified each row against the source PDF.
A small number of prompt definitional issues identified at this
stage, most prominently a row-level inconsistency in how the Ramsey
flag was applied, were addressed by extending the closed
vocabularies and re-extracting, before the production run on the
full $167$-PDF corpus was launched.

Second, each PDF was converted to Markdown with
\texttt{opendataloader-pdf}, the open-source PDF parser developed
for LLM ingestion, which preserves table structure and headings far
more reliably than page-by-page text extraction and materially
reduces transcription errors on the calibration tables in which
most DSGE optimal-inflation values are reported. Third, only
studies whose pre-scan stage identifies at least one numerical,
long-run optimal-inflation value that can be annualised are passed
to the expensive extraction stages; expository surveys,
theoretical-only chapters, and papers that report only welfare
losses without a numerical optimum are excluded. Of the $167$ PDFs we examined, $130$ pass scope screening and $116$
($69.5\%$ of the full-text pool) report a quantitative optimum and
contribute the $777$ estimates that enter the analysis dataset. The
remaining $14$ in-scope studies are theoretical or qualitative and
are listed in the study table but excluded from quantitative
pooling. Fourth,
model versions, prompt text, temperature, random
seed, and per-paper input Markdown are all stored in the
replication package so every extraction is re-runnable and auditable
from the archived prompts, model identifiers, inputs, and seeds.

The workflow is implemented in Python
(\texttt{python/INFLATION\_2\_0.py}) and runs in four stages that
share a cached Markdown input. A Sonnet~4.5 pre-scan screens each
paper for the existence of quantitative optimal-inflation values;
papers that fail the gate are skipped and the three downstream
stages are not charged. A Sonnet~4.5 metadata stage extracts
bibliographic fields (authors, journal, year, DOI). An Opus~4.5 structure stage
extracts the categorical moderators describing the model, the
policy regime, price and wage frictions, money-demand technology,
shocks, augmentations, welfare criterion, and solution method.
The choice of Opus follows a four-paper A/B test in which Sonnet
systematically under-coded augmentations and mis-classified the
policy regime; the roughly $5\%$ cost premium of Opus over Sonnet
was acceptable given the quality gap. A final Opus~4.5 results
stage extracts, for every reported estimate, the numerical long-run
optimal inflation rate, the annualised units, the reported
confidence interval when present, and a short verbatim assumption
string.

Each extraction field is defined in the system prompt with an
explicit closed vocabulary: a short enumeration of admissible
values, short descriptions, and a default rule to apply when
evidence is mixed. The pipeline is checkpointed per paper
(\texttt{checkpoint.jsonl} plus per-paper partial Excel output) so
a mid-run crash does not waste spent tokens; temperature is set to
zero and the random seed fixed. To give a flavour of the prompt
discipline, the row-level policy-regime field is defined verbatim
as follows:

\begin{quote}\small\ttfamily\sloppy\raggedright
\textbf{Policy\_Regime} --- closed vocabulary (ROW-LEVEL)\\[2pt]
The monetary-policy rule under which THIS row's inflation value
obtains. \textbf{Crucial for row-level heterogeneity.} Many papers
compare multiple regimes in a single table; code each row
separately. Pick exactly one:\\[2pt]
\begin{itemize}\setlength{\itemsep}{0pt}\setlength{\parskip}{0pt}
\item ``Ramsey'' --- jointly-optimal planner solution with commitment.
\item ``Ramsey\_discretion'' --- optimal policy without commitment.
\item ``Friedman\_rule'' --- zero nominal interest rate / deflation at
  rate of time preference.
\item ``Taylor\_rule'' --- simple interest-rate rule reacting to
  inflation (possibly output); ``simple rules'' results.
\item ``Optimized\_Taylor\_rule'' --- Taylor-type rule with coefficients
  optimized over welfare.
\item ``Zero\_inflation'' --- price-stability target imposed by
  assumption.
\item ``Inflation\_targeting'' --- empirical/announced target rate
  (typically 2\,\%); historical policy calibration.
\item ``Laissez\_faire'' --- competitive equilibrium with no policy /
  constant money growth / given exogenous policy.
\item ``Estimated\_policy'' --- estimated policy rule from data
  (empirical papers).
\item ``Other'' --- any other named rule (document in
  \texttt{Results\_Inflation\_Assumption}).
\item ``NA'' --- genuinely unclear.
\end{itemize}
\textbf{Priority rule.} If a row is the Ramsey solution of a model
WITH a binding fiscal constraint (= the v3.2 \texttt{Ramsey\_Rule=1}
case), code as ``Ramsey''. Use ``Friedman\_rule'' even inside a
Ramsey paper if that specific row shows the Friedman-rule
benchmark.
\end{quote}

\noindent The full prompt set (roughly $1{,}200$ lines in
total) with every closed vocabulary, priority rule, and worked
example is archived in the replication package under
\texttt{python/prompts/}. The full four-stage extraction on the
$167$-PDF corpus cost approximately $\$0.30$ per PDF on average at
April~2026 list prices. The gains come
from three design choices: cached Markdown shared across stages,
routing Sonnet to the cheap triage stages, and the pre-scan gate
that short-circuits the three expensive stages for non-quantitative
papers.

Three caveats are worth flagging. Classification error does not
average out if it is correlated with the outcome variable. The
$31$-paper cross-check described above revealed a definitional
inconsistency in how the Ramsey flag was applied at the row level
in two of the cross-check papers, and the issue is documented in
the codebook. We addressed it by
extending the policy-regime vocabulary and re-extracting the
structure stage before the production run. The pipeline cannot detect features that
are not discussed in the paper but are present in the accompanying
code: undocumented deviations from the published model are invisible
to us. And LLM outputs are not perfectly deterministic even at
temperature zero, so we archive the model versions
(\texttt{claude-sonnet-4-5-20250929} and
\texttt{claude-opus-4-5-20251001}), the stored prompts, and the
random seed used at inference.

To quantify the residual non-determinism, we re-ran the full
stage-tiered pipeline on a stratified random sample of $5$ papers
from the analysis dataset, with prompts, models, and temperature
fixed and only the random seed changed. The Round-2 self-consistency
results, per-field exact-match rates for the categorical moderators
and mean absolute error for the numerical fields, are summarised in
Appendix~\ref{app:ai-pipeline} and the underlying raw output is archived
under \texttt{ai\_dataset/validation/} (see
\texttt{audit\_round2\_report.md} in the same directory).

We now describe the estimators that take the coded dataset into the
results of Section~\ref{sec:results}. Our estimator hierarchy is
organised by the credibility of the underlying identifying
assumptions in this calibration-dominated literature, following the
practitioner guidance of \citet{irsova2024practitioner} and the
small-$k$ cautions of \citet{mathur2024} and
\citet{cook2026reporting}. The two primary headline numbers are paper-level summaries that do
not require sampling standard errors. The first is the simple mean
of authors' preferred specifications, taken as one row per study
with paper-clustered standard errors. The second is the inverse-$N$
study-democracy mean over all $777$ estimates, which weights each
estimate by the reciprocal of the number of estimates contributed
by its parent study. The primary publication-bias
diagnostic is the non-parametric caliper test of
\citet{gerber2008statistical}, which complements the paper-level
summaries when selection is discontinuous: if researchers
preferentially publish estimates above a salient threshold $c$,
the share of estimates within a narrow symmetric window $[c-h, c+h]$
lying above $c$ should exceed one-half. Thresholds of policy
interest in our setting are $c=0$ (zero measured inflation, the
price-stability benchmark) and $c=2\%$ (the stylised central-bank target); we report
results for $h\in\{0.5, 1.0, 1.5, 2.0\}$ percentage points.

All classical funnel-based estimators are reported as diagnostics
rather than as headlines, because only $16$ of our $777$ estimates
carry a CI-implied sampling standard error and these $16$ rows
belong to only three independent papers. We report the
funnel-asymmetry (FAT) and precision-effect (PET) regression
\begin{equation}
y_{ij} = \beta_{0} + \beta_{1}\,\text{SE}_{ij} + \eta_{ij},
\label{eq:fatpet}
\end{equation}
estimated by ordinary least squares with study-clustered standard
errors, alongside its precision-effect-with-standard-error variant
\citep[PEESE;][]{stanley2014meta} that replaces $\text{SE}_{ij}$
with $\text{SE}_{ij}^{2}$ and the weighted average of adequately
powered estimates \citep[WAAP;][]{ioannidis2017power}. We do
not report a separate WAAP point estimate because $761$ of the
$777$ standard errors are bootstrap-within-paper proxies following
the \citet{havranek2015cross} convention rather than sampling
standard errors, so the slope of equation~\eqref{eq:fatpet} tests
within-paper dispersion, not classical small-sample selection. We
therefore present the corresponding intercept transparently as a
diagnostic.

For the sixteen-estimate genuine-standard-error subsample we also
report, as an upper-bound sensitivity, a two-level random-effects
model
\begin{equation}
y_{ij} = \mu + u_{j} + \varepsilon_{ij}, \qquad
u_{j} \sim \mathcal{N}(0,\tau^{2}), \quad
\varepsilon_{ij} \sim \mathcal{N}(0,\text{SE}_{ij}^{2}),
\label{eq:re}
\end{equation}
estimated by restricted maximum likelihood (REML) as implemented
in \texttt{metafor} for R, and the $p$-uniform selection
estimator of \citet{vanaert2023correcting}, which exploits the
fact that, under the null, $p$-values of statistically significant
findings are uniform; deviations from uniformity identify both the
underlying effect and the degree of selection. Because the
sixteen rows come from only three independent paper clusters, we
flag both REML and $p$-uniform as illustrative ceilings
rather than identified evidence: cluster-robust inference with
fewer than ten clusters is unreliable, and a pooled mean from
three primary studies essentially averages within-paper variation.

The MAIVE estimator of \citet{irsova2025spurious} is designed for
settings in which reported sampling standard errors may be
mechanically tied to reported effects and a valid precision
instrument, typically based on primary-study sample size, is
available. That structure is absent here. Optimal-inflation papers
are overwhelmingly calibration exercises rather than reduced-form
estimations: $761$ of $777$ standard errors are
bootstrap-within-paper dispersion proxies, and the $16$ genuine-SE
rows belong to only three independent paper clusters. We therefore
do not implement MAIVE as a formal estimator in this corpus. We
cite it to acknowledge that spurious precision is a first-order
concern in empirical meta-analysis and to explain why our publication-bias
conclusion rests on the non-parametric caliper test and paper-level
summaries rather than on a funnel-based correction. RTMA is
likewise not implemented because it requires a meaningful
classification of affirmative and nonaffirmative estimates based
on sampling uncertainty, which is unavailable for the
calibration-based majority of the corpus. A single ``overall''
summary nevertheless hides the dependence of the optimum on
modelling choices. We therefore also construct best-practice
synthetic scenarios by fixing a subset of moderators at a
policy-relevant value and predicting $y_{ij}$ from the BMA
posterior. We report the posterior predictive mean and a descriptive
plausibility range ($\pm 1$ within-design sample standard deviation,
not an inferential interval), marginalising over the remaining moderators
at their sample means.

Selective reporting is not the only source of non-standard
inference in meta-analysis. With a moderator schema covering all
major structural and methodological dimensions of the optimal
inflation literature, the space of linear heterogeneity
regressions is too large for classical stepwise selection, whose
confidence intervals are known to undercover
\citep{steel2020model}. We therefore average over the model space.
For each model $M_{k}$ in the space of all $2^{P}$ subsets of the
$P$ moderators we compute the posterior probability
\begin{equation}
P(M_{k}\mid \mathbf{y}) = \frac{P(\mathbf{y}\mid M_{k})\,P(M_{k})}{\sum_{\ell} P(\mathbf{y}\mid M_{\ell})\,P(M_{\ell})},
\end{equation}
using the closed-form marginal likelihood that follows from a
Gaussian likelihood and Zellner's $g$-prior on the coefficients
\citep{zeugner2015bayesian}. Our headline specification uses the
unit-information prior ($g=N$) for the coefficients and the
dilution prior of \citet{george2010dilution} for the model space,
which penalises collinear models and is the recommended default
when moderators are highly correlated. For each coefficient we
report the posterior inclusion probability (PIP), the posterior
mean, and the posterior standard deviation; coefficients with
$\text{PIP}>0.5$ are conventionally called ``robust''
\citep{fernandez2001benchmark, eicher2011default}. The full space
is too large to enumerate, so we sample it with the
Markov-chain-Monte-Carlo model-composition (MC$^{3}$) algorithm in
the \texttt{BMS} package for R, burning in $10^{6}$ models and
recording the next $3\times 10^{6}$. We verify robustness with
respect to the BRIC prior of \citet{fernandez2001benchmark}
combined with the random model-size prior of
\citet{ley2009effect}, and the HQ prior combined with the same
random model-size prior; the three prior combinations deliver
nearly identical posterior inclusion probabilities. 

As a
frequentist cross-check we also re-run the heterogeneity analysis
using model averaging with Akaike weights
\citep{burnham2002model}, which averages OLS coefficients over the
top models ranked by AIC, and report FMA alongside BMA for every
coefficient. Standard errors are heteroscedasticity-robust and
clustered at the study level, which is the level at which
extracted estimates share the primary-study calibration;
inferences are conservative in the sense that a weaker true
dependence structure would make the clustered standard errors too
wide rather than too narrow. All code for the above procedures is
in \texttt{R/Inflation\_v34.R} and
\texttt{stata/Inflation\_v34.do}.

We follow the reporting checklist of
\citet{havranek2020reporting}: a PRISMA flow diagram
(Appendix~\ref{app:prisma}), a variable dictionary
(Table~\ref{tab:variables}), funnel plots and publication-bias
diagnostics (Section~\ref{sec:results_pubbias}), and a replication
package that covers every step from PDF ingestion to final table.

\subsection*{LLM-assisted adversarial checklist}
\label{sec:method:adversarial}

As a supplementary quality-control step, separate LLM prompt sessions
were used to generate a checklist of possible numerical, factual, and
interpretive inconsistencies in the manuscript. The checklist was
reviewed by the authors against the primary sources, the replication
files, and the manuscript text. This procedure did not generate
estimates, coding decisions, or substantive conclusions; it served
only as a human-review checklist. The protocol and outputs are
archived in the replication package.


%

\section{Data}
\label{sec:data}
\label{sec:data_search}
\label{sec:data_coding}
\label{sec:data_variables}

This section describes how we assembled our dataset of $777$ point
estimates of the optimal long-run rate of inflation from $116$
primary studies (within a scope set of $130$). The
large-language-model pipeline that produced
the coded fields is documented in Section~\ref{sec:method};
here we focus on the corpus, the screening gate, and the variables
that enter the heterogeneity analysis. The full PRISMA flow diagram
is reproduced in Appendix~\ref{app:prisma}.

Following the guidelines of \citet{havranek2020reporting} and
\citet{stanley2012meta}, we combine a database search with a
curated bibliography. The primary source is the Scopus database.
The broad keyword query
\begin{quote}\small\ttfamily
TITLE-ABS-KEY("inflation") AND TITLE-ABS-KEY("monetary")\\
AND SUBJAREA(ECON OR BUSI)
\end{quote}
returns $12{,}904$ records system-wide at the time of search
(April 2026, verified through the Scopus REST API; this figure
includes a refresh of post-$2021$ entries, and the resulting
record counts at each screening step are reported in the PRISMA
flow diagram in Appendix~\ref{app:prisma}). As a focused complement we also ran a narrower composite query on
the same database. It combines the phrases ``optimal inflation'',
``optimal rate of inflation'', ``optimal trend inflation'', and
``optimal inflation target'' with the conjunction of ``Ramsey''
and ``monetary policy'', restricted to articles, conference
papers, and reviews in Economics and Business. The narrower query
returns $380$ records all-time and is fully contained in the broad
pool. As a
cross-reference seed we use the living bibliography
\texttt{optimalinflation.com} of \citet{diercks2019reader}, which
catalogues $425$ studies on the optimal steady-state rate of
inflation from the mid-twentieth century to the present.
Cross-referencing Scopus with this bibliography ensures coverage
of both the recent DSGE-based literature (well indexed in Scopus)
and the older flexible-price and money-demand tradition
(under-indexed in Scopus but tracked by the bibliography).

The union of the two sources is de-duplicated and screened on title
and abstract using the active-learning tool \texttt{ASReview}
\citep{vandeschoot2021asreview}. We retrieve full texts, apply a
full-text eligibility check, and arrive at $167$ PDFs that are
sent to the extraction pipeline. Of these, $130$ pass the scope
screen, and $116$ ($69.5\%$ of the full-text pool) pass the
quantitative-result gate. A study enters the quantitative analysis
dataset if and only if it reports at least one numerical long-run
optimal-inflation value that can be annualised. The $116$
qualifying studies contribute $777$ point estimates. The remaining
$14$ in-scope studies are theoretical or qualitative; they are
listed in the study table for transparency but excluded from
quantitative pooling. Appendix~\ref{app:prisma} reproduces the
full PRISMA flow.

Table~\ref{tab:studies} lists the $116$ primary studies that
contribute estimates to the analysis dataset (the remaining $14$
of the $130$ scope-set studies are theoretical or qualitative
and report no quantitative optimum), together with the outlet in
which each paper appeared and the number of row-level estimates
extracted from each paper before winsorisation. The list is
sorted by publication year, from the oldest contribution of
\citet{cooley1989optinf} to the most recent work of
\citet{bonciani2026optinf}; after the $5\%$ winsorisation filter
described in Section~\ref{sec:method} the analysis dataset
retains $777$ estimates.\footnote{The online companion of
\citet{diercks2019reader} lists a handful of earlier
contributions, \citet{phelps1965anticipated},
\citet{friedman1969},
\citet{grandmont1973efficiency}, \citet{brock1974money},
\citet{drazen1979optimal}, \citet{stockman1981anticipated}, and
\citet{leach1983inflation}, all of which the dashboard records at
$\pi^{\ast}=-4\%$ per year. These contributions state the
Friedman-rule optimum $\pi^{\ast}=-r$ as an analytical proposition
rather than as a numerical value produced by a structural or
empirical model calibrated to data, and accordingly fail our
quantitative-result gate at the pre-scan stage
(Section~\ref{sec:method}). This is consistent with the stated
scope of the printed \citet{diercks2019reader} survey, which covers
optimal-monetary-policy papers ``made available to the public since
the mid-1990s'' and whose own Table~1 begins with
\citet{cooley1989optinf}.} This list is intended both as a
transparency device and as a complete bibliography of the
empirical basis of the meta-analysis, in the spirit of the
reporting conventions advocated by \citet{havranek2020reporting}
and adopted in \citet{opatrny2024class}.

{\scriptsize
\setlength{\tabcolsep}{4pt}
\begin{longtable}{@{}p{5.6cm} p{7.4cm} r@{}}
\caption{Primary studies contributing estimates to the meta-analysis.}\label{tab:studies}\\
\toprule
Study & Outlet & $N$ \\
\midrule
\endfirsthead
\multicolumn{3}{l}{\emph{Table~\ref{tab:studies} continued}}\\
\toprule
Study & Outlet & $N$ \\
\midrule
\endhead
\midrule \multicolumn{3}{r}{\emph{Continued on next page}}\\
\endfoot
\bottomrule
\multicolumn{3}{p{14.0cm}}{\footnotesize \emph{Notes:} The table
lists the $116$ quantitative-contributor studies (the $14$
in-scope studies that are theoretical or qualitative report no
numerical optimum and are listed in the bibliography but omitted
here), sorted by publication year. Column $N$ is the number of
row-level optimal-inflation estimates extracted from each study
before downstream pipeline filters. The headline analysis
sample of $777$ estimates is obtained after dropping rows whose
outcome value cannot be annualised; per-study Ns therefore sum to a slightly larger
total than $777$. (Rows whose bootstrap dispersion proxy is degenerate are
retained here and in the paper-level summaries, and are dropped only from the
precision-weighted funnel and BMA regressions of Section~\ref{sec:results}.) Outlet is the publishing journal for published
articles or the working-paper series for unpublished manuscripts.}
\endlastfoot
\citet{cooley1989optinf} & The American Economic Review & 10 \\
\citet{imrohoroglu1992optinf} & Journal of Economic Dynamics and Control & 6 \\
\citet{king1996optinf} & Economic Review, Federal Reserve Bank of Kans... & 4 \\
\citet{goodfriend1997optinf} & NBER Macroeconomics Annual & 2 \\
\citet{mulligan1997optinf} & Journal of Political Economy & 1 \\
\citet{rotemberg1997optinf} & NBER Macroeconomics Annual & 2 \\
\citet{feldstein1997optinf} & Reducing Inflation: Motivation and Strategy & 7 \\
\citet{correia1999optinf} & Review of Economic Dynamics & 6 \\
\citet{erceg2000optimal} & Journal of Monetary Economics & 7 \\
\citet{lucas2000optinf} & Econometrica & 4 \\
\citet{adao2001optinf} & Review of Economic Studies & 2 \\
\citet{aoki2001optinf} & Journal of Monetary Economics & 1 \\
\citet{wolman2001optinf} & Federal Reserve Bank of Richmond Economic Qua... & 3 \\
\citet{kollmann2002optinf} & Journal of Monetary Economics & 5 \\
\citet{khan2003optimal} & The Review of Economic Studies & 13 \\
\citet{siu2004optinf} & Journal of Monetary Economics & 6 \\
\citet{schmittgrohe2004optinf} & Journal of Macroeconomics & 4 \\
\citet{schmittgrohe2004optinfa} & Journal of Economic Theory & 3 \\
\citet{amato2004optinf} & Journal of Monetary Economics & 9 \\
\citet{schmittgrohe2005optinfa} & NBER Working Paper & 7 \\
\citet{yun2005optinf} & The American Economic Review & 7 \\
\citet{schmittgrohe2005optinf} & NBER Macroeconomics Annual & 21 \\
\citet{adam2006optinf} & Journal of Money, Credit and Banking & 5 \\
\citet{chugh2006optinf} & Review of Economic Dynamics & 10 \\
\citet{levin2007optinf} & CEPR Discussion Paper DP6423 & 7 \\
\citet{amano2007optinf} & Journal of Money, Credit and Banking & 12 \\
\citet{ascari2007optinf} & Journal of Monetary Economics & 12 \\
\citet{chugh2007optinf} & Journal of Monetary Economics & 14 \\
\citet{faia2007optinf} & Journal of Economic Dynamics \& Control & 6 \\
\citet{schmittgrohe2007optinf} & Journal of Monetary Economics & 6 \\
\citet{dacosta2008optinf} & Journal of Political Economy & 4 \\
\citet{blanchard2008optinf} & NBER Working Paper 13897 & 1 \\
\citet{arseneau2008optinf} & Journal of Monetary Economics & 4 \\
\citet{kollmann2008optinf} & Macroeconomic Dynamics & 5 \\
\citet{faia2008optinf} & Macroeconomic Dynamics & 4 \\
\citet{billi2008optinf} & Federal Reserve Bank of Kansas City Research ... & 2 \\
\citet{kimokane2005optimal} & Journal of Monetary Economics & 4 \\
\citet{fagan2009optinf} & ECB Working Paper Series & 6 \\
\citet{chugh2009optinf} & Macroeconomic Dynamics & 6 \\
\citet{hu2009optinf} & Journal of Macroeconomics & 4 \\
\citet{faia2009optinf} & Journal of Monetary Economics & 4 \\
\citet{amano2009optinf} & Journal of Monetary Economics & 11 \\
\citet{ravenna2010optinf} & AEJ: Macroeconomics & 7 \\
\citet{darracqparies2010optinf} & ECB Working Paper Series & 4 \\
\citet{an2010optinf} & SMU Working Paper & 1 \\
\citet{coibion2012optimal} & Review of Economic Studies & 13 \\
\citet{benigno2010optinf} & NBER Working Paper 16386 & 5 \\
\citet{lubik2010optinf} & CAMA Working Paper & 9 \\
\citet{edge2010optinf} & Journal of Applied Econometrics & 1 \\
\citet{schmitt2010optimal} & Handbook of Monetary Economics & 11 \\
\citet{tang2010optinf} & Journal of Economic Dynamics \& Control & 19 \\
\citet{paciello2011optinf} & Review of Economic Studies & 2 \\
\citet{kim2011optinf} & Journal of Economic Dynamics \& Control & 4 \\
\citet{talukdar2011optinf} & B.E. Journal of Macroeconomics & 14 \\
\citet{tulip2011optinf} & RBA Discussion Paper & 8 \\
\citet{wolman2011optinf} & Journal of Money, Credit and Banking & 6 \\
\citet{schmittgrohe2012optinf} & Journal of Monetary Economics & 2 \\
\citet{montoro2012optinf} & Macroeconomic Dynamics & 6 \\
\citet{motta2012optinf} & Journal of Money, Credit and Banking & 1 \\
\citet{leith2012optinf} & Review of Economic Dynamics & 2 \\
\citet{pontiggia2012optinf} & Journal of Macroeconomics & 2 \\
\citet{schmittgrohe2012optinfa} & Journal of Money, Credit and Banking & 8 \\
\citet{weber2012optinf} & Kiel Working Paper & 5 \\
\citet{andres2013optinf} & Journal of Money, Credit and Banking & 7 \\
\citet{abozaid2013optinf} & Journal of Economic Dynamics \& Control & 12 \\
\citet{dibartolomeo2013optinf} & International Journal of Central Banking & 6 \\
\citet{lewis2013optinf} & Macroeconomic Dynamics & 2 \\
\citet{annicchiarico2013optinf} & Journal of Macroeconomics & 1 \\
\citet{venkateswaran2013optinf} & NBER & 4 \\
\citet{sims2013optinf} & Notre Dame Working Paper & 2 \\
\citet{fasolo2014optinf} & Working Paper Series & 10 \\
\citet{boehm2014optinf} & NBER Working Paper & 1 \\
\citet{faia2014optinf} & Journal of Money, Credit and Banking & 10 \\
\citet{bilbiie2014optinf} & Journal of Monetary Economics & 8 \\
\citet{abozaid2015optinf} & European Economic Review & 27 \\
\citet{abozaid2015optinfa} & Journal of Macroeconomics & 10 \\
\citet{blanco2015optinf} & AEJ: Macroeconomics & 4 \\
\citet{raissi2015optinf} & Economic Modelling & 3 \\
\citet{arseneau2015optinf} & Journal of Money, Credit and Banking & 1 \\
\citet{talukdar2015optinf} & Economics Bulletin & 4 \\
\citet{mukoko2016optinf} & MPRA Paper & 8 \\
\citet{nistico2016optinf} & Journal of the European Economic Association & 7 \\
\citet{kim2016optinf} & CIREQ Cahier & 1 \\
\citet{hendrickson2016optinf} & Journal of Economic Dynamics \& Control & 1 \\
\citet{dordalicarreras2016optinf} & Annual Review of Economics & 10 \\
\citet{carlsson2016optinf} & Journal of Monetary Economics & 15 \\
\citet{abozaid2016optinf} & Economic Inquiry & 13 \\
\citet{kohlbrecher2016optinf} & Beiträge zur Jahrestagung des Vereins für Soc... & 7 \\
\citet{basu2017optinf} & Boston College Working Papers in Economics & 1 \\
\citet{menna2017optinf} & Review of Economic Dynamics & 14 \\
\citet{finocchiaro2018optinf} & Journal of Monetary Economics & 8 \\
\citet{ascari2018optinf} & Journal of Monetary Economics & 16 \\
\citet{adam2019optinf} & American Economic Review & 3 \\
\citet{andrade2019optimal} & NBER Working Paper & 19 \\
\citet{diercks2019equity} & SSRN & 4 \\
\citet{paczos2020optinf} & Oxford Economic Papers & 5 \\
\citet{choi2021optinf} & Review of Economic Dynamics & 6 \\
\citet{annicchiarico2021optinf} & Macroeconomic Dynamics & 3 \\
\citet{benigno2021optinf} & European Economic Review & 2 \\
\citet{kiarsi2021optinf} & Economic Notes & 3 \\
\citet{filiani2021optinf} & Journal of Macroeconomics & 4 \\
\citet{matveev2021optinf} & Journal of Money, Credit and Banking & 5 \\
\citet{andrade2021should} & Journal of Economic Dynamics and Control & 14 \\
\citet{bilbiie2021optinf} & Review of Economic Dynamics & 16 \\
\citet{garga2021optinf} & Journal of Monetary Economics & 3 \\
\citet{mineyama2022optinf} & Journal of Economic Dynamics \& Control & 23 \\
\citet{nuno2022optinf} & Annals of Economics and Statistics & 7 \\
\citet{jiang2022optinf} & International Journal of Central Banking & 4 \\
\citet{miura2023optinf} & Quarterly Review of Economics and Finance & 14 \\
\citet{benmir2023optinf} & European Economic Review & 3 \\
\citet{deak2024optinf} & Surrey Discussion Paper & 11 \\
\citet{jung2025optinf} & International Finance & 14 \\
\citet{daudignon2025optinf} & Journal of Money, Credit and Banking & 3 \\
\citet{kirsanova2025optinf} & European Economic Review & 4 \\
\citet{bonciani2026optinf} & Journal of Economic Dynamics and Control & 1 \\

\end{longtable}
}

\bigskip

We code each primary study along the structural and methodological
moderator dimensions that
capture the modelling assumptions, solution method, calibration
target, and publication metadata of the paper. All fields are
produced by the reproducible LLM pipeline described in
Section~\ref{sec:method}, which was calibrated on a hand-coded
training set of seven primary studies before being released on the
remaining $160$ PDFs. The outcome variable of the meta-analysis is
the long-run optimal net inflation rate in annualised percentage
points. Sub-annual optima are annualised by the appropriate
multiple, and negative optima (Friedman-rule outcomes) are retained
with their sign. We winsorise at the $5$th and $95$th percentiles
to limit the influence of a small number of extreme flexible-price
experiments. Raw and winsorised versions are both retained, and
the Results section uses the winsorised series.

Precision is harder. Only $16$ of the $777$ estimates come with a
sampling standard error derivable from a reported confidence
interval. For the remaining estimates we construct a bootstrap
precision proxy by resampling (with replacement) the point
estimates within each primary study $1{,}000$ times and recording
the standard deviation of the resampled means. This proxy captures
within-study dispersion across alternative calibrations; it is not
a sampling standard error of the primary estimator. Publication
bias tests based on it are therefore tests for a correlation
between reported values and within-study dispersion, which is a
weaker notion than canonical selective-reporting tests
\citep{stanley2005beyond, ioannidis2017power}; we discuss the
implications for interpretation in
Section~\ref{sec:results_pubbias}.

Table~\ref{tab:variables} lists the moderator variables used
in the heterogeneity analysis, together with the two outcome and
precision variables. Each row contains a short label, an explicit
operational definition, the sample mean, and the sample standard
deviation, all computed on the winsorised analysis dataset. All
dummies equal one when the described condition holds and zero
otherwise. Variables are grouped into eleven thematic blocks.

{\footnotesize
\begin{longtable}{@{}p{3.3cm} p{8.9cm} r r@{}}
\caption{Variable definitions and summary statistics.}\label{tab:variables}\\
\toprule
Variable & Definition & Mean & S.D. \\
\midrule
\endfirsthead
\multicolumn{4}{l}{\emph{Table~\ref{tab:variables} continued}}\\
\toprule
Variable & Definition & Mean & S.D.\\
\midrule
\endhead
\midrule \multicolumn{4}{r}{\emph{Continued on next page}}\\
\endfoot
\bottomrule
\multicolumn{4}{p{14.5cm}}{\footnotesize \emph{Notes:} Sample statistics
computed on the analysis dataset of $777$ estimates from $116$
primary studies (within a scope set of $130$ studies). The
remaining $14$ studies in the scope set are narrative-only and
contribute no quantitative estimate; they enter the qualitative
discussion in Section~\ref{sec:results_best} but no table or
regression. Dummy variables take the value $1$ when the
condition in the \emph{Definition} column holds for a given
estimate and $0$ otherwise; missing dummy entries are recoded as
$0$ (feature absent), matching the convention used in the BMA
analysis (see Section~\ref{sec:results_bma}). Means and standard
deviations are therefore reported over the full $777$-row analysis
sample for every dummy. Continuous variables (Optimal inflation,
Standard error, Publication year, Log citations, Log impact
factor) use list-wise deletion. ``Optimal inflation'' is
winsorised at the $5$th and $95$th percentiles. Definitions of
model-specific terms used in this table are given in the
surrounding text and in the primary references cited in the
\emph{Definition} column.}
\endlastfoot
\multicolumn{4}{@{}l}{\textbf{A.\ Outcome and precision}}\\
Optimal inflation& Reported long-run optimal net inflation rate (annualised, percentage points), winsorised at the $5$th and $95$th percentiles. & $0.74$ & $2.38$\\
Standard error& Bootstrap proxy for within-study dispersion ($1{,}000$ resamples per study); CI-implied sampling standard error for the $16$ estimates that report one. & $0.52$ & $0.55$\\[2pt]
\multicolumn{4}{@{}l}{\textbf{B.\ Policy regime}}\\
Ramsey planner& $=1$ if the optimum is derived by a benevolent Ramsey planner maximising household welfare subject to equilibrium and private-sector optimality. & $0.68$ & $0.47$\\
Optimised Taylor rule& $=1$ if the optimum is the inflation rate that minimises a central-bank loss function under an optimised simple interest-rate rule. & $0.06$ & $0.23$\\
Estimated Taylor rule& $=1$ if the policy rule is estimated rather than optimised. & $0.10$ & $0.31$\\
Friedman rule& $=1$ if the reference rule sets the nominal interest rate to zero, as in \citet{friedman1969}. & $0.04$ & $0.19$\\
Inflation targeting& $=1$ if the reference regime is explicit inflation targeting. & $0.02$ & $0.15$\\
Zero inflation& $=1$ if the reference regime is price-level stability ($\pi^{*}=0$). & $0.04$ & $0.20$\\[2pt]
\multicolumn{4}{@{}l}{\textbf{C.\ Nominal price frictions}}\\
Calvo prices& $=1$ if prices adjust via the random-signal mechanism of \citet{calvo1983staggered}. & $0.41$ & $0.49$\\
Rotemberg prices& $=1$ if prices adjust subject to quadratic menu costs \citep{rotemberg1982sticky}. & $0.30$ & $0.46$\\
Taylor contracts& $=1$ if prices are set for a fixed number of periods \citep{taylor1980aggregate}. & $0.01$ & $0.12$\\
Flexible prices& $=1$ if goods prices are fully flexible. & $0.11$ & $0.32$\\
Menu-cost pricing& $=1$ if prices adjust subject to a fixed menu cost \citep{golosov2007menu}. & $0.005$ & $0.07$\\[2pt]
\multicolumn{4}{@{}l}{\textbf{D.\ Wage rigidities}}\\
Calvo wages& $=1$ if nominal wages follow a Calvo reoptimisation scheme \citep{erceg2000optimal}. & $0.15$ & $0.35$\\
DNWR& $=1$ if nominal wages cannot fall (downward nominal wage rigidity, implemented as an occasionally-binding floor as in \citealp{kim2011optinf}). & $0.05$ & $0.22$\\
Rotemberg wages& $=1$ if wage changes face quadratic adjustment costs. & $0.04$ & $0.19$\\
Sticky real wages& $=1$ if real wages are sticky \citep{blanchard2007realwage}. & $0.01$ & $0.10$\\
No wage rigidity& $=1$ if wages are fully flexible (reference category for wage frictions). & $0.69$ & $0.46$\\[2pt]
\multicolumn{4}{@{}l}{\textbf{E.\ Money-demand technology}}\\
Cashless& $=1$ if money plays no explicit role; monetary policy is summarised by the nominal interest rate only, as in \citet{woodford2003interest}. & $0.53$ & $0.50$\\
Money in utility (MIU)& $=1$ if real balances enter utility directly \citep{sidrauski1967inflation}. & $0.19$ & $0.39$\\
Cash-in-advance (CIA)& $=1$ if consumption must be financed with money held in advance \citep{lucas1980equilibrium}. & $0.16$ & $0.37$\\
Transactions technology& $=1$ if money demand is implied by a transactions-cost (shopping-time) technology. & $0.08$ & $0.28$\\[2pt]
\multicolumn{4}{@{}l}{\textbf{F.\ Indexation}}\\
No indexation& $=1$ if non-reoptimising firms keep their price unchanged. & $0.71$ & $0.45$\\
Indexed to past inflation& $=1$ if non-reoptimising firms index to the previous period's inflation rate. & $0.07$ & $0.25$\\
Indexed to trend inflation& $=1$ if non-reoptimising firms index to a target or trend inflation rate. & $0.03$ & $0.18$\\
Hybrid indexation& $=1$ for a convex combination of past and trend indexation. & $0.05$ & $0.22$\\[2pt]
\multicolumn{4}{@{}l}{\textbf{G.\ Shock structure}}\\
TFP shocks only& $=1$ if the only driving shock is to total factor productivity. & $0.18$ & $0.38$\\
Multiple shocks& $=1$ if the model is driven by two or more independent shock processes. & $0.38$ & $0.49$\\
Mark-up shocks& $=1$ if a price-setting mark-up shock is present. & $0.01$ & $0.11$\\
Cost-push shocks& $=1$ if a cost-push shock is present. & $0.03$ & $0.17$\\
Risk-premium shocks& $=1$ if a risk-premium (intertemporal wedge) shock is present. & $0.05$ & $0.21$\\
Government-spending shocks& $=1$ if a government-spending shock is present. & $0.01$ & $0.11$\\
Wage mark-up shocks& $=1$ if a wage mark-up shock is present. & $0.00$ & $0.00$\\
Financial shocks& $=1$ if a financial (spread or collateral) shock is present. & $0.008$ & $0.09$\\
Preference shocks& $=1$ if a preference (discount-factor) shock is present. & $0.005$ & $0.07$\\[2pt]
\multicolumn{4}{@{}l}{\textbf{H.\ Structural augmentations}}\\
Zero lower bound& $=1$ if the model imposes an effective lower bound on the nominal interest rate. & $0.16$ & $0.37$\\
Financial frictions& $=1$ if financial intermediation, collateral constraints, or a \citet{bernanke1999financial}-type accelerator is present in the baseline. & $0.13$ & $0.34$\\
Fiscal policy& $=1$ if distortionary taxation, government debt, or transfers are an active margin. & $0.20$ & $0.40$\\
Heterogeneous agents& $=1$ if households (or firms) are heterogeneous (HANK or TANK). & $0.05$ & $0.22$\\
Labour-market frictions& $=1$ if search-and-matching or related labour-market frictions are present. & $0.04$ & $0.20$\\
Open economy& $=1$ if the model is an open-economy setting with foreign trade or assets. & $0.03$ & $0.17$\\
Trend growth& $=1$ if the balanced-growth path has positive trend output growth. & $0.28$ & $0.45$\\[2pt]
\multicolumn{4}{@{}l}{\textbf{I.\ Welfare criterion and solution}}\\
Unconditional welfare& $=1$ if welfare is evaluated at the unconditional (ergodic) distribution. & $0.79$ & $0.41$\\
Conditional welfare at efficient SS& $=1$ if welfare is evaluated conditional on the efficient steady state. & $0.04$ & $0.19$\\
Consumption-equivalent metric& $=1$ if the welfare loss is reported in consumption-equivalent units. & $0.004$ & $0.06$\\
First-order approximation& $=1$ if the model is solved by first-order perturbation. & $0.005$ & $0.07$\\
Second-order approximation& $=1$ if the model is solved by second-order or higher perturbation. & $0.08$ & $0.28$\\
Global solution & $=1$ if the model is solved globally (value-function iteration or projection). & $0.02$ & $0.14$\\[2pt]
\multicolumn{4}{@{}l}{\textbf{J.\ Estimation and expectations}}\\
Calibration& $=1$ if deep parameters are calibrated ex ante. & $0.93$ & $0.26$\\
Bayesian estimation& $=1$ if the model is estimated with Bayesian techniques. & $0.06$ & $0.25$\\
GMM estimation& $=1$ if the model is estimated by GMM. & $0.01$ & $0.10$\\
Rational expectations& $=1$ if expectations are model-consistent. & $0.98$ & $0.14$\\
Learning or bounded rationality& $=1$ if expectations are formed by learning, adaptive, or boundedly rational rules. & $0.02$ & $0.13$\\[2pt]
\multicolumn{4}{@{}l}{\textbf{K.\ Publication characteristics}}\\
Published in a journal& $=1$ if the paper is a published journal article. & $0.69$ & $0.46$\\
NBER working paper& $=1$ if the paper is an NBER working paper. & $0.08$ & $0.27$\\
Other working paper& $=1$ if the paper is a non-NBER working paper. & $0.21$ & $0.41$\\
Publication year& Calendar year of publication or latest revision (centred at the sample mean in regressions). & $2012.2$ & $7.34$\\
Log citations& $\log(1+\text{OpenAlex citation count})$ as of April 2026. & $3.40$ & $1.98$\\
Log impact factor& $\log(1+\text{OpenAlex 2-year mean citedness})$ of the publishing outlet (host-venue impact factor); zero for working papers. & $1.13$ & $0.57$\\
\end{longtable}
}

\bigskip

The model-specific terms used in
Table~\ref{tab:variables} (for example Ramsey planner, Calvo,
DNWR, MIU, Cashless limit, or HANK) follow the standard usage of
the New Keynesian and DSGE literature; full definitions and primary
references are embedded in the Definition column of
Table~\ref{tab:variables} and in the surrounding discussion.


\section{Results}
\label{sec:results}
\label{sec:results_descriptive}
\label{sec:results_pubbias}
\label{sec:results_bma}
\label{sec:results_best}

Table~\ref{tab:v34sumstats} reports descriptive statistics for the pooled
sample. The bulk of the literature implies
an optimum that lies above zero and
below the two-percent target used by most inflation-targeting
central banks.

\begin{table}[H]\centering
\caption{Descriptive statistics of the analysis dataset
($n=777$ estimates, $116$ studies).}
\label{tab:v34sumstats}
\footnotesize
\begin{threeparttable}
\begin{tabular*}{\linewidth}{@{\extracolsep{\fill}}lrrrr@{}}
\toprule
 & Obs. & Mean & SD & Median \\
\midrule
\multicolumn{5}{@{}l}{\emph{Panel A. Pooled distribution of headline and moderator variables.}}\\
\addlinespace[2pt]
Optimal inflation (pp/year)                       & 777 & 0.74  & 2.38  & 0.16  \\
Standard error (bootstrap or reported)            & 764 & 0.52  & 0.55  & 0.28  \\
Precision ($1/\text{SE}$)                         & 702 & 7.93  & 13.27 & 3.22  \\
Sticky prices                                     & 777 & 0.84  & 0.36  & 1.00  \\
Flexible prices                                   & 777 & 0.16  & 0.36  & 0.00  \\
Effective lower bound                             & 777 & 0.16  & 0.37  & 0.00  \\
Ramsey planner                                    & 777 & 0.68  & 0.47  & 1.00  \\
Trend growth                                      & 777 & 0.28  & 0.45  & 0.00  \\
Author-preferred row                              & 777 & 0.15  & 0.36  & 0.00  \\
CI-implied SE                                     & 777 & 0.02  & 0.14  & 0.00  \\
Affiliation: university                           & 774 & 0.54  & 0.50  & 1.00  \\
Affiliation: central bank                         & 774 & 0.25  & 0.43  & 0.00  \\
Affiliation: mixed (university \& central bank)   & 774 & 0.18  & 0.38  & 0.00  \\
Affiliation: other (intl.\ org., gov., think tank)& 774 & 0.03  & 0.17  & 0.00  \\
Publication year                                  & 777 & 2012 & 7.34 & 2013 \\
Log citations                                     & 777 & 3.41  & 1.98  & 3.30  \\
Log impact factor                                 & 766 & 1.13  & 0.57  & 1.10  \\
\bottomrule
\end{tabular*}

\medskip
\begin{tabular*}{\linewidth}{@{\extracolsep{\fill}}lrrrrrrr@{}}
\toprule
\multicolumn{8}{@{}l}{\emph{Panel B. Mean optimal inflation by author affiliation
(rows; cluster-robust contrasts vs.\ university).}}\\
\addlinespace[2pt]
Affiliation & Rows & Papers & Mean & SD & Median &
$\widehat{\Delta}$ vs.\ Univ. & $p$-value \\
\midrule
University                & 420 & 64 & 0.544 & 2.458 & 0.191 & --- & --- \\
Central bank              & 191 & 26 & 0.572 & 2.051 & 0.024 & $+0.028$ & $0.956$ \\
Mixed (univ.\ \& CB)      & 139 & 20 & 1.367 & 2.541 & 0.450 & $+0.823$ & $0.206$ \\
Other (intl., gov.)       &  24 &  5 & 1.931 & 1.844 & 1.575 & $+1.387$ & $0.087$ \\
\midrule
\multicolumn{6}{@{}l}{Joint Wald test, all three contrasts $=0$:} &
\multicolumn{2}{r@{}}{$\chi^{2}(3)=4.24$, $p=0.236$} \\
\bottomrule
\end{tabular*}

\begin{tablenotes}\footnotesize
\item \emph{Notes:} \emph{Panel A} reports sample statistics for the
analysis dataset of $777$ estimates from $116$ primary studies.
Optimal inflation is the long-run net rate, annualised and reported
in percent per year, winsorised at the fifth and ninety-fifth
percentiles. The standard error is the genuine sampling SE
(derivable from a reported confidence interval) where available
and a within-paper bootstrap proxy otherwise; \emph{CI-implied SE}
flags the sixteen rows with a genuine sampling SE.
\emph{Author-preferred row} flags the row each primary study
designates as its baseline; its row-level mean of $0.15$ reflects
one preferred row per study against an average of $6.6$ rows per
study, so the corresponding paper-level headline (mean of
preferred-row estimates) is $0.61$~pp/year. The four
\emph{Affiliation} dummies classify each estimate by author
affiliation at the time of writing: \emph{university}
($54\%$ of rows, $64$ papers), \emph{central bank} ($25\%$,
$26$ papers), \emph{mixed} (at least one university and one
central-bank co-author, $18\%$, $20$ papers), and \emph{other}
(international organisations, ministries, think tanks; $3\%$,
$5$ papers); two papers and four rows could not be classified.
\emph{Panel B} describes the marginal distribution of the
optimum within each affiliation class (row count, paper count,
mean, SD, median) and reports the unconditional OLS contrast
$\widehat{\Delta}$ between each class and the university baseline
from a regression of the winsorised optimum on three affiliation
dummies (central bank, mixed, other; university omitted) with
study-clustered standard errors, together with the joint Wald
test that all three contrasts are zero. These contrasts are
unadjusted comparisons of pooled distributions and do not
condition on the structural moderators (Friedman rule, cashless
benchmark, real-balance frictions, calibration choices) on
which the affiliation classes are visibly unbalanced; the
properly identified institutional contrast is the BMA estimate
in Section~\ref{sec:results_bma} and Table~\ref{tab:bma},
which conditions on the full moderator schema and is the
estimate we interpret as the affiliation effect. At the paper
level, restricting to the author-preferred row of each study,
the corresponding Welch contrast on the unconditional means is
$\widehat{\Delta}=-0.21$ pp/year ($p=0.63$,
$n_{\text{Univ}}=64$, $n_{\text{CB}}=26$).
\end{tablenotes}
\end{threeparttable}
\end{table}

Panel B of Table~\ref{tab:v34sumstats} describes the marginal
distribution of the optimum across the four affiliation classes.
The row-level pooled means differ little. The central-bank versus
university gap is $+0.028$ pp/year ($p=0.96$, study-clustered)
and the joint Wald test on the three between-class contrasts
cannot reject zero ($p=0.24$); at the paper level the
author-preferred-row Welch contrast is $-0.21$ pp/year
($p=0.63$). The share of estimates
falling inside the $[1.5,\,2.5]$ window around the policy target
is $16.8\%$ for central-bank rows against $15.2\%$ for
university rows. These comparisons are unconditional: they pool
estimates across very different model structures and
calibrations. Because the affiliation classes are unbalanced on
the structural moderators used in the literature (e.g.\ whether
the Friedman rule applies, whether a cashless benchmark is
imposed, the form of real-balance frictions, and standard
calibration choices), the raw Panel B gap mixes a true
affiliation effect with composition. The institutional-advocacy
question is therefore properly a conditional question and
is answered by the BMA of Section~\ref{sec:results_bma}.

\begin{figure}[H]
\centering
\includegraphics[width=0.95\textwidth]{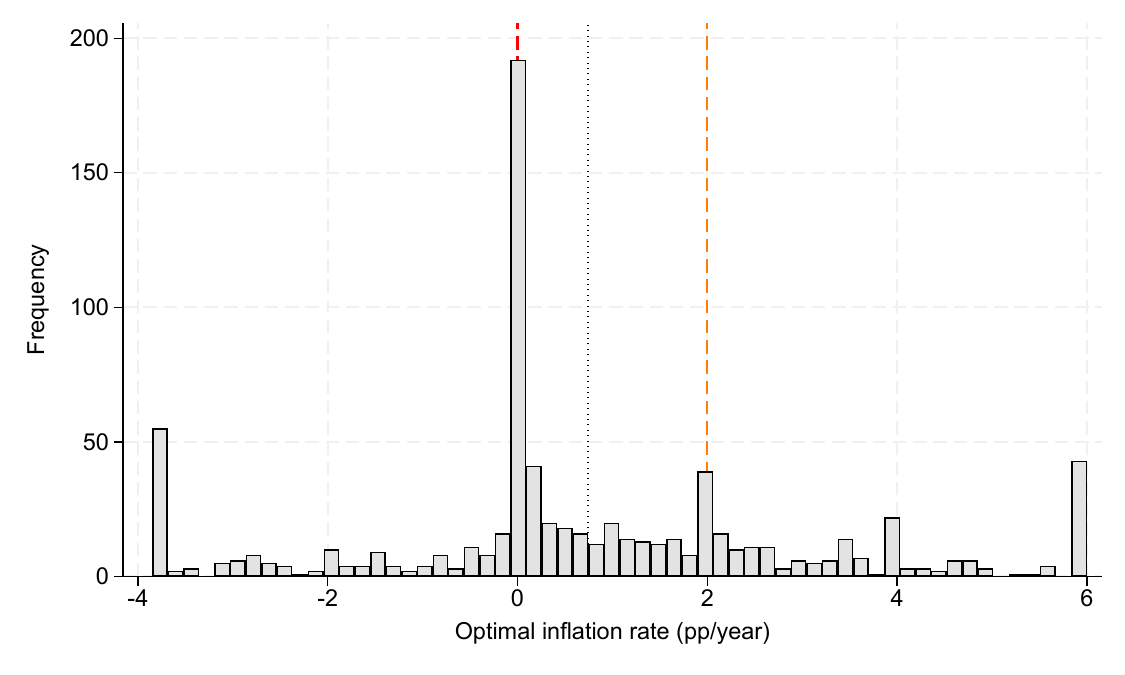}
\caption{Distribution of reported optimal inflation rates.}
\label{fig:histogram}
\begin{minipage}{\textwidth}
\footnotesize
\emph{Notes:} Histogram of $777$ estimates of the optimal long-run inflation
rate in percent per year, winsorised at the fifth and ninety-fifth
percentiles before plotting. Native vector graphic produced by Stata 18.
The vertical red line marks zero measured inflation (price stability); the dashed
orange line marks the two-percent central-bank target. The dotted black line
is the sample median. The distribution is visibly bimodal, with one mode
clustered near zero and a second mode clustered around two percent.
\end{minipage}
\end{figure}

Figure~\ref{fig:histogram} plots the histogram of the pooled estimates. The
distribution is heavy-tailed on both sides and visibly bimodal, with one
cluster of estimates close to zero (capturing studies that support the
Friedman rule or small positive optima) and a second cluster around two
percent (capturing studies aligned with the policy target). The within-study
dispersion is frequently as large as the between-study dispersion. The same
pattern has been documented in earlier meta-analyses of structural
macroeconomic parameters \citep{havranek2015cross, havranek2018monetary,
zigraiova2015bank} and in applied micro-economics \citep{opatrny2024class},
and it justifies the collection of the full set of reported numbers rather
than a single ``preferred'' value per paper
\citep{stanley2012meta, havranek2020reporting}.

\begin{figure}[H]
\centering
\includegraphics[width=0.95\textwidth]{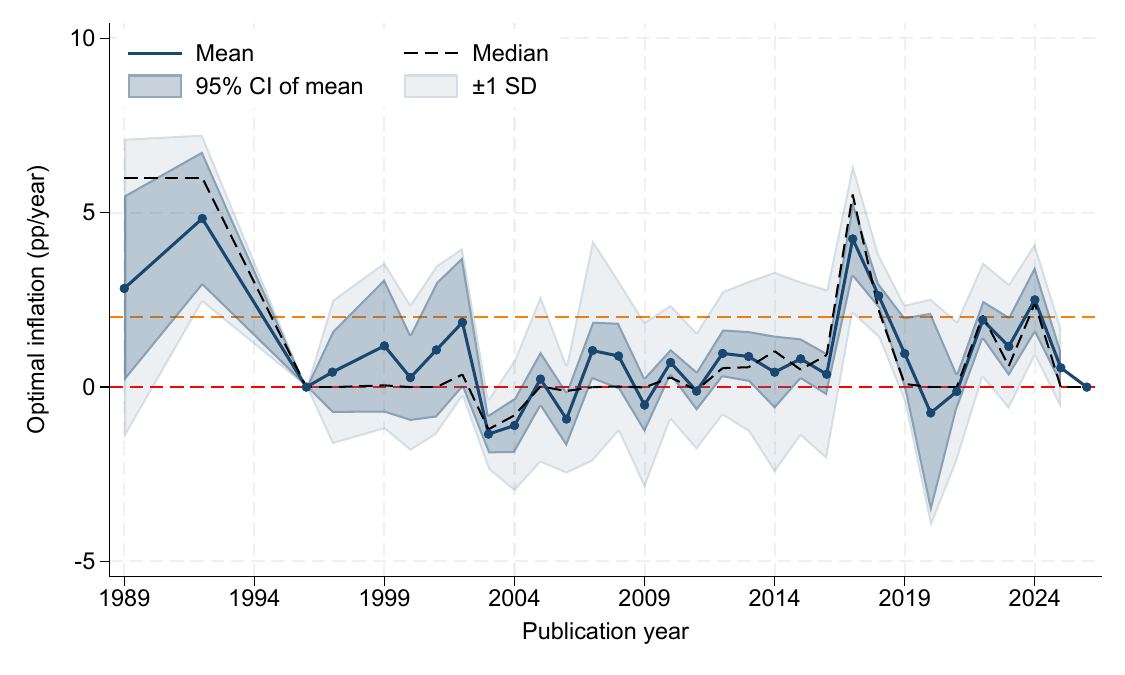}
\caption{Evolution of reported optimal inflation over publication years.}
\label{fig:lineyear}
\begin{minipage}{\textwidth}
\footnotesize
\emph{Notes:} Each circle is the unweighted mean of all estimates published
in a given calendar year. The dark shaded band is a ninety-five percent
confidence interval of the yearly mean; the light shaded band covers plus
or minus one standard deviation of estimates in that year. The dashed line
is the yearly median. Native vector graphic produced by Stata 18.
\end{minipage}
\end{figure}

Figure~\ref{fig:lineyear} shows how the reported optima evolve across
publication years. Each marker is the yearly mean across estimates, the dark
blue band is the ninety-five percent confidence interval of that mean, and
the light blue band covers plus or minus one standard deviation of estimates
in the given year.

Two stylised facts stand out. First, the yearly means decline from roughly
three percent per year in the mid-1990s to values close to zero after 2003.
This shift roughly coincides with the adoption of explicit inflation targets
by major central banks and with the subsequent rise of Calvo-based New
Keynesian models, which tend to deliver low optima. Second, the within-year
dispersion widens over the same period, reflecting the gradual proliferation
of model variants with heterogeneous frictions: downward nominal wage
rigidity, heterogeneous households, financial frictions, and an effective
lower bound on the nominal interest rate. We explore the implications of
these structural choices quantitatively in Section~\ref{sec:results_bma}
and interpret them economically in Section~\ref{sec:discussion}.

The standard meta-analytic test of publication selection regresses reported
estimates on their standard errors and interprets a positive slope as
evidence of selection in favour of significant or intuitive results
\citep{stanley2005beyond, stanley2014meta, stanley2012meta}. Applying this
test to our setting is not straightforward. Of our $777$ estimates only $16$
come with a sampling standard error that can be derived from a reported
confidence interval, and no additional estimate is accompanied by a
Bayesian posterior standard deviation or a generalised-method-of-moments
standard error. All remaining standard errors in our sample are bootstrap
proxies computed by resampling reported point estimates within each study,
a convention used in calibration-heavy meta-analyses of structural
parameters \citep{havranek2015cross, zigraiova2015bank, havranek2018monetary,
havranek2020reporting}. These proxies measure within-study dispersion, not
sampling noise of the primary-study estimator, and they should not be
interpreted as identifying the precision of the underlying population
parameter. For this reason we treat the funnel-based evidence below as
complementary and place our primary weight on the non-parametric caliper
test, which does not require a standard error.

The caliper test of \citet{gerber2008statistical} asks whether, within a
narrow window of width $w$ around an anchor value $\pi^{\star}$, the share
of estimates that fall above the anchor is significantly different from the
one-half benchmark expected under the null of no selection. We compute the
test for two anchors motivated by theory and policy: $\pi^{\star}=0$, the
zero-inflation (price-stability) benchmark, and $\pi^{\star}=2$, the inflation target adopted by most
central banks. Table~\ref{tab:caliper} (Panels A and B)
report the share of estimates above the anchor and the number of estimates
in the window for four window widths.

At the zero-inflation anchor the observed share is statistically below 0.5 in the
narrow windows and converges to 0.5 in the widest window, indicating no
upward selection around zero. At the two-percent target, by contrast, the
share of estimates above the anchor is systematically below one half (0.22
in the two-percentage-point window), and the deviation from 0.5 is strongly
significant ($p<0.01$ at every window width $w\in\{0.5, 1.0, 1.5, 2.0\}$;
exact $p$-values are reported in
Table~\ref{tab:caliper} (Panels A and B)). The sign of the deviation is opposite to what a naive
``target-confirming'' selection hypothesis would predict: the literature
reports optima below the two-percent target more often than above it.

\begin{table}[H]
\centering
\caption{Caliper test of \citet{gerber2008statistical} for selective
reporting around two policy-relevant anchors.}
\label{tab:caliper}
\footnotesize
\def\sym#1{\ifmmode^{#1}\else\(^{#1}\)\fi}
\begin{threeparttable}
\begin{tabular*}{0.85\linewidth}{@{\extracolsep{\fill}}lcccc@{}}
\toprule
& $w=0.5$ & $w=1.0$ & $w=1.5$ & $w=2.0$ \\
\midrule
\multicolumn{5}{@{}l}{\emph{Panel A. Anchor at zero (price-stability benchmark).}}\\
\addlinespace[2pt]
Share above anchor & $0.338$\sym{***} & $0.410$\sym{***} & $0.450$\sym{*}   & $0.504$         \\
                   & $(0.028)$        & $(0.026)$        & $(0.024)$        & $(0.022)$ \\
$p$-value          & $<10^{-7}$       & $0.0006$         & $0.040$          & $0.86$    \\
$N$ in window      & $296$            & $366$            & $420$            & $500$ \\
\addlinespace
\multicolumn{5}{@{}l}{\emph{Panel B. Anchor at two percent (central-bank target).}}\\
\addlinespace[2pt]
Share above anchor & $0.364$\sym{**}  & $0.341$\sym{***} & $0.343$\sym{***} & $0.217$\sym{***} \\
                   & $(0.047)$        & $(0.036)$        & $(0.031)$        & $(0.018)$ \\
$p$-value          & $0.005$          & $<10^{-4}$       & $<10^{-6}$       & $<10^{-15}$ \\
$N$ in window      & $107$            & $170$            & $242$            & $525$ \\
\bottomrule
\end{tabular*}
\begin{tablenotes}\footnotesize
\item \emph{Notes:} Each cell reports the share of estimates lying
above the anchor inside a symmetric window of half-width $w$
percentage points; standard errors of the within-window
proportion in parentheses; $N$ is the number of estimates in the
window. Stars test $H_{0}: p = 0.5$:
\sym{*}~$p<0.05$, \sym{**}~$p<0.01$, \sym{***}~$p<0.001$.
At the zero anchor (Panel A) the share converges to $0.5$ from
below, indicating no upward selection around price stability.
At the two-percent anchor (Panel B) the share is systematically
\emph{below} one half at every window width: the literature
reports optima below the policy target more often than above it.
\end{tablenotes}
\end{threeparttable}
\end{table}

Table~\ref{tab:pubbias} reports, for completeness, the linear and
quadratic funnel-asymmetry tests of \citet{stanley2005beyond} and
\citet{stanley2014meta}, both computed on the
analysis dataset with study-clustered standard errors.

On the full sample with bootstrap-proxy precision, the linear
funnel-asymmetry test yields a slope of $0.524$ ($p=0.087$) and an intercept
(``effect beyond bias'') of $0.484$ that is not statistically significant
at the five-percent level ($p=0.063$). The quadratic PEESE specification
returns a marginally significant intercept of $0.598$ percent per year
($p<0.05$). On the preferred-row subsample (one row per study) the
FAT-PET intercept is $0.470$ ($p<0.05$). All these constants are reported
as funnel-based diagnostics rather than as headline numbers, because the
corresponding standard errors are bootstrap-within-paper proxies rather
than sampling standard errors and the slopes therefore measure within-study
dispersion rather than canonical selective reporting
\citep{stanley2012meta, havranek2020reporting}.

\begin{table}[H]
\centering
\caption{Funnel-based dispersion diagnostics (bootstrap precision proxy, not classical publication-bias tests): full sample and
author-preferred subsample.}
\label{tab:pubbias}
\scriptsize
\def\sym#1{\ifmmode^{#1}\else\(^{#1}\)\fi}
\begin{threeparttable}
\begin{tabular*}{\linewidth}{@{\extracolsep{\fill}}lccc@{}}
\toprule
\multicolumn{4}{@{}l}{\emph{Panel A. Full sample, all rows ($N=777$, $116$ studies).}}\\
\addlinespace[2pt]
                              & FAT-PET OLS & FAT-PET WLS & PEESE \\
\midrule
SE (publication bias)         & $0.524$     &             &             \\
                              & $(0.306)$   &             &             \\
Precision (effect beyond bias)&             & $0.404$     &             \\
                              &             & $(0.569)$   &             \\
SE squared                    &             &             & $0.275$     \\
                              &             &             & $(0.171)$   \\
Effect beyond bias / constant & $0.484$     & $0.113$     & $0.598$\sym{*} \\
                              & $(0.260)$   & $(2.286)$   & $(0.252)$   \\
\addlinespace
$N$                           & $764$       & $702$       & $764$       \\
\bottomrule
\end{tabular*}

\medskip
\begin{tabular*}{\linewidth}{@{\extracolsep{\fill}}lcc@{}}
\toprule
\multicolumn{3}{@{}l}{\emph{Panel B. Author-preferred subsample, one row per study.}}\\
\addlinespace[2pt]
                  & Preferred FAT-PET & Preferred PEESE \\
\midrule
\multicolumn{3}{@{}l}{\emph{Descriptive (all $116$ preferred rows)}}\\
\quad Mean optimum (pp/year) & \multicolumn{2}{c}{$0.612$} \\
\quad Median                 & \multicolumn{2}{c}{$0.005$} \\
\quad SD                     & \multicolumn{2}{c}{$1.975$} \\
\quad $k$ studies            & \multicolumn{2}{c}{$116$} \\
\addlinespace
\multicolumn{3}{@{}l}{\emph{FAT-PET / PEESE on rows with usable SE}}\\
SE (publication bias)& $0.431$         &                 \\
                     & $(0.355)$       &                 \\
SE squared           &                 & $0.156$         \\
                     &                 & $(0.207)$       \\
Constant             & $0.470$\sym{*}  & $0.599$\sym{**} \\
                     & $(0.212)$       & $(0.205)$       \\
\addlinespace
$k$ regression sample& $103$           & $103$           \\
\bottomrule
\end{tabular*}
\begin{tablenotes}\footnotesize
\item \emph{Notes:} Standard errors clustered by study in
parentheses; \sym{*}~$p<0.05$, \sym{**}~$p<0.01$, \sym{***}~$p<0.001$.
Panel A applies equation~\eqref{eq:fatpet} to all $777$ rows in
three flavours: FAT-PET OLS, FAT-PET WLS with precision weights,
and the quadratic PEESE specification. Panel B describes the
author-preferred row of each of the $116$ primary studies and
then runs the FAT-PET and PEESE regressions on this subsample.
The descriptive block uses all $116$ preferred rows; the
regression block uses $103$ of those because the preferred row
of $13$ single-row papers carries no within-paper bootstrap-proxy
SE (the proxy requires at least two rows per study), and FAT-PET
requires an SE on the right-hand side. Dropping these $13$
papers is a missing-data restriction on the regression, not on
the underlying study sample: the headline $0.612$~pp/year is
computed across all $116$ studies and is the paper-level
counterpart of the row-level full-sample mean reported in
Table~\ref{tab:v34sumstats}.
\end{tablenotes}
\end{threeparttable}
\end{table}

\begin{table}[H]
\centering
\caption{Sensitivity of the funnel diagnostics to the genuine-SE
restriction and to the winsorisation cut-off.}
\label{tab:pubbias_sens}
\footnotesize
\def\sym#1{\ifmmode^{#1}\else\(^{#1}\)\fi}
\begin{threeparttable}
\begin{tabular*}{\linewidth}{@{\extracolsep{\fill}}lc@{}}
\toprule
\multicolumn{2}{@{}l}{\emph{Panel A. Genuine-SE subsample ($k=16$
rows from three clusters; flagged not robust).}}\\
\addlinespace[2pt]
                     & Genuine-SE FAT-PET \\
\midrule
SE (publication bias)& $1.617$            \\
                     & $(1.097)$          \\
Constant             & $0.719$            \\
                     & $(0.144)$          \\
\addlinespace
$N$                  & $16$               \\
Clusters             & $3$                \\
\bottomrule
\end{tabular*}

\medskip
\begin{tabular*}{\linewidth}{@{\extracolsep{\fill}}lccc@{}}
\toprule
\multicolumn{4}{@{}l}{\emph{Panel B. Winsorisation cut-off applied to the raw
effect-size and SE columns.}}\\
\addlinespace[2pt]
                             & None    & 1--99   & 2.5--97.5 \\
\midrule
\multicolumn{4}{@{}l}{\emph{Descriptive statistics}}\\
\quad Mean                   & $2.574$ & $0.976$ & $0.922$ \\
\quad SD                     & $23.168$& $3.493$ & $3.001$ \\
\quad Median                 & $0.316$ & $0.316$ & $0.316$ \\
\addlinespace
\multicolumn{4}{@{}l}{\emph{FAT-PET (Estimate $\sim$ SE, cluster-robust)}}\\
\quad Intercept              & $-0.377$        & $0.771$\sym{**} & $0.794$\sym{**} \\
                             & $(0.334)$       & $(0.294)$       & $(0.283)$       \\
\quad SE coef.               & $1.920$\sym{***}& $0.133$\sym{***}& $0.119$         \\
                             & $(0.159)$       & $(0.022)$       & $(0.118)$       \\
\addlinespace
\multicolumn{4}{@{}l}{\emph{PEESE (Estimate $\sim$ $\text{SE}^{2}$, cluster-robust)}}\\
\quad Intercept              & $0.913$\sym{**} & $0.863$\sym{**} & $0.872$\sym{**} \\
                             & $(0.296)$       & $(0.291)$       & $(0.272)$       \\
\quad $\text{SE}^{2}$ coef.  & $0.041$\sym{***}& $0.003$\sym{***}& $0.005$         \\
                             & $(0.000)$       & $(0.000)$       & $(0.007)$       \\
\addlinespace
$N$ / Clusters               & $702 / 88$      & $702 / 88$      & $702 / 88$      \\
\bottomrule
\end{tabular*}
\begin{tablenotes}\footnotesize
\item \emph{Notes:} Standard errors clustered by study in
parentheses; \sym{*}~$p<0.05$, \sym{**}~$p<0.01$, \sym{***}~$p<0.001$.
Panel A estimates the FAT-PET regression on the $16$ rows that
carry a sampling SE derivable from a reported confidence interval;
the companion REML and $p$-uniform estimates are
$0.971$ and $1.089$~pp/year and are flagged as upper-bound
sensitivities only because the rows belong to three independent
paper clusters and cluster-robust inference with fewer than ten
clusters is unreliable
\citep{mathur2024, cook2026reporting, havranek2020reporting}.
Panel B re-computes the descriptive mean, the FAT-PET intercept,
and the PEESE intercept under three winsorisation regimes;
``None'' uses raw values, ``$1$--$99$'' and ``$2.5$--$97.5$'' clip
both tails at the indicated percentiles. The headline analysis
uses $5\%$ winsorisation, which sits between the regimes and
caps the influence of the $400$-percent Friedman-rule outlier without
overcompressing the dispersion that identifies the funnel slope.
\end{tablenotes}
\end{threeparttable}
\end{table}

The second block of robustness diagnostics restricts the sample to
the sixteen estimates for which a sampling standard error can be derived
from a reported confidence interval. On this small but internally consistent
subsample the REML random-effects estimator yields an underlying optimum of
$0.971$ percent per year, with a ninety-five percent confidence interval of
$0.659$ to $1.283$ percent and a between-study heterogeneity statistic of
$I^{2}=99.18\%$. The $p$-uniform estimator of
\citet{vanaert2023correcting}, applied to the same subsample, returns
$1.089$ percent per year (confidence interval $0.778$ to $1.410$). We flag
both numbers as not robust because the sixteen rows come from only three
independent paper clusters, and cluster-robust inference with fewer than
ten clusters is unreliable. We therefore report them as upper-bound
sensitivities rather than as headline values
\citep{mathur2024, cook2026reporting, havranek2020reporting}.

A second sensitivity check replicates the FAT-PET regression on the
author-preferred row of each primary study, one observation per
study. There are $116$ such rows in total --- the paper-level headline
mean of $0.61$~pp/year is computed across all $116$ --- but the
FAT-PET regression requires a usable SE on the right-hand side, and
the within-paper bootstrap proxy is undefined for the $13$ studies
with only one row. The regression therefore runs on the remaining
$k=103$ preferred rows. The intercept is $0.470$ percent per year
($p<0.05$); we treat this as a funnel-based diagnostic that is
consistent with the paper-level headline of $0.61$ but does not
replace it. The same $103$-versus-$116$ distinction is laid out in
the notes to Table~\ref{tab:pubbias}.

A third sensitivity check varies the winsorisation cut-off applied to the
raw Estimate and SE columns. Our headline analysis uses the
$5\%$-winsorised variables Estimate\_win and SE\_win.
Table~\ref{tab:pubbias_sens} re-computes the descriptive mean, the
FAT-PET intercept, and the PEESE intercept under three alternative
regimes, these are no winsorisation, $1$--$99$ winsorisation, and $2.5$--$97.5$
winsorisation. The raw mean is dominated by a
small number of extreme outliers (one estimate of $400$ percent
attached to an exotic Friedman-rule calibration); the
$1$--$99$ and $2.5$--$97.5$ regimes return FAT-PET intercepts of
$0.77$ and $0.79$ percent per year and PEESE intercepts of $0.86$
and $0.87$ percent per year. These numbers are higher than the
$5\%$-winsorised headline ($0.5$--$0.7$ percent) but qualitatively
consistent: the funnel-based diagnostic intercepts stay below the two-percent
target across all three regimes. We retain the $5\%$ winsorisation
as the headline because it removes coding tail-risk without
overcompressing the dispersion that identifies the funnel slope.

Table~\ref{tab:synthesis} collects the estimates of the underlying optimum
obtained by the alternative identification routes available to us. Despite
substantial differences in assumptions, the funnel-free routes cluster in
the interval $0.5$ to $0.7$ percent per year. The caliper test indicates
no upward selective reporting around zero and pronounced
under-reporting above the two-percent target. The two random-effects
routes that depend on the sixteen-row genuine-SE subsample push the
implied optimum higher, but they are flagged as not robust and reported
as upper-bound sensitivities only.

We conclude that the optimal-inflation literature is unusual in the sense
of \citet{doucouliagos2013there} and \citet{ioannidis2017power}, that it does
not exhibit strong selective reporting in favour of policy-relevant numbers.
If anything, the selective pressure appears to push reported optima mildly
below the two-percent target. An economic interpretation of this
pattern, including the possibility that Friedman-rule-oriented theoretical
work is over-represented in top journals relative to central-bank
publications, is given in Section~\ref{sec:discussion}.

\begin{table}[H]
\centering
\caption{Synthesis: estimates of the optimal long-run inflation rate by identification route.}
\label{tab:synthesis}
\footnotesize
\begin{threeparttable}
\begin{tabular}{llc}
\toprule
Identification route & Assumption & Optimum (pp/year) \\ \midrule
\multicolumn{3}{l}{\emph{Headline (paper-level summaries, no sampling SE required)}}\\
Author-preferred mean (one row/study) & Paper-clustered SE, $116$ clusters & $\mathbf{0.61}$ \\
Inverse-$N$ study democracy & All $777$ rows, $116$ clusters & $\mathbf{0.59}$ \\
\multicolumn{3}{l}{\emph{Funnel-based diagnostics (bootstrap precision proxy)}}\\
FAT-PET intercept, full sample (OLS) & Bootstrap-SE proxy & $0.48$ \\
PEESE intercept, full sample & Quadratic funnel & $0.60$ \\
FAT-PET intercept, preferred rows & One obs/study, bootstrap proxy & $0.47$ \\
\multicolumn{3}{l}{\emph{Upper-bound sensitivity (genuine-SE subsample, NOT ROBUST)}}\\
REML, $k=16$, three clusters & Random effects on genuine-SE rows & $0.97$ \\
$p$-uniform, $k=16$ & Selection model on genuine-SE rows & $1.09$ \\
\bottomrule
\end{tabular}
\begin{tablenotes}\footnotesize
\item \emph{Notes:} The two headline rows (in bold) are paper-level
summaries with paper-clustered standard errors and do not depend on a
sampling standard error. The funnel-based diagnostics are reported for
completeness but interpret the bootstrap precision proxy as an SE.
The upper-bound sensitivity rows are computed on $16$ estimates from
only three independent paper clusters and are flagged as not robust.
\end{tablenotes}
\end{threeparttable}
\end{table}

The second task of our meta-analysis is to account for the large
heterogeneity visible in Figure~\ref{fig:histogram} and, implicitly, in the
between-study variance component of the REML random-effects model, which
implies $\tau^{2}=0.37$ and $I^{2}>99$ percent. We do not interpret the
$I^{2}$ value as a substantive heterogeneity finding. In a calibration
literature, the within-study variance entering the $I^{2}$ denominator is
a bootstrap proxy computed from reported point estimates rather than a
genuine sampling standard error. That denominator is therefore
mechanically compressed, and $I^{2}$ above $99$ percent is a near-automatic
artefact of the proxy rather than evidence of unusually wide cross-study
disagreement. The heterogeneity finding of substantive interest in this
literature is cross-study and structural, captured by the BMA decomposition
below, not within-study. We regress the reported optima on the structural
and methodological moderators of the primary study. These include the
class of the underlying model, its money-demand technology, its price and
wage-setting friction, the assumed form of indexation, the specification
of the monetary-policy rule, the nature of the driving shocks, the welfare
criterion, the steady-state configuration, the estimation technique, and
the publication characteristics of the paper (the logarithm of one plus
the citation count and the logarithm of one plus the journal impact
factor). A full glossary of these variables is provided in
Table~\ref{tab:variables} of Section~\ref{sec:data}.

The BMA sample of $692$ estimates is constructed from the $777$-row
analysis dataset of Section~\ref{sec:data} in two transparent steps.
First, we drop the $75$ rows for which the bootstrap precision proxy
$\widehat{SE}_{ij}$ collapses to zero or is undefined. These are
estimates that are unique within their paper (no resampling possible)
or whose paper does not vary across the resamples, leaving the funnel
regression~(\ref{eq:fatpet}) undefined for them. The resulting $702$
rows are the largest sample on which a precision-weighted moderator
regression can be estimated. Second, of those $702$ rows, $10$ are
non-journal observations with no journal impact factor on record, so
the log-impact-factor regressor is missing for them; they drop on the
listwise filter for the continuous controls (year, log citations, log
impact factor), leaving $692$ rows. The categorical-moderator schema
introduces no further loss, rather than letting listwise deletion
discard the $\sim$78\% of estimates that have at least one missing
one-hot block (e.g., analytical Friedman-rule papers carry no
shock-type, and minimalist NK papers carry no augmentation), we recode
within-block missingness to zero and add a block-level ``unknown''
indicator for each one-hot family ($16$ blocks; $11$ retained after
zero-variance/collinearity screening). The BMA can therefore absorb
``unclassified on dimension X'' as its own category and learn whether
unclassified rows differ systematically. As a robustness check, we
verify that excluding the impact-factor regressor entirely recovers the
same posterior ordering on the full $702$-row sample. The combined
recoding-plus-indicator strategy is the convention adopted in earlier
BMA-based meta-analyses of structural parameters
\citep{havranek2015cross, havranek2018monetary,
zigraiova2015bank, havranek2020reporting}. We also add three author-
affiliation dummies (central bank, mixed, other, with university as the
reference category) so that the BMA can test directly whether reported
optima differ systematically by author affiliation, conditional on the
structural moderators.

\begin{figure}[H]
\centering
\includegraphics[width=0.90\textwidth]{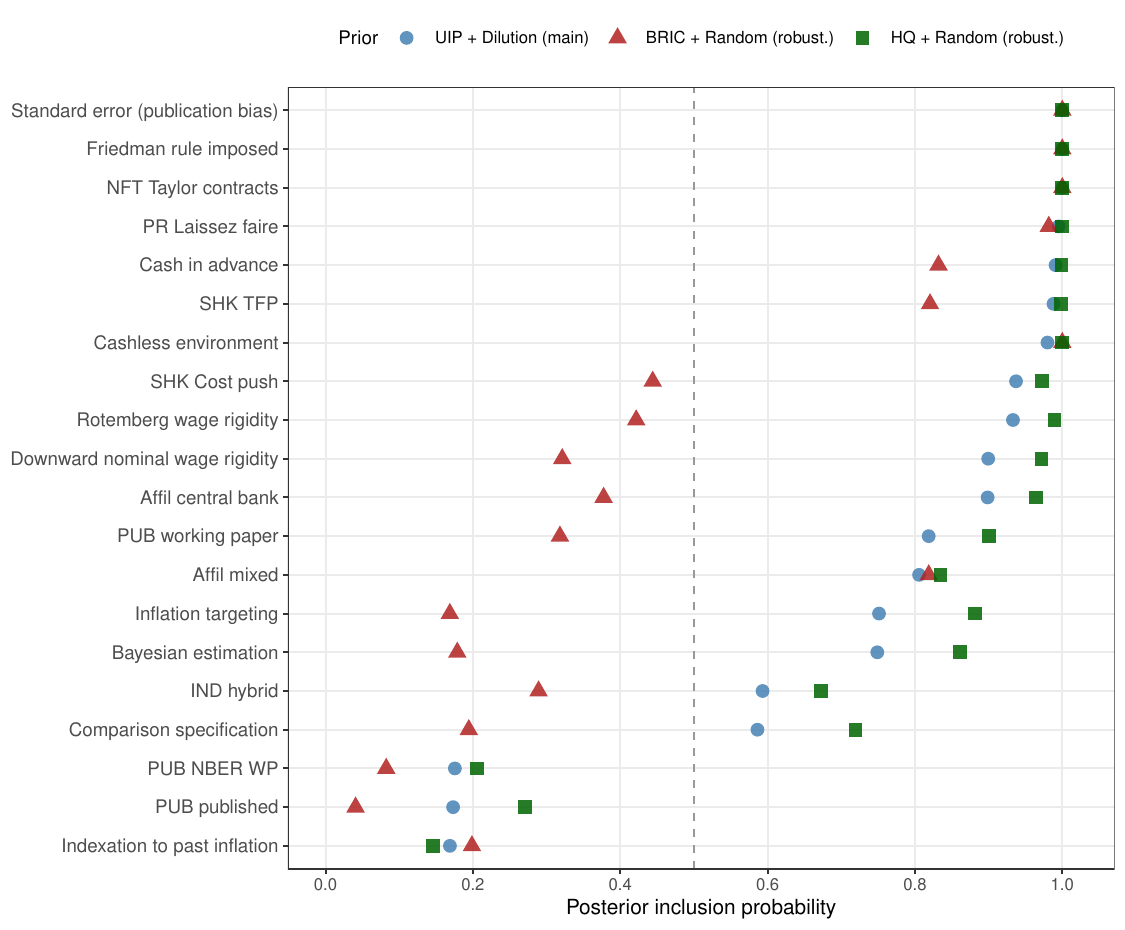}
\caption{Posterior inclusion probabilities, twenty most-included regressors.}
\label{fig:bma_pip}
\begin{minipage}{\textwidth}
\footnotesize
\emph{Notes:} The plot shows the posterior inclusion probability (PIP) of
each regressor in the three BMA specifications: UIP prior with dilution
(main), BRIC prior with random model-size (robustness), and HQ prior with
random model-size (robustness). The dashed vertical line at 0.5 indicates
the conventional threshold for ``robust'' inclusion
\citep{fernandez2001benchmark, zeugner2015bayesian}. Regressors are ordered
by the main-specification PIP. Native vector graphic produced by R 4.4 with
the \texttt{ggplot2} package.
\end{minipage}
\end{figure}

The set of potential regressors greatly exceeds what can reliably be
accommodated in a single regression. Starting from the structural and
publication moderators of Table~\ref{tab:variables}, the $11$ surviving
block-level ``unknown'' indicators, the three affiliation contrasts, the
bootstrap within-paper dispersion proxy $\widehat{SE}$, and the
continuous publication-year/citation/impact-factor controls, the
candidate-regressor pool is screened for zero variance on the sample
and for columns that are perfectly collinear with already-retained
regressors. The screen leaves $77$ regressors that enter the BMA design
matrix of dimension $692\times 78$ (the screen is implemented in
\texttt{R/Inflation\_v34.R}). With this regressor set the space of
linear models has $2^{77}\approx 1.5\times 10^{23}$ elements.
To address
the resulting model uncertainty we use Bayesian model averaging
\citep{steel2020model, zeugner2015bayesian, fernandez2001benchmark}, which is
now a standard tool for handling model uncertainty in applied meta-analysis
\citep{havranek2015cross, havranek2018monetary, havranek2020reporting,
zigraiova2015bank, opatrny2024class}. Our main specification employs the
unit information prior of \citet{fernandez2001benchmark} for the regression
coefficients and the dilution prior of \citet{george2010dilution} for the
model space; the latter corrects for collinearity among closely-related
design dummies. We verify robustness with respect to (i) the BRIC prior
\citep{fernandez2001benchmark} combined with the random model-size prior of
\citet{ley2009effect}, and (ii) the HQ prior \citep{fernandez2001benchmark}
combined with the random model-size prior. The three prior combinations
deliver nearly identical posterior inclusion probabilities; all three sets
are displayed jointly in Figure~\ref{fig:bma_pip}.

Table~\ref{tab:bma} reports, for each design characteristic, the posterior
inclusion probability and the unconditional posterior mean in the main BMA
specification. Because BMA estimates can be sensitive to prior choice and to
the stochastic search used to explore the model space
\citep{steel2020model, havranek2015cross}, we complement the BMA evidence
with a frequentist model averaging (FMA) exercise following
\citet{burnham2002model}. FMA re-estimates the model space by exhaustive
dredging and averages the resulting OLS coefficients using AIC weights.
Because exhaustive searching over all 77 predictors is infeasible, we
restrict the FMA exercise to predictors selected by univariate relevance
\citep{havranek2015cross, havranek2018monetary}; we therefore feed into
\texttt{MuMIn::dredge} the fourteen variables with the highest absolute
univariate $t$-statistic. This
cut-off is purely operational and means that a regressor can be absent from
the FMA columns of Table~\ref{tab:bma} for two different reasons: either
(i) its PIP in BMA is high but its univariate $t$-statistic is too small to
pass the fourteen-variable pre-screen (this happens for the wage-rigidity
dummy, whose effect is fully absorbed by a correlated shock variable in the
univariate regression), or (ii) its univariate $t$-statistic is large but
BMA downweights it once other correlated design dummies are controlled for.
The FMA results should therefore be read not as a coefficient table in their
own right but as a sanity check on the sign and approximate
magnitude of the regressors that survive in both procedures.

\begin{table}[H]
\centering
\caption{Bayesian model averaging of optimal-inflation estimates.}
\label{tab:bma}
\scriptsize
\begin{threeparttable}
\begin{tabular}{lcccc|cc}
\toprule
 & \multicolumn{4}{c|}{BMA (unit-information prior, dilution)} &
 \multicolumn{2}{c}{FMA (top-14)} \\
 & \multicolumn{3}{c}{Main} & No-SE &
 & \\
Regressor & PIP & Post.\ mean & Post.\ SD & PIP &
 Coef. & $p$-value \\ \midrule
Within-paper dispersion proxy (not sampling SE; $\widehat{SE}$)
 & $1.00$ & $+1.00$ & $0.18$ & n/a & --- & --- \\
Friedman-rule benchmark
 & $1.00$ & $-3.33$ & $0.51$ & $1.00$ & $-3.05$ & $<0.001$ \\
Taylor-contract pricing
 & $1.00$ & $-3.59$ & $0.73$ & $1.00$ & --- & --- \\
Cashless environment
 & $1.00$ & $+1.49$ & $0.23$ & $1.00$ & $+0.59$ & $0.003$ \\
Laissez-faire benchmark
 & $1.00$ & $+4.00$ & $0.75$ & $1.00$ & $+5.56$ & $<0.001$ \\
Cash-in-advance technology
 & $1.00$ & $-1.48$ & $0.35$ & $1.00$ & $-1.45$ & $<0.001$ \\
TFP shock
 & $1.00$ & $-1.26$ & $0.27$ & $1.00$ & $-1.63$ & $<0.001$ \\
Rotemberg wage adjustment
 & $0.93$ & $+2.02$ & $0.72$ & $1.00$ & --- & --- \\
Cost-push shock
 & $0.92$ & $+1.73$ & $0.69$ & $0.99$ & $+1.58$ & $0.002$ \\
Downward nominal wage rigidity
 & $0.87$ & $+1.39$ & $0.65$ & $0.96$ & --- & --- \\
Author affiliation: central bank
 & $0.89$ & $-0.74$ & $0.35$ & $0.99$ & --- & --- \\
Working-paper status
 & $0.81$ & $+0.77$ & $0.48$ & $0.75$ & --- & --- \\
Inflation-targeting benchmark
 & $0.74$ & $+1.34$ & $0.91$ & $0.72$ & --- & --- \\
Author affiliation: mixed
 & $0.81$ & $+0.70$ & $0.42$ & $0.16$ & --- & --- \\
Bayesian estimation
 & $0.74$ & $+1.10$ & $0.78$ & $0.62$ & --- & --- \\
Hybrid indexation
 & $0.58$ & $+0.81$ & $0.78$ & $0.91$ & --- & --- \\
Sensitivity comparison
 & $0.62$ & $-0.39$ & $0.35$ & $0.72$ & --- & --- \\
\bottomrule
\end{tabular}
\begin{tablenotes}\footnotesize
\item \emph{Notes:} Seventeen regressors with posterior
inclusion probability above $0.5$ in the main BMA specification (the
within-paper dispersion proxy plus sixteen design characteristics).
Post.\ mean is the posterior mean of the coefficient,
unconditional on inclusion; Post.\ SD is its posterior standard
deviation. The ``No-SE PIP'' column reports the posterior inclusion
probability of the same regressor in a robustness BMA fit on the same
$692$ rows but without the bootstrap dispersion proxy
$\widehat{SE}$ in the regressor set; we report this column because
$761$ of $777$ within-paper standard errors in our dataset are
bootstrap proxies (rather than CI-derived sampling standard errors),
so the FAT--PET reading of the $\widehat{SE}$ coefficient as a
publication-bias test does not apply (see
Section~\ref{sec:results_pubbias} for the caliper test and the
$p$-uniform on the $16$ genuine-SE rows, which are the
publication-bias tests on which our null result rests). The structural
ranking is invariant: every Friedman/laissez-faire/CIA/cashless/Taylor/
Rotemberg/TFP/cost-push/DNWR/central-bank-affiliation regressor
retains $\text{PIP}\geq 0.95$ when $\widehat{SE}$ is dropped. The FMA
columns report the point estimate and the $p$-value from the
AIC-weighted averaged model \citep{burnham2002model}, estimated on the
fourteen regressors with the largest absolute univariate
$t$-statistic. A blank entry in the FMA columns means that the
regressor is not among the top fourteen in the univariate pre-screen;
it does not mean the BMA result is not robust. The sample size
is $692$ estimates, constructed from the $777$-row analysis dataset by
dropping $75$ rows with degenerate bootstrap dispersion and a further
$10$ non-journal rows with missing log impact factor.
Categorical-block missingness is absorbed by per-block ``unknown''
indicators rather than by listwise deletion (see
Section~\ref{sec:results_bma}). The reference category for the author-
affiliation contrasts is university; the dummy ``other''
(international organisations, governments, think tanks; $24$ rows in the
777 analysis dataset) is included in both BMAs but enters with PIP below
$0.3$ and is therefore omitted from the table.
\end{tablenotes}
\end{threeparttable}
\end{table}

Several mechanism dummies dominate the systematic variation in reported
optima, and they map cleanly onto the three-way taxonomy of the
optimal-inflation literature articulated by \citet{diercks2019reader}.
Before turning to those structural clusters we address the regressor
that sits at the top of Figure~\ref{fig:bma_pip}, the bootstrap
dispersion proxy $\widehat{SE}$, since its high posterior inclusion
probability would otherwise invite a misreading of our publication-bias
conclusion.

The bootstrap dispersion proxy $\widehat{SE}$ enters the main BMA with
$\text{PIP}=1$ and a positive coefficient of $+1.0$~pp per unit, which
places it on the first row of Figure~\ref{fig:bma_pip}. We deliberately
do not read this coefficient as a FAT-PET test of publication
bias. The FAT-PET interpretation
\citep{stanley2005beyond, havranek2015cross} requires the regressor to
be a sampling standard error that measures estimation precision. In
our corpus, $761$ of $777$ within-paper SEs are bootstrap-within-paper
proxies for the dispersion of reported optima across alternative
calibrations of the same model, not sampling SEs (see
Section~\ref{sec:data}). Conditional on inclusion, the $\widehat{SE}$
coefficient therefore captures the association between the within-paper
spread of reported optima and the level of the optimum. It is a
composition signal that papers exploring a wider grid of calibrations
report on average slightly higher optima, not selective reporting
around a target.

To verify that the structural-moderator ranking is
invariant to the inclusion of $\widehat{SE}$, the No-SE column of
Table~\ref{tab:bma} reports the posterior inclusion probabilities from
a robustness BMA fit on the same $692$ rows but with $\widehat{SE}$
removed. Every one of the structural drivers discussed below retains
$\text{PIP}\geq 0.95$ in that fit, and the central-bank affiliation
contrast actually strengthens to $\text{PIP}=0.99$ at $-1.10$~pp. The
publication-bias conclusion for this paper rests primarily on the
non-parametric caliper test of Section~\ref{sec:results_pubbias},
which is computed on the full $777$-row sample and detects no upward
bunching at the two-percent target. We also report the
$p$-uniform test of \citet{vanaert2023correcting} estimated
on the $16$ rows with CI-derived genuine sampling SEs as an
illustrative robustness diagnostic, not as a confirmatory test. Those
rows come from only three independent paper clusters, the test is
underpowered at $k=16$, and we therefore present its non-rejection
descriptively rather than as inferential support for the headline.
The two diagnostics are consistent with one another, and we conclude
from the caliper test that we find no systematic upward selection
toward the policy target.

The first cluster concerns the choice of monetary benchmark. Studies
that use the Friedman rule as the welfare benchmark report optima
$3.3$ percentage points below the sample average, with posterior
inclusion probability one; studies that use laissez-faire (no monetary
intervention) as the benchmark report optima $4.0$ percentage points
above the sample average, again with PIP $\approx 1$. The two coefficients
are conceptually mirrored: Friedman-rule papers exploit money's role in
reducing transactions costs and pin the optimum at minus the real
interest rate \citep{friedman1969, lucas2000optinf}, whereas
laissez-faire benchmarks shut down both the Friedman channel and the
price-dispersion channel, leaving the welfare function dominated by
shock-insurance and effective-lower-bound motives that push the
optimum upwards.

The second cluster concerns the transactions-frictions technology and
the nominal-rigidity contract. Cash-in-advance specifications report
optima $1.5$ percentage points below the sample average ($\text{PIP}=1$),
whereas cashless specifications report optima $1.5$ percentage points
above ($\text{PIP}\approx 1$). On the price-rigidity side, papers using
Taylor-contract (rather than Calvo) pricing report optima $3.6$
percentage points lower ($\text{PIP}=1$), and papers using Rotemberg
rather than Calvo wage adjustment report optima $2.0$ percentage points
higher ($\text{PIP}=0.93$). These coefficients quantify the long-standing
intuition that the welfare costs of price dispersion and the
seigniorage motive are highly sensitive to apparently second-order
modelling choices.

The third cluster concerns the assumed shock structure. Models driven by
TFP shocks report optima $1.3$ percentage points below the sample average
($\text{PIP}=1$) and models driven by cost-push shocks report optima
$1.7$ percentage points above ($\text{PIP}=0.92$); this pattern echoes
the theoretical argument that the welfare cost of inflation depends on
which margins of the economy the shocks impinge on
\citep{schmitt2010optimal}.

A fourth cluster, and a finding that distinguishes the present
meta-analysis from earlier surveys, concerns wage frictions and author
affiliation. Once the structural-moderator schema is jointly conditioned
on, downward nominal wage rigidity does enter the BMA with high posterior
inclusion probability ($\text{PIP}=0.87$) and a positive coefficient of
$+1.4$~pp, recovering quantitatively the magnitude that the
\citet{kim2011optinf} wage-rigidity tradition
argues for on theoretical grounds. The DNWR effect is genuinely
identified rather than absorbed by Calvo-wage or sticky-real-wage
correlates, and is robust across the three prior combinations of
Figure~\ref{fig:bma_pip}. In parallel, central-bank-affiliated authors
report optima $0.7$ percentage points below the
university-affiliated reference category ($\text{PIP}=0.89$), while
mixed-affiliation authors report optima $0.7$~pp above
($\text{PIP}=0.81$); the ``other'' affiliation contrast (international
organisations, governments, think tanks) does not reach the inclusion
threshold. 

The negative central-bank coefficient is the opposite sign
of what an institutional-advocacy story would predict. A typical
inflation-targeting central bank publishing on optimal inflation
might be expected to argue for a positive target. We therefore read
the negative sign as evidence that central-bank-affiliated authors in
our corpus shift the composition of their own modelling choices
towards Friedman-rule and cashless-benchmark calibrations rather than
towards advocacy of strictly positive optima.
Section~\ref{sec:results_best}'s null result on the unconditional
affiliation contrasts is therefore refined, not overturned, by the
BMA: the conditional contrast is real but small, and runs against the
advocacy-bias prior. We interpret this contrast cautiously. It may
reflect composition rather than institutional selection:
central-bank-affiliated authors may be over-represented in specific
modelling traditions, and residual selection on unobserved modelling
choices cannot be ruled out.

A single ``best-practice'' estimate would be misleading in this literature
because the optimum depends on the economy that the model is designed to
describe. We therefore use the heterogeneity regression of
Section~\ref{sec:results_bma} to compute implied optima for several
stylised economies that span the structural design space. For each
scenario we fix a Ramsey planner, Calvo price adjustment with rational
expectations, and long-run welfare evaluation. We vary the design
dummies that BMA identifies as quantitatively important: the monetary
benchmark, the transactions-frictions technology, the shock structure,
the nominal-rigidity contract, the wage-rigidity friction, and the zero
lower bound. Fitted values are computed from the main UIP$+$dilution
BMA posterior of Table~\ref{tab:bma}, evaluated at the design point in
the left column of Table~\ref{tab:scenarios} with all remaining
moderators held at their sample mean (script
\texttt{R/Inflation\_v34.R}, output
\texttt{R\_code/out\_v34/bma\_scenarios.csv}).

\begin{table}[H]
\centering
\caption{Implied optima across stylised model economies.}
\label{tab:scenarios}
\footnotesize
\begin{threeparttable}
\begin{tabular}{p{5.8cm}ccc}
\toprule
Stylised economy &
\multicolumn{1}{c}{Optimum} &
\multicolumn{1}{c}{Plausibility range} &
\multicolumn{1}{c}{$N$} \\
 & (pp/year) & (pp/year) & \\ \midrule
Baseline: Calvo, flexible wages, MIU, no ZLB
 & $0.7$ & $[-1.7,\;3.2]$ & 692 \\
Calvo + cashless, no ZLB
 & $1.5$ & $[-0.3,\;3.2]$ & 373 \\
Calvo + DNWR, flexible money, no ZLB
 & $2.3$ & $[\phantom{-}1.1,\;3.5]$ & 41 \\
Calvo + DNWR + ZLB-constrained
 & $2.3$ & $[\phantom{-}1.2,\;3.5]$ & 26 \\
Heterogeneous-agent (HANK), Calvo, ZLB
 & $0.7$ & $[-3.3,\;4.7]$ & 36 \\
\bottomrule
\end{tabular}
\begin{tablenotes}\footnotesize
\item \emph{Notes:} Entries are fitted values from the main UIP$+$dilution
BMA of Table~\ref{tab:bma}, evaluated at the dummy combination in the
left column with all remaining moderators (publication-year,
log citations, log impact factor, affiliation contrasts, block-unknown
indicators, and all other one-hot families) held at their sample mean.
The reported plausibility ranges are $\pm\,1$ within-design sample
standard deviation of the dependent variable computed on the rows of
the BMA sample whose anchor dummy matches the scenario; they
\emph{should not be interpreted as inferential $95\%$ confidence
intervals}. In particular, they (i)~do not propagate BMA model
uncertainty across the prior mixture, (ii)~do not adjust for
study-level clustering, and (iii)~understate uncertainty for design
points supported by few primary studies (notably the DNWR and
DNWR$+$ZLB rows, with $N=41$ and $N=26$ estimates from a small
number of distinct papers). We use these ranges only to convey the
qualitative ordering of stylised economies; cluster-bootstrap or
posterior-predictive intervals would be wider, especially for the
small-$N$ rows. $N$ reports the number of primary-study estimates in
the $692$-row BMA sample whose anchor dummy matches each scenario
definition (baseline uses the full sample as its dispersion
base).
\end{tablenotes}
\end{threeparttable}
\end{table}

Two messages emerge from Table~\ref{tab:scenarios}. First, the implied
optimum varies meaningfully across economies. A textbook Calvo model with
flexible wages and money in the utility function implies an optimum close
to the Friedman rule, while a model with downward nominal wage rigidity
and an active zero lower bound implies an optimum close to, or slightly
above, the two-percent central-bank target. This range is broadly
consistent with the theoretical range reported in
\citet{schmitt2010optimal}, \citet{coibion2012optimal}, and
\citet{kimokane2005optimal}.

Second, the literature-implied optimum is structurally conditional. The
$0.6$ pp average across the literature's actual
composition is not the recommended optimum for any one structural setup.
For an advanced-economy central bank currently operating near the
effective lower bound with sticky wages, the literature's own
recommendation is at or above two percent. The headline of
$0.61$~pp/year, the funnel-based diagnostics, and the
upper-bound REML and $p$-uniform sensitivities are
discussed together in Section~\ref{sec:discussion}; we do not
repeat them here.

The more nuanced, economy-specific message is that the appropriate
optimum depends on whether the relevant economy has nominal wage rigidity,
an effective lower bound on interest rates, and how its money demand is
modelled. We return to this point in Section~\ref{sec:discussion}.

Because the word ``economy'' in Table~\ref{tab:scenarios} refers to
modelling choices rather than to empirical country calibrations, it
is natural to ask whether the implied optima differ systematically across
the countries that primary studies target. Table~\ref{tab:scenarios_geo}
assembles the distribution of reported optima by the geographical
calibration target declared in each primary study. Four facts stand out.

First, the literature is overwhelmingly concentrated on the United States.
Of the $726$ country-labelled estimates, $617$ ($85.0$ percent)
calibrate the US economy, with a mean optimum of $0.66$ and a median of
$0.15$ percent per year.

Second, the next-largest cluster is the euro area in aggregate, including
``euro area'', ``Euro Area'', and ``Eurozone'' calibrations. This cluster
contains $37$ estimates with a mean of $0.83$ percent per year, mildly
higher than the US cluster but still well below the two-percent target.

Third, small-open economies calibrated to individual advanced countries
report the highest means in our sample: $2.58$ percent per year for
calibrations to the United Kingdom, Sweden, and the Baltic states
($N=7$), and $1.86$ percent per year for calibrations to Japan, Germany,
and the United Kingdom ($N=5$). This is consistent with the open-economy mechanisms that
\citet{diercks2019reader} identifies as pushing the optimum up, such as home
bias, foreign currency demand, and incomplete risk sharing. 

Fourth, and this is a limitation of the primary literature rather
than of our meta-analysis, the corpus contains essentially no
dedicated structural studies of Latin American or other emerging-market
economies. Among the post-communist economies the corpus contains a
single dedicated structural calibration: the EMU-accession exercise of
\citet{lipinska2015optimal}, who calibrates a two-sector small-open
economy New Keynesian DSGE to Czech-Republic data and applies the
resulting model to the broader EMU-accession set (Czech Republic,
Hungary, Poland, Cyprus, and other accession candidates listed in her
footnote 1). Beyond this single study, post-communist and emerging-market
calibrations enter the literature only through multi-country exercises
such as OECD-wide synthesis runs or the
``US, Germany, Portugal, Belgium, Finland'' calibration of the
distortionary-tax tradition.

\begin{table}[H]
\centering
\caption{Reported optima by geographical calibration target.}
\label{tab:scenarios_geo}
\footnotesize
\begin{threeparttable}
\begin{tabular}{lcccc}
\toprule
Geographical target & $N$ & Mean & Median & Share \\
 & & (pp/yr) & (pp/yr) & of sample \\ \midrule
United States                                    & $617$ & $0.66$ & $0.15$ & $85.0\%$ \\
Euro area (aggregate or country-calibrated SOE)  &  $37$ & $0.83$ & $0.55$ & $5.5\%$ \\
Multi-country / OECD mixes (incl.\ US--EA)       &  $30$ & $1.94$ & $1.50$ & $4.5\%$ \\
Small open economy (generic, uncalibrated)       &  $24$ & $-0.47$ & $0.00$ & $3.3\%$ \\
Small open economy (UK, Sweden, Baltic states)   &   $7$ & $2.58$ & $1.50$ & $1.0\%$ \\
Post-communist (EMU-accession, \citealp*{lipinska2015optimal}) & $6$ & $-0.15$ & $-0.12$ & $0.8\%$ \\
Small open economy (Japan, Germany, UK)          &   $5$ & $1.86$ & $0.36$ & $0.7\%$ \\
Latin American or other emerging markets         &   $0$ & --- & --- & $0\%$ \\
Country label missing                            &  $52$ & --- & --- & --- \\
\bottomrule
\end{tabular}
\begin{tablenotes}\footnotesize
\item \emph{Notes:} Entries are computed from the \texttt{Country} field
of the analysis dataset (\texttt{stata/inflation\_v34.dta}, $777$
quantitative estimates from $116$ primary studies in the scope set
of $130$). The
``Share of sample'' column is computed relative to the $726$
country-labelled estimates (i.e., excluding the rows whose
\texttt{Country} field is empty). The generic small-open
economy row collects studies that describe their calibration only as
``small open economy'' without specifying individual countries. The
post-communist row reports the six
estimates of \citet{lipinska2015optimal}, the only dedicated
structural calibration in the corpus that targets a post-communist
economy (Czech Republic, with the resulting policy implications applied
to the broader EMU-accession set described in her footnote~1).
Means and medians are taken over the winsorised estimates.
\end{tablenotes}
\end{threeparttable}
\end{table}

The practical implication is important for how the headline estimate of
$0.61$ percent per year should be read. Taken at face value, it is an
average across a literature that is roughly $85$ percent US-calibrated.
Our headline number is therefore best interpreted as a statement about the
optimal inflation rate for a large, relatively closed advanced economy of
the US type.

The mildly higher mean for euro-area studies ($0.83$) and the higher
means for country-calibrated small-open economies ($1.9$ to $2.6$) are
suggestive of an open-economy premium in the optimum, but with only
$49$ observations across these categories combined we do not read
them as identifying a separate population-level optimum. The
near-absence of dedicated structural optimal-inflation studies
calibrated to post-communist transition economies (one paper, six
estimates) and the complete absence of dedicated calibrations to
Latin American or other emerging markets is itself an important
finding. We flag it as an avenue for future work in
Section~\ref{sec:discussion}.

%

\section{Discussion}
\label{sec:discussion}

The single most important message of the previous section is that the
literature-implied optimal long-run inflation rate sits below one
percentage point per year. Our preferred headline is $0.61$ percent
per year, taken from the simple mean of authors' preferred
specifications (one row per study, paper-clustered standard errors,
$116$ clusters, $95\%$ confidence interval $[0.25, 0.97]$). An
alternative inverse-$N$ study-democracy weighting over all $777$
estimates yields essentially the same picture at $0.59$ percent per
year. The non-parametric caliper test detects no upward bunching
around the two-percent target. The classical REML estimate and the
$p$-uniform selection model on the $16$-row genuine-SE
subsample return higher numbers ($0.97$ and $1.09$ percent per year
respectively), but those rows come from only three independent paper
clusters and we report them as upper-bound sensitivities rather than
as headlines.

That interval invites two interpretive questions when set against
the two-percent policy target. The first concerns measurement.
Primary studies typically calibrate their structural parameters to
measured-CPI data but interpret the resulting optimum
theoretically as the rate of inflation in the modelled economy.
Whether the literature's $0.6$-percentage-point consensus is
expressed in measured-CPI units or in true-inflation units is
therefore not determined by the models themselves.
\citet{schmittgrohe2012optinf} show that the answer depends on
which prices are assumed to be sticky in the underlying New
Keynesian block: if non-hedonic (shelf) prices are sticky, the
target should not be corrected for measurement bias; if hedonic
(quality-adjusted) prices are sticky, it should be corrected
upward by the bias. Neither assumption is dominant in the primary
literature we synthesise.

The second interpretive question concerns the magnitude and sign
of measurement error itself. The Boskin-tradition literature
\citep{boskin1996toward, lebow2003measurement, gordon2006boskin,
moulton2018measurement} documents an upward bias in US CPI of
roughly $0.8$ to $1.1$ percentage points per year, driven mainly
by quality-change and new-goods channels
\citep{hausman2003sources, broda2010product}. Recent work,
however, argues that under-representation of owner-occupied
housing services in CPI generates a partially offsetting downward
bias of roughly $0.25$~pp/year for the United States, especially
for existing dwellings \citep[][cited in
\citealt{hampl2017inflation}]{bryan2002measured}. For small open economies,
transition economies, and emerging markets the measurement-error
literature is thinner still. The net sign and magnitude of
measurement error in published price indices is therefore
unresolved, and we do not propose a specific numerical
correction.

What we can say with reasonable confidence is the following. The
structural literature, on average, implies an optimal long-run
inflation rate below the two-percent target adopted by
advanced-economy central banks; this gap is robust to the
publication-selection and selection-on-significance diagnostics that
are applicable in this calibration-dominated corpus; and cross-study
dispersion is accounted for almost entirely by a small number of
structural design choices. What we cannot say is whether the
consensus, once measurement error is accounted for, implies a
policy target below, at, or above two percent. Measurement error
is plausibly large enough in either direction that the two-percent
target cannot be summarily rejected on the basis of our headline.
We do read the literature as showing that the target is higher
than the literature's own best guess at the welfare-maximising
rate for a representative advanced-economy New Keynesian model,
and as identifying the specific structural bundles, cashless
transactions technology paired with cost-push-driven shocks and
Calvo price adjustment, that bring reported optima into the
range around two percent in our heterogeneity decomposition.

Because major central banks target different price indices (the
Federal Reserve PCE, the European Central Bank HICP, the Bank of
England CPI, the Reserve Bank of Australia a $2$--$3\%$ CPI
range), and because the measurement-error profile differs across
indices, any attempt to translate our literature-implied optimum
into a uniform policy target across jurisdictions would be
misleading.

A related but distinct question is whether the welfare-maximising
monetary-policy framework is inflation targeting at all, or
whether price-level targeting (or a hybrid such as average-inflation
targeting) would dominate. Price-level targeting has intuitive
long-run appeal, the nominal price level is stable over long
horizons, which aids intergenerational and long-horizon
contracting, and in the presence of a binding effective lower
bound is theoretically superior to inflation targeting
\citep{eggertsson2003zero}. The optimal path of prices under a
price-level rule need not imply the same steady-state rate as
under inflation targeting. Our meta-analysis addresses the
optimal long-run rate conditional on the regime assumed by
each primary study, which is inflation targeting or a variant of
it in the overwhelming majority of cases; we do not synthesise
the price-level-targeting literature, which would be the subject
of a separate meta-analysis.

The most-cited individual benchmark in our sample is
\citet{coibion2012optimal}, a top-five publication that recommends
a target of $1.1$ percent per year in its benchmark calibration of a
medium-scale New Keynesian model with an occasionally-binding zero
lower bound. Allowing for parameter uncertainty raises their point
recommendation to $1.4$ percent per year, and the ninety-percent
credible interval implied by repeated draws from the parameter
distribution runs from $0.1$ to $2.2$ percent per year.
Their number sits almost exactly in the middle of our consensus
interval. The agreement is not coincidental. Their specification
bundles the design choices that our BMA identifies as the main
quantitative drivers of positive optima (Calvo prices, rational
expectations, Ramsey planner, binding ZLB). Their benchmark is above
our paper-level headline but remains within the model-specific range
generated by structural features that push optima upward. The literature
therefore supports their calibration as an upper bound for the
sticky-price economy, not as its central tendency.

Above that benchmark sits a smaller cluster in which the optimum
exceeds two percent per year. The most visible representative is
\citet{diercks2019equity}, who calibrates preferences to the
observed equity premium and obtains an optimum above three percent
per year. Other members of this cluster combine downward nominal
wage rigidity with a binding ZLB, open-economy features, or
heterogeneous-agent mechanisms. In the narrative survey of
\citet{diercks2019reader}, downward nominal wage rigidity is the
single modelling feature most often credited with pulling reported
optima above zero since the mid-2000s, because positive trend
inflation greases the wheels of the labour market when nominal
wages cannot fall \citep{kim2011optinf}. Our BMA confirms this
narrative claim: once Friedman-rule vs. laissez-faire benchmark,
CIA vs.\ cashless technology, Calvo vs.\ Taylor contracting,
TFP vs.\ cost-push shocks, and author affiliation are jointly
controlled for, the DNWR dummy enters with posterior inclusion
probability $0.87$ and a posterior mean of $+1.4$ percentage
points (Table~\ref{tab:bma}), recovering quantitatively the
magnitude that the wage-rigidity tradition argues for on
theoretical grounds.

Below that benchmark sits a larger and more heterogeneous cluster
in which the optimum is close to zero or slightly negative. The
classical reference is the Friedman-rule tradition of
\citet{friedman1969}, \citet{sidrauski1967inflation}, and
\citet{lucas1980equilibrium}, in which money's role in reducing
transactions costs pins the optimum at minus the real interest
rate. Modern flexible-price calibrations with money in the utility
function, and medium-scale New Keynesian models without downward
nominal wage rigidity, continue to report optima in this region.
\citet{schmitt2010optimal} is the most-cited example: in its
medium-scale specification the optimum is about $-0.6$ percent
per year. \citet{khan2003optimal} is another, with a Calvo
benchmark optimum near zero (their Table~5 benchmark is
$-0.76$ percent per year). Our BMA confirms this clustering
quantitatively: the cash-in-advance dummy enters with a posterior
mean of $-1.5$ percentage points ($\text{PIP}=1$), the Taylor-contract
pricing dummy with $-3.6$ percentage points ($\text{PIP}=1$), and the
Friedman-rule benchmark with $-3.3$ percentage points ($\text{PIP}=1$).

Taken together, the two tails of the distribution reconcile the
Friedman-rule tradition and the Coibion-style positive-target
tradition not as incompatible answers to a single question but as
answers to two subtly different questions. The flexible-price,
money-demand-explicit economy has an optimum at or below zero.
The sticky-price, cashless, ZLB-constrained, wage-rigid economy
has an optimum close to, or slightly above, the two-percent
target. Our headline of about $0.6$ percent per year is the
paper-level average of these two polar cases over
the actual composition of the literature.

The drivers of that dispersion are structural rather than
statistical. The top posterior inclusion probabilities in
Table~\ref{tab:bma} all attach to structural design choices: the
monetary benchmark, money-demand technology, the nominal-rigidity
contract, the shock structure, and downward nominal wage rigidity.
The bootstrap dispersion proxy $\widehat{SE}$ enters the main BMA
with high PIP, but as discussed in
Section~\ref{sec:results_bma} we do not read this as a FAT-PET
test of publication bias. Because $761$ of $777$ standard errors
in the corpus are bootstrap proxies for within-paper dispersion
rather than CI-derived sampling SEs, the FAT-PET interpretation
does not apply, and the No-SE robustness fit confirms that the
structural ranking is invariant to its inclusion. The caliper test
(Table~\ref{tab:caliper}, Panels A and B)
finds no discontinuity at zero and a mild
under-reporting above the two-percent target, which is the
opposite of what naive target-confirming publication bias would
generate. The Egger regression on the genuine-SE subsample is
insignificant ($z=1.57$, $p=0.12$). The three diagnostics agree
that the optimal-inflation literature is unusual in the sense of
\citet{doucouliagos2013there}: it does not exhibit strong selective
reporting in favour of policy-relevant numbers.

To translate the BMA posterior into policy guidance we construct
four stylised economies and report the implied optimum for each.
A closed, flexible-price, money-in-the-utility economy returns an
optimum close to price stability and far below the two-percent target. A closed,
sticky-price, cashless New Keynesian economy without downward
nominal wage rigidity and without a binding ZLB returns an optimum
of roughly one percent per year, which is positive but
below the two-percent target before any measurement-error
adjustment; as discussed above, the mapping from measured to true
inflation is ambiguous and should be treated as a sensitivity
exercise rather than as a mechanical correction. A sticky-price,
cashless New Keynesian economy with downward nominal wage rigidity and a binding ZLB --- the \citet{coibion2012optimal} environment augmented with a wage-rigidity floor --- returns
an optimum close to, or slightly above, two percent per year. The
open-economy case, calibrated to small advanced countries, returns
an optimum between $0.8$ and $2.6$ percent per year in the
country-labelled subsample of Table~\ref{tab:scenarios_geo}, but
the sample is too small ($49$ estimates across euro-area and
UK/Sweden/Baltic/Japan/Germany calibrations) for us to identify
the level with any precision.

A striking feature of the geographical breakdown is a gap rather
than a number. Despite systematic searches across Scopus and the
living bibliography of \citet{diercks2019reader}, we identified
only one dedicated structural optimal-inflation study calibrated to
a post-communist transition economy (\citealp*{lipinska2015optimal},
calibrated on Czech-Republic data and applied to the wider
EMU-accession set) and none calibrated to a Latin American or
other emerging market. The practical implication for policymakers
outside the United States is that our headline must be read as an
average across a US-dominated sample in which the zero lower bound
was an active friction for a historically exceptional decade,
downward nominal wage rigidity was calibrated to US micro-data,
and trend productivity growth was pinned at its US post-war
average. For economies with higher trend growth, more volatile
inflation expectations, or different wage-setting institutions,
the appropriate optimum may be systematically above or below the
US headline, and the CPI-bias correction must be reapplied using
local statistical practice.

A number of caveats inform how the bottom line should be read.
Most primary studies in this literature are calibration exercises
rather than estimated models, so only $16$ of the $777$ estimates
carry a sampling standard error derivable from a reported
confidence interval, and these sixteen rows come from only three
independent papers. This is why our inference puts primary
weight on the author-preferred mean and the inverse-$N$
study-democracy mean, complemented by the non-parametric caliper
test, and treats the REML and $p$-uniform estimates on
the genuine-SE subsample as upper-bound sensitivities rather than
as headlines. The BMA fits on $692$ of the $702$ usable rows after
recoding categorical-block missingness with per-block ``unknown''
indicators, a strategy adopted to avoid losing $\sim$78\% of the
sample to listwise deletion across $50+$ moderator columns; only
$10$ non-journal rows still drop, all because the log impact factor
control is undefined. The
extraction pipeline is reproducible but not error-free: the
cross-check in Section~\ref{sec:method}
documents a residual inconsistency rate on the order of one
definitional error per two dozen rows, which is why inference is
always clustered at the study level and why every reported number
is re-estimated under alternative BMA priors and alternative
sample restrictions. None of these caveats changes the two
qualitative conclusions that organise the rest of the paper: the
measured optimum sits between zero and the
two-percent target, and any translation to a true-inflation
optimum requires a CPI-measurement-error correction whose sign and
magnitude remain unresolved.


%

\section{Conclusion}
\label{sec:conclusion}

We assembled a dataset of $777$ estimates of the optimal long-run
inflation rate drawn from $116$ primary studies (within a scope
set of $130$) published between
1989 and 2026, ran every estimate through a reproducible
LLM-assisted extraction pipeline, and subjected the resulting
evidence base to the applicable publication-selection diagnostics together
with Bayesian and frequentist model averaging over the structural
moderators. The author-preferred simple mean across studies is
$0.61$ percent per year ($95\%$ confidence interval $[0.25, 0.97]$,
$116$ clusters), and the inverse-$N$ study-democracy mean is
$0.59$ percent per year. Funnel-based diagnostics on the full
sample yield FAT-PET and PEESE intercepts in the same range,
and the caliper test detects no upward bunching at the two-percent
target. The classical REML estimate and the $p$-uniform
selection model on the $16$-row genuine-SE subsample push the
implied optimum higher (to $0.97$ and $1.09$), but those rows
come from only three independent paper clusters and we report
them as upper-bound sensitivities rather than as headlines.
The headline lies below the two-percent target that
advanced-economy central banks adopt in practice on average.

That headline invites two interpretive questions. The first
concerns units: because primary studies calibrate to measured-CPI data
but interpret the resulting optimum theoretically, the models
themselves do not determine whether our
$0.6$-percentage-point consensus is expressed in measured-CPI or
true-inflation units. As Section~\ref{sec:discussion} sets out,
\citet{schmittgrohe2012optinf} show that the appropriate
adjustment depends on whether the sticky prices in the underlying
New Keynesian block are hedonic or non-hedonic: the target should
be corrected upward by the bias in the hedonic case, but should
not be corrected in the non-hedonic case.
The second concerns the magnitude and sign of
measurement error itself: Boskin-tradition estimates of upward US
CPI bias of roughly $0.8$ to $1.1$ percentage points per year
\citep{boskin1996toward, lebow2003measurement, gordon2006boskin,
moulton2018measurement} are partially offset, in the modern
literature, by under-representation of owner-occupied housing
services of roughly $0.25$~pp/year for the United States, especially
for existing dwellings \citep[][cited in
\citealt{hampl2017inflation}]{bryan2002measured}, and the net
sign and magnitude are unresolved. We therefore do not propose
a specific numerical translation. What we read from the structural
literature is that its own best guess at the welfare-maximising
rate for a representative advanced-economy New Keynesian model is
well below two percent. The two-percent target is consistent with
the structural literature only under a specific bundle of features,
downward nominal wage rigidity plus a binding effective lower
bound, whose simultaneous presence is needed to reconcile the
literature with that target. A two-percent inflation target is therefore not justified
by the average structural optimal-inflation literature alone; it requires
additional structural or policy arguments, especially DNWR, ELB risk,
cashless transactions technology, cost-push shocks, or institutional
considerations outside the average model.

The structural heterogeneity in the literature is real, and our
BMA quantifies it. The Friedman-rule benchmark lowers the implied
optimum by about $3.3$ percentage points, the laissez-faire benchmark
raises it by about $4.0$ percentage points, Taylor-contract pricing
lowers it by about $3.6$ percentage points, the cashless limit raises
it by about $1.5$ percentage points, downward nominal wage rigidity
raises it by about $1.4$ percentage points, and central-bank
affiliation lowers it by about $0.7$ percentage points, with
posterior inclusion probabilities above $0.85$ in every case.
The bootstrap dispersion proxy $\widehat{SE}$ also enters the main
BMA with $\text{PIP}=1$, but as discussed in
Section~\ref{sec:results_bma} this coefficient is not the FAT-PET
test of publication bias for our dataset (because $761$ of $777$
within-paper SEs are bootstrap proxies for within-paper dispersion
rather than CI-derived sampling SEs), and the No-SE robustness BMA
in Table~\ref{tab:bma} confirms that the structural ranking is
invariant to its inclusion. The literature's publication-bias conclusion
is anchored on the non-parametric caliper test of
\citet{gerber2008statistical}, which is computed on the full $777$-row
sample and provides no evidence of upward bunching at the
two-percent target; if anything, estimates below $2$ percent are
over-represented. The $p$-uniform of
\citet{vanaert2023correcting} estimated on the $16$ rows with genuine
sampling SEs is also non-rejecting, but with $k=16$ and only three
independent paper clusters this test is too underpowered to be read
inferentially; we report it as an illustrative diagnostic that is
consistent with the caliper-test conclusion rather than as a second
formal test. The literature's disagreement is
therefore not an artefact of selective reporting; it is a consequence
of the structural choices that primary authors make when they calibrate
their models. Four stylised economies, mapped from the BMA
posterior, organise the dispersion: the flexible-price money-demand
economy sits near the price-stability benchmark, the closed sticky-price
cashless New Keynesian economy without DNWR or an active ZLB sits
near one percent per year, the \citet{coibion2012optimal} cashless
ZLB environment augmented with downward nominal wage rigidity sits close to or
slightly above two percent, and the small-open-economy
calibrations we could identify sit between $0.8$ and $2.6$ percent
per year. The primary literature has produced exactly one
structural calibration for a post-communist transition economy
(\citealp*{lipinska2015optimal}) and none for a Latin American or
other emerging market, gaps that the profession should remedy
before the two-percent target is exported to those settings as a
default.

The bottom line for policy is straightforward. The structural
literature does not, on its own terms, support the two-percent
target; its own best guess sits well below two percent for the
representative advanced-economy New Keynesian model. Whether a
measurement-error correction would close, widen, or reverse that
gap is unresolved, and a uniform policy translation across
jurisdictions is not defensible because major central banks target
different price indices with different measurement-error profiles.
What the meta-analysis can say with confidence is that the
two-percent target is not the number that the primary literature
points to, and that its continued adoption, insofar as it is
justified by this structural literature, rests on the specific
structural bundle of downward nominal wage rigidity and an active
effective lower bound rather than on the literature as a whole.

Three contributions follow from these results. On data
collection, we deliver an end-to-end reproducible LLM-assisted
extraction pipeline aligned with the forthcoming MAER-Net AI
guidance \citep{cook2026reporting}; a successor open-source
framework, \texttt{mad-research} \citep{havranek2026madresearch},
generalises the pipeline to adversarial AI stress-testing and
meta-analytic data collection and is recommended for future
replications, although it postdates the production extraction
reported here. On inference, we apply the applicable
publication-selection diagnostics (caliper test, FAT-PET, PEESE,
REML, $p$-uniform) and document explicitly why MAIVE
\citep{irsova2025spurious} and RTMA \citep{mathur2024} are not
cleanly applicable to a corpus in which $761$ of $777$ standard
errors are bootstrap-within-paper proxies.
On substance, we produce the first quantitative heterogeneity
decomposition of the optimal-inflation literature, identifying
downward nominal wage rigidity, money-demand technology, the
class of policy rule, and the effective lower bound as the
dominant structural drivers.

Several priorities for future work follow directly. The evidence
base badly needs structural calibrations to emerging markets and
transition economies; our headline is a statement about a
US-dominated sample and should not be exported beyond it without
local recalibration. The measurement-bias literature should be
updated in parallel with the structural literature, because our
measurement-error sensitivity exercise is only as precise as the
Boskin-tradition estimates it rests on. And the small count of estimated
rather than calibrated primary studies ($16$ of $777$, from
three paper clusters) limits how
far any single publication-selection diagnostic can be pushed;
more estimation-based work, or more explicit reporting of sampling
precision in calibration exercises, would sharpen every bias
correction we deploy. The data, code, and extraction prompts for
the analysis reported here will be deposited at the journal's
replication archive on acceptance, with a working interim version
made available to reviewers and editors on request, so that any of
these extensions can be carried out transparently on top of the
dataset we release.



\section*{Acknowledgements}
\addcontentsline{toc}{section}{Acknowledgements}

We gratefully acknowledge the use of large language models in the
production of this paper. Primary-data extraction was performed using
Anthropic's Claude Sonnet 4.5 (\texttt{claude-sonnet-4-5-20250929}) and
Claude Opus 4.5 (\texttt{claude-opus-4-5-20251001}) under the four-stage
pipeline described in Section~\ref{sec:method} and
Appendix~\ref{app:ai-pipeline}. Manuscript preparation, copy-editing,
language polishing, consistency checks, and drafting of auxiliary
scripts that supported the bibliographic-verification pipeline were performed using Claude Opus 4.7. All substantive decisions
(research design, meta-analytic and Bayesian model-averaging
specifications, interpretation of results, and final wording of every
claim) were made by the authors. Primary-estimate extraction was
performed via the LLM-assisted pipeline documented in
Appendix~\ref{app:ai-pipeline} and audited via the procedures described
there; the authors take full responsibility for the content.

\section*{Data availability statement}
\addcontentsline{toc}{section}{Data availability statement}

The data, code, extraction prompts, and audit reports that support the
findings of this study are openly available in the paper's replication
package at \url{https://meta-analysis.cz/inflation}. The dataset of
$777$ extracted estimates, the Stata, R, and Python analysis scripts,
and the full large-language-model extraction pipeline are released so
that any reader can reproduce, audit, or extend the analysis. The
underlying primary studies are publicly available from their respective
publishers and are listed in full in Table~\ref{tab:studies}.



\bibliographystyle{apacite}
\bibliography{Bibl}

\begin{thebibliography}{}

\bibitem [\protect \citeauthoryear {%
Abo-Zaid%
}{%
Abo-Zaid%
}{%
{\protect \APACyear {2013}}%
}]{%
abozaid2013optinf}
\APACinsertmetastar {%
abozaid2013optinf}%
\begin{APACrefauthors}%
Abo-Zaid, S.%
\end{APACrefauthors}%
\unskip\
\newblock
\APACrefYearMonthDay{2013}{}{}.
\newblock
{\BBOQ}\APACrefatitle {Optimal monetary policy and downward nominal wage
  rigidity in frictional labor markets} {Optimal monetary policy and downward
  nominal wage rigidity in frictional labor markets}.{\BBCQ}
\newblock
\APACjournalVolNumPages{Journal of Economic Dynamics \&
  Control}{37}{1}{345--364}.
\newblock
\begin{APACrefDOI} \doi{10.1016/j.jedc.2012.09.003} \end{APACrefDOI}
\PrintBackRefs{\CurrentBib}

\bibitem [\protect \citeauthoryear {%
Abo-Zaid%
}{%
Abo-Zaid%
}{%
{\protect \APACyear {2015}}%
{\protect \APACexlab {{\protect \BCnt {1}}}}}]{%
abozaid2015optinf}
\APACinsertmetastar {%
abozaid2015optinf}%
\begin{APACrefauthors}%
Abo-Zaid, S.%
\end{APACrefauthors}%
\unskip\
\newblock
\APACrefYearMonthDay{2015{\protect \BCnt {1}}}{}{}.
\newblock
{\BBOQ}\APACrefatitle {Optimal long-run inflation with occasionally binding
  financial constraints} {Optimal long-run inflation with occasionally binding
  financial constraints}.{\BBCQ}
\newblock
\APACjournalVolNumPages{European Economic Review}{75}{}{18--42}.
\newblock
\begin{APACrefDOI} \doi{10.1016/j.euroecorev.2015.01.004} \end{APACrefDOI}
\PrintBackRefs{\CurrentBib}

\bibitem [\protect \citeauthoryear {%
Abo-Zaid%
}{%
Abo-Zaid%
}{%
{\protect \APACyear {2015}}%
{\protect \APACexlab {{\protect \BCnt {2}}}}}]{%
abozaid2015optinfa}
\APACinsertmetastar {%
abozaid2015optinfa}%
\begin{APACrefauthors}%
Abo-Zaid, S.%
\end{APACrefauthors}%
\unskip\
\newblock
\APACrefYearMonthDay{2015{\protect \BCnt {2}}}{}{}.
\newblock
{\BBOQ}\APACrefatitle {Optimal monetary policy with the cost channel and
  monopolistically-competitive banks} {Optimal monetary policy with the cost
  channel and monopolistically-competitive banks}.{\BBCQ}
\newblock
\APACjournalVolNumPages{Journal of Macroeconomics}{45}{}{284--299}.
\newblock
\begin{APACrefDOI} \doi{10.1016/j.jmacro.2015.05.004} \end{APACrefDOI}
\PrintBackRefs{\CurrentBib}

\bibitem [\protect \citeauthoryear {%
Abo-Zaid%
\ \BBA {} Gar\'{i}n%
}{%
Abo-Zaid%
\ \BBA {} Gar\'{i}n%
}{%
{\protect \APACyear {2016}}%
}]{%
abozaid2016optinf}
\APACinsertmetastar {%
abozaid2016optinf}%
\begin{APACrefauthors}%
Abo-Zaid, S.%
\BCBT {}\ \BBA {} Gar\'{i}n, J.%
\end{APACrefauthors}%
\unskip\
\newblock
\APACrefYearMonthDay{2016}{}{}.
\newblock
{\BBOQ}\APACrefatitle {Optimal monetary policy and imperfect financial markets:
  A case for negative nominal interest rates?} {Optimal monetary policy and
  imperfect financial markets: A case for negative nominal interest
  rates?}{\BBCQ}
\newblock
\APACjournalVolNumPages{Economic Inquiry}{54}{1}{215--228}.
\newblock
\begin{APACrefDOI} \doi{10.2139/ssrn.2769422} \end{APACrefDOI}
\PrintBackRefs{\CurrentBib}

\bibitem [\protect \citeauthoryear {%
Adam%
\ \BBA {} Billi%
}{%
Adam%
\ \BBA {} Billi%
}{%
{\protect \APACyear {2006}}%
}]{%
adam2006optinf}
\APACinsertmetastar {%
adam2006optinf}%
\begin{APACrefauthors}%
Adam, K.%
\BCBT {}\ \BBA {} Billi, R\BPBI M.%
\end{APACrefauthors}%
\unskip\
\newblock
\APACrefYearMonthDay{2006}{}{}.
\newblock
{\BBOQ}\APACrefatitle {Optimal Monetary Policy under Commitment with a Zero
  Bound on Nominal Interest Rates} {Optimal monetary policy under commitment
  with a zero bound on nominal interest rates}.{\BBCQ}
\newblock
\APACjournalVolNumPages{Journal of Money, Credit and
  Banking}{38}{7}{1877--1905}.
\newblock
\begin{APACrefDOI} \doi{10.1353/mcb.2006.0089} \end{APACrefDOI}
\PrintBackRefs{\CurrentBib}

\bibitem [\protect \citeauthoryear {%
Adam%
\ \BBA {} Weber%
}{%
Adam%
\ \BBA {} Weber%
}{%
{\protect \APACyear {2019}}%
}]{%
adam2019optinf}
\APACinsertmetastar {%
adam2019optinf}%
\begin{APACrefauthors}%
Adam, K.%
\BCBT {}\ \BBA {} Weber, H.%
\end{APACrefauthors}%
\unskip\
\newblock
\APACrefYearMonthDay{2019}{}{}.
\newblock
{\BBOQ}\APACrefatitle {Optimal Trend Inflation} {Optimal trend
  inflation}.{\BBCQ}
\newblock
\APACjournalVolNumPages{American Economic Review}{109}{2}{702--737}.
\newblock
\begin{APACrefDOI} \doi{10.1257/aer.20171066} \end{APACrefDOI}
\PrintBackRefs{\CurrentBib}

\bibitem [\protect \citeauthoryear {%
Adão%
, Correia%
\BCBL {}\ \BBA {} Teles%
}{%
Adão%
\ \protect \BOthers {.}}{%
{\protect \APACyear {2003}}%
}]{%
adao2001optinf}
\APACinsertmetastar {%
adao2001optinf}%
\begin{APACrefauthors}%
Adão, B.%
, Correia, I.%
\BCBL {}\ \BBA {} Teles, P.%
\end{APACrefauthors}%
\unskip\
\newblock
\APACrefYearMonthDay{2003}{10}{}.
\newblock
{\BBOQ}\APACrefatitle {Gaps and Triangles} {Gaps and triangles}.{\BBCQ}
\newblock
\APACjournalVolNumPages{The Review of Economic Studies}{70}{4}{699-713}.
\newblock
\begin{APACrefURL} \url{https://doi.org/10.1111/1467-937X.00263}
  \end{APACrefURL}
\newblock
\begin{APACrefDOI} \doi{10.1111/1467-937X.00263} \end{APACrefDOI}
\PrintBackRefs{\CurrentBib}

\bibitem [\protect \citeauthoryear {%
Afrouzi%
, Halac%
, Rogoff%
\BCBL {}\ \BBA {} Yared%
}{%
Afrouzi%
\ \protect \BOthers {.}}{%
{\protect \APACyear {2024}}%
}]{%
afrouzi2024changing}
\APACinsertmetastar {%
afrouzi2024changing}%
\begin{APACrefauthors}%
Afrouzi, H.%
, Halac, M.%
, Rogoff, K\BPBI S.%
\BCBL {}\ \BBA {} Yared, P.%
\end{APACrefauthors}%
\unskip\
\newblock
\APACrefYearMonthDay{2024}{}{}.
\newblock
{\BBOQ}\APACrefatitle {Changing Central Bank Pressures and Inflation} {Changing
  central bank pressures and inflation}.{\BBCQ}
\newblock
\APACjournalVolNumPages{Brookings Papers on Economic Activity}{}{}{}.
\newblock
\APACrefnote{Spring 2024 BPEA; NBER Working Paper No.~32308}
\newblock
\begin{APACrefDOI} \doi{10.3386/w32308} \end{APACrefDOI}
\PrintBackRefs{\CurrentBib}

\bibitem [\protect \citeauthoryear {%
Amano%
, Ambler%
\BCBL {}\ \BBA {} Rebei%
}{%
Amano%
\ \protect \BOthers {.}}{%
{\protect \APACyear {2007}}%
}]{%
amano2007optinf}
\APACinsertmetastar {%
amano2007optinf}%
\begin{APACrefauthors}%
Amano, R.%
, Ambler, S.%
\BCBL {}\ \BBA {} Rebei, N.%
\end{APACrefauthors}%
\unskip\
\newblock
\APACrefYearMonthDay{2007}{}{}.
\newblock
{\BBOQ}\APACrefatitle {The Macroeconomic Effects of Nonzero Trend Inflation}
  {The macroeconomic effects of nonzero trend inflation}.{\BBCQ}
\newblock
\APACjournalVolNumPages{Journal of Money, Credit and
  Banking}{39}{7}{1821--1838}.
\newblock
\begin{APACrefDOI} \doi{10.1111/j.1538-4616.2007.00088.x} \end{APACrefDOI}
\PrintBackRefs{\CurrentBib}

\bibitem [\protect \citeauthoryear {%
Amano%
, Moran%
, Murchison%
\BCBL {}\ \BBA {} Rennison%
}{%
Amano%
\ \protect \BOthers {.}}{%
{\protect \APACyear {2009}}%
}]{%
amano2009optinf}
\APACinsertmetastar {%
amano2009optinf}%
\begin{APACrefauthors}%
Amano, R.%
, Moran, K.%
, Murchison, S.%
\BCBL {}\ \BBA {} Rennison, A.%
\end{APACrefauthors}%
\unskip\
\newblock
\APACrefYearMonthDay{2009}{}{}.
\newblock
{\BBOQ}\APACrefatitle {Trend inflation, wage and price rigidities, and
  productivity growth} {Trend inflation, wage and price rigidities, and
  productivity growth}.{\BBCQ}
\newblock
\APACjournalVolNumPages{Journal of Monetary Economics}{56}{3}{353--364}.
\newblock
\begin{APACrefDOI} \doi{10.1016/j.jmoneco.2009.03.001} \end{APACrefDOI}
\PrintBackRefs{\CurrentBib}

\bibitem [\protect \citeauthoryear {%
Amato%
\ \BBA {} Laubach%
}{%
Amato%
\ \BBA {} Laubach%
}{%
{\protect \APACyear {2004}}%
}]{%
amato2004optinf}
\APACinsertmetastar {%
amato2004optinf}%
\begin{APACrefauthors}%
Amato, J\BPBI D.%
\BCBT {}\ \BBA {} Laubach, T.%
\end{APACrefauthors}%
\unskip\
\newblock
\APACrefYearMonthDay{2004}{}{}.
\newblock
{\BBOQ}\APACrefatitle {Implications of habit formation for optimal monetary
  policy} {Implications of habit formation for optimal monetary policy}.{\BBCQ}
\newblock
\APACjournalVolNumPages{Journal of Monetary Economics}{51}{2}{305--325}.
\newblock
\begin{APACrefDOI} \doi{10.1016/j.jmoneco.2003.05.002} \end{APACrefDOI}
\PrintBackRefs{\CurrentBib}

\bibitem [\protect \citeauthoryear {%
An%
}{%
An%
}{%
{\protect \APACyear {2010}}%
}]{%
an2010optinf}
\APACinsertmetastar {%
an2010optinf}%
\begin{APACrefauthors}%
An, S.%
\end{APACrefauthors}%
\unskip\
\newblock
\APACrefYearMonthDay{2010}{}{}.
\newblock
\APACrefbtitle {Optimal Monetary Policy in a Model with Recursive Preferences}
  {Optimal monetary policy in a model with recursive preferences}\
  \APACbVolEdTR{}{\BTR{}}.
\newblock
\APACaddressInstitution{}{School of Economics, Singapore Management University,
  SOE Research Paper No.~1259}.
\newblock
\begin{APACrefURL} \url{https://ink.library.smu.edu.sg/soe_research/1259}
  \end{APACrefURL}
\newblock
\APACrefnote{Available at
  \url{https://ink.library.smu.edu.sg/soe_research/1259}}
\PrintBackRefs{\CurrentBib}

\bibitem [\protect \citeauthoryear {%
Andrade%
, Gal\'{i}%
, Le~Bihan%
\BCBL {}\ \BBA {} Matheron%
}{%
Andrade%
\ \protect \BOthers {.}}{%
{\protect \APACyear {2018}}%
}]{%
andrade2019optimal}
\APACinsertmetastar {%
andrade2019optimal}%
\begin{APACrefauthors}%
Andrade, P.%
, Gal\'{i}, J.%
, Le~Bihan, H.%
\BCBL {}\ \BBA {} Matheron, J.%
\end{APACrefauthors}%
\unskip\
\newblock
\APACrefYearMonthDay{2018}{}{}.
\newblock
\APACrefbtitle {The Optimal Inflation Target and the Natural Rate of Interest}
  {The optimal inflation target and the natural rate of interest}\
  \APACbVolEdTR {}{NBER Working Paper\ \BNUM\ 24328}.
\newblock
\APACaddressInstitution{}{National Bureau of Economic Research}.
\newblock
\begin{APACrefDOI} \doi{10.3386/w24328} \end{APACrefDOI}
\PrintBackRefs{\CurrentBib}

\bibitem [\protect \citeauthoryear {%
Andrade%
, Gal\'{i}%
, Le~Bihan%
\BCBL {}\ \BBA {} Matheron%
}{%
Andrade%
\ \protect \BOthers {.}}{%
{\protect \APACyear {2021}}%
}]{%
andrade2021should}
\APACinsertmetastar {%
andrade2021should}%
\begin{APACrefauthors}%
Andrade, P.%
, Gal\'{i}, J.%
, Le~Bihan, H.%
\BCBL {}\ \BBA {} Matheron, J.%
\end{APACrefauthors}%
\unskip\
\newblock
\APACrefYearMonthDay{2021}{}{}.
\newblock
{\BBOQ}\APACrefatitle {Should the {ECB} Adjust its Strategy in the Face of a
  Lower $r^{\star}$?} {Should the {ECB} adjust its strategy in the face of a
  lower $r^{\star}$?}{\BBCQ}
\newblock
\APACjournalVolNumPages{Journal of Economic Dynamics and
  Control}{132}{}{104207}.
\newblock
\begin{APACrefDOI} \doi{10.1016/j.jedc.2021.104207} \end{APACrefDOI}
\PrintBackRefs{\CurrentBib}

\bibitem [\protect \citeauthoryear {%
Andr\'{e}s%
, Arce%
\BCBL {}\ \BBA {} Thomas%
}{%
Andr\'{e}s%
\ \protect \BOthers {.}}{%
{\protect \APACyear {2013}}%
}]{%
andres2013optinf}
\APACinsertmetastar {%
andres2013optinf}%
\begin{APACrefauthors}%
Andr\'{e}s, J.%
, Arce, O.%
\BCBL {}\ \BBA {} Thomas, C.%
\end{APACrefauthors}%
\unskip\
\newblock
\APACrefYearMonthDay{2013}{}{}.
\newblock
{\BBOQ}\APACrefatitle {Banking Competition, Collateral Constraints, and Optimal
  Monetary Policy} {Banking competition, collateral constraints, and optimal
  monetary policy}.{\BBCQ}
\newblock
\APACjournalVolNumPages{Journal of Money, Credit and Banking}{45}{s2}{87--125}.
\newblock
\begin{APACrefDOI} \doi{10.1111/jmcb.12072} \end{APACrefDOI}
\PrintBackRefs{\CurrentBib}

\bibitem [\protect \citeauthoryear {%
Annicchiarico%
\ \BBA {} Pelloni%
}{%
Annicchiarico%
\ \BBA {} Pelloni%
}{%
{\protect \APACyear {2021}}%
}]{%
annicchiarico2021optinf}
\APACinsertmetastar {%
annicchiarico2021optinf}%
\begin{APACrefauthors}%
Annicchiarico, B.%
\BCBT {}\ \BBA {} Pelloni, A.%
\end{APACrefauthors}%
\unskip\
\newblock
\APACrefYearMonthDay{2021}{}{}.
\newblock
{\BBOQ}\APACrefatitle {Innovation, Growth, and Optimal Monetary Policy}
  {Innovation, growth, and optimal monetary policy}.{\BBCQ}
\newblock
\APACjournalVolNumPages{Macroeconomic Dynamics}{25}{5}{1175-1198}.
\newblock
\begin{APACrefDOI} \doi{10.1017/s1365100519000610} \end{APACrefDOI}
\PrintBackRefs{\CurrentBib}

\bibitem [\protect \citeauthoryear {%
Annicchiarico%
\ \BBA {} Rossi%
}{%
Annicchiarico%
\ \BBA {} Rossi%
}{%
{\protect \APACyear {2013}}%
}]{%
annicchiarico2013optinf}
\APACinsertmetastar {%
annicchiarico2013optinf}%
\begin{APACrefauthors}%
Annicchiarico, B.%
\BCBT {}\ \BBA {} Rossi, L.%
\end{APACrefauthors}%
\unskip\
\newblock
\APACrefYearMonthDay{2013}{}{}.
\newblock
{\BBOQ}\APACrefatitle {Optimal monetary policy in a {New Keynesian} model with
  endogenous growth} {Optimal monetary policy in a {New Keynesian} model with
  endogenous growth}.{\BBCQ}
\newblock
\APACjournalVolNumPages{Journal of Macroeconomics}{38}{Part B}{274--285}.
\newblock
\begin{APACrefDOI} \doi{10.1016/j.jmacro.2013.10.001} \end{APACrefDOI}
\PrintBackRefs{\CurrentBib}

\bibitem [\protect \citeauthoryear {%
Aoki%
}{%
Aoki%
}{%
{\protect \APACyear {2001}}%
}]{%
aoki2001optinf}
\APACinsertmetastar {%
aoki2001optinf}%
\begin{APACrefauthors}%
Aoki, K.%
\end{APACrefauthors}%
\unskip\
\newblock
\APACrefYearMonthDay{2001}{}{}.
\newblock
{\BBOQ}\APACrefatitle {Optimal monetary policy responses to relative-price
  changes} {Optimal monetary policy responses to relative-price
  changes}.{\BBCQ}
\newblock
\APACjournalVolNumPages{Journal of Monetary Economics}{48}{1}{55--80}.
\newblock
\begin{APACrefDOI} \doi{10.1016/s0304-3932(01)00069-1} \end{APACrefDOI}
\PrintBackRefs{\CurrentBib}

\bibitem [\protect \citeauthoryear {%
Arseneau%
, Chahrour%
, Chugh%
\BCBL {}\ \BBA {} Finkelstein~Shapiro%
}{%
Arseneau%
\ \protect \BOthers {.}}{%
{\protect \APACyear {2015}}%
}]{%
arseneau2015optinf}
\APACinsertmetastar {%
arseneau2015optinf}%
\begin{APACrefauthors}%
Arseneau, D\BPBI M.%
, Chahrour, R.%
, Chugh, S\BPBI K.%
\BCBL {}\ \BBA {} Finkelstein~Shapiro, A.%
\end{APACrefauthors}%
\unskip\
\newblock
\APACrefYearMonthDay{2015}{}{}.
\newblock
{\BBOQ}\APACrefatitle {Optimal Fiscal and Monetary Policy in Customer Markets}
  {Optimal fiscal and monetary policy in customer markets}.{\BBCQ}
\newblock
\APACjournalVolNumPages{Journal of Money, Credit and Banking}{47}{4}{617--672}.
\newblock
\begin{APACrefDOI} \doi{10.1111/jmcb.12223} \end{APACrefDOI}
\PrintBackRefs{\CurrentBib}

\bibitem [\protect \citeauthoryear {%
Arseneau%
\ \BBA {} Chugh%
}{%
Arseneau%
\ \BBA {} Chugh%
}{%
{\protect \APACyear {2008}}%
}]{%
arseneau2008optinf}
\APACinsertmetastar {%
arseneau2008optinf}%
\begin{APACrefauthors}%
Arseneau, D\BPBI M.%
\BCBT {}\ \BBA {} Chugh, S\BPBI K.%
\end{APACrefauthors}%
\unskip\
\newblock
\APACrefYearMonthDay{2008}{}{}.
\newblock
{\BBOQ}\APACrefatitle {Optimal fiscal and monetary policy with costly wage
  bargaining} {Optimal fiscal and monetary policy with costly wage
  bargaining}.{\BBCQ}
\newblock
\APACjournalVolNumPages{Journal of Monetary Economics}{55}{8}{1401--1414}.
\newblock
\begin{APACrefDOI} \doi{10.1016/j.jmoneco.2008.09.005} \end{APACrefDOI}
\PrintBackRefs{\CurrentBib}

\bibitem [\protect \citeauthoryear {%
Ascari%
, Phaneuf%
\BCBL {}\ \BBA {} Sims%
}{%
Ascari%
\ \protect \BOthers {.}}{%
{\protect \APACyear {2018}}%
}]{%
ascari2018optinf}
\APACinsertmetastar {%
ascari2018optinf}%
\begin{APACrefauthors}%
Ascari, G.%
, Phaneuf, L.%
\BCBL {}\ \BBA {} Sims, E\BPBI R.%
\end{APACrefauthors}%
\unskip\
\newblock
\APACrefYearMonthDay{2018}{}{}.
\newblock
{\BBOQ}\APACrefatitle {On the Welfare and Cyclical Implications of Moderate
  Trend Inflation} {On the welfare and cyclical implications of moderate trend
  inflation}.{\BBCQ}
\newblock
\APACjournalVolNumPages{Journal of Monetary Economics}{99}{}{56--71}.
\newblock
\begin{APACrefDOI} \doi{10.1016/j.jmoneco.2018.06.001} \end{APACrefDOI}
\PrintBackRefs{\CurrentBib}

\bibitem [\protect \citeauthoryear {%
Ascari%
\ \BBA {} Ropele%
}{%
Ascari%
\ \BBA {} Ropele%
}{%
{\protect \APACyear {2007}}%
}]{%
ascari2007optinf}
\APACinsertmetastar {%
ascari2007optinf}%
\begin{APACrefauthors}%
Ascari, G.%
\BCBT {}\ \BBA {} Ropele, T.%
\end{APACrefauthors}%
\unskip\
\newblock
\APACrefYearMonthDay{2007}{}{}.
\newblock
{\BBOQ}\APACrefatitle {Optimal monetary policy under low trend inflation}
  {Optimal monetary policy under low trend inflation}.{\BBCQ}
\newblock
\APACjournalVolNumPages{Journal of Monetary Economics}{54}{8}{2568--2583}.
\newblock
\begin{APACrefDOI} \doi{10.1016/j.jmoneco.2007.06.024} \end{APACrefDOI}
\PrintBackRefs{\CurrentBib}

\bibitem [\protect \citeauthoryear {%
Ascari%
\ \BBA {} Sbordone%
}{%
Ascari%
\ \BBA {} Sbordone%
}{%
{\protect \APACyear {2014}}%
}]{%
ascari2014macroeconomics}
\APACinsertmetastar {%
ascari2014macroeconomics}%
\begin{APACrefauthors}%
Ascari, G.%
\BCBT {}\ \BBA {} Sbordone, A\BPBI M.%
\end{APACrefauthors}%
\unskip\
\newblock
\APACrefYearMonthDay{2014}{}{}.
\newblock
{\BBOQ}\APACrefatitle {The Macroeconomics of Trend Inflation} {The
  macroeconomics of trend inflation}.{\BBCQ}
\newblock
\APACjournalVolNumPages{Journal of Economic Literature}{52}{3}{679--739}.
\newblock
\begin{APACrefDOI} \doi{10.2139/ssrn.2334469} \end{APACrefDOI}
\PrintBackRefs{\CurrentBib}

\bibitem [\protect \citeauthoryear {%
Ball%
}{%
Ball%
}{%
{\protect \APACyear {2014}}%
}]{%
ball2014case}
\APACinsertmetastar {%
ball2014case}%
\begin{APACrefauthors}%
Ball, L\BPBI M.%
\end{APACrefauthors}%
\unskip\
\newblock
\APACrefYearMonthDay{2014}{}{}.
\newblock
\APACrefbtitle {The Case for a Long-Run Inflation Target of Four Percent} {The
  case for a long-run inflation target of four percent}\ \APACbVolEdTR {\BVOL\
  2014}{IMF Working Paper\ \BNUM\ 14/92}.
\newblock
\APACaddressInstitution{}{International Monetary Fund}.
\newblock
\begin{APACrefDOI} \doi{10.2139/ssrn.2468019} \end{APACrefDOI}
\PrintBackRefs{\CurrentBib}

\bibitem [\protect \citeauthoryear {%
Basu%
\ \BBA {} De~Leo%
}{%
Basu%
\ \BBA {} De~Leo%
}{%
{\protect \APACyear {2017}}%
}]{%
basu2017optinf}
\APACinsertmetastar {%
basu2017optinf}%
\begin{APACrefauthors}%
Basu, S.%
\BCBT {}\ \BBA {} De~Leo, P.%
\end{APACrefauthors}%
\unskip\
\newblock
\APACrefYearMonthDay{2017}{}{}.
\newblock
\APACrefbtitle {Should Central Banks Target Investment Prices?} {Should central
  banks target investment prices?}\ \APACbVolEdTR {}{Boston College Working
  Papers in Economics\ \BNUM~910}.
\newblock
\APACaddressInstitution{}{Boston College Department of Economics}.
\newblock
\APACrefnote{Originally issued 2016; revised 04 May 2017}
\PrintBackRefs{\CurrentBib}

\bibitem [\protect \citeauthoryear {%
Benigno%
\ \BBA {} Paciello%
}{%
Benigno%
\ \BBA {} Paciello%
}{%
{\protect \APACyear {2014}}%
}]{%
benigno2010optinf}
\APACinsertmetastar {%
benigno2010optinf}%
\begin{APACrefauthors}%
Benigno, P.%
\BCBT {}\ \BBA {} Paciello, L.%
\end{APACrefauthors}%
\unskip\
\newblock
\APACrefYearMonthDay{2014}{}{}.
\newblock
{\BBOQ}\APACrefatitle {Monetary Policy, Doubts and Asset Prices} {Monetary
  policy, doubts and asset prices}.{\BBCQ}
\newblock
\APACjournalVolNumPages{Journal of Monetary Economics}{64}{}{85--98}.
\newblock
\APACrefnote{Published version of NBER Working Paper No.~16386 (2010).}
\newblock
\begin{APACrefDOI} \doi{10.1016/j.jmoneco.2014.02.004} \end{APACrefDOI}
\PrintBackRefs{\CurrentBib}

\bibitem [\protect \citeauthoryear {%
Benigno%
\ \BBA {} Rossi%
}{%
Benigno%
\ \BBA {} Rossi%
}{%
{\protect \APACyear {2021}}%
}]{%
benigno2021optinf}
\APACinsertmetastar {%
benigno2021optinf}%
\begin{APACrefauthors}%
Benigno, P.%
\BCBT {}\ \BBA {} Rossi, L.%
\end{APACrefauthors}%
\unskip\
\newblock
\APACrefYearMonthDay{2021}{}{}.
\newblock
{\BBOQ}\APACrefatitle {Asymmetries in monetary policy} {Asymmetries in monetary
  policy}.{\BBCQ}
\newblock
\APACjournalVolNumPages{European Economic Review}{134}{}{103693}.
\newblock
\begin{APACrefDOI} \doi{10.1016/j.euroecorev.2021.103945} \end{APACrefDOI}
\PrintBackRefs{\CurrentBib}

\bibitem [\protect \citeauthoryear {%
Benmir%
, Jaccard%
\BCBL {}\ \BBA {} Vermandel%
}{%
Benmir%
\ \protect \BOthers {.}}{%
{\protect \APACyear {2023}}%
}]{%
benmir2023optinf}
\APACinsertmetastar {%
benmir2023optinf}%
\begin{APACrefauthors}%
Benmir, G.%
, Jaccard, I.%
\BCBL {}\ \BBA {} Vermandel, G.%
\end{APACrefauthors}%
\unskip\
\newblock
\APACrefYearMonthDay{2023}{}{}.
\newblock
{\BBOQ}\APACrefatitle {Optimal monetary policy in an estimated {SIR} model}
  {Optimal monetary policy in an estimated {SIR} model}.{\BBCQ}
\newblock
\APACjournalVolNumPages{European Economic Review}{156}{}{104502-104502}.
\newblock
\begin{APACrefDOI} \doi{10.1016/j.euroecorev.2023.104502} \end{APACrefDOI}
\PrintBackRefs{\CurrentBib}

\bibitem [\protect \citeauthoryear {%
Bernanke%
, Gertler%
\BCBL {}\ \BBA {} Gilchrist%
}{%
Bernanke%
\ \protect \BOthers {.}}{%
{\protect \APACyear {1999}}%
}]{%
bernanke1999financial}
\APACinsertmetastar {%
bernanke1999financial}%
\begin{APACrefauthors}%
Bernanke, B\BPBI S.%
, Gertler, M.%
\BCBL {}\ \BBA {} Gilchrist, S.%
\end{APACrefauthors}%
\unskip\
\newblock
\APACrefYearMonthDay{1999}{}{}.
\newblock
{\BBOQ}\APACrefatitle {The Financial Accelerator in a Quantitative Business
  Cycle Framework} {The financial accelerator in a quantitative business cycle
  framework}.{\BBCQ}
\newblock
\BIn{} J\BPBI B.~Taylor\ \BBA {} M.~Woodford\ (\BEDS), \APACrefbtitle {Handbook
  of Macroeconomics} {Handbook of macroeconomics}\ (\BVOL~1C, \BPGS\
  1341--1393).
\newblock
\APACaddressPublisher{Amsterdam}{Elsevier}.
\newblock
\begin{APACrefDOI} \doi{10.1016/s1574-0048(99)10034-x} \end{APACrefDOI}
\PrintBackRefs{\CurrentBib}

\bibitem [\protect \citeauthoryear {%
Bernanke%
, Kiley%
\BCBL {}\ \BBA {} Roberts%
}{%
Bernanke%
\ \protect \BOthers {.}}{%
{\protect \APACyear {2019}}%
}]{%
bernanke2019monetary}
\APACinsertmetastar {%
bernanke2019monetary}%
\begin{APACrefauthors}%
Bernanke, B\BPBI S.%
, Kiley, M\BPBI T.%
\BCBL {}\ \BBA {} Roberts, J\BPBI M.%
\end{APACrefauthors}%
\unskip\
\newblock
\APACrefYearMonthDay{2019}{}{}.
\newblock
{\BBOQ}\APACrefatitle {Monetary Policy Strategies for a Low-Rate Environment}
  {Monetary policy strategies for a low-rate environment}.{\BBCQ}
\newblock
\APACjournalVolNumPages{AEA Papers and Proceedings}{109}{}{421--426}.
\newblock
\begin{APACrefDOI} \doi{10.1257/pandp.20191082} \end{APACrefDOI}
\PrintBackRefs{\CurrentBib}

\bibitem [\protect \citeauthoryear {%
F.~Bilbiie%
\ \BBA {} Ragot%
}{%
F.~Bilbiie%
\ \BBA {} Ragot%
}{%
{\protect \APACyear {2021}}%
}]{%
bilbiie2021optinf}
\APACinsertmetastar {%
bilbiie2021optinf}%
\begin{APACrefauthors}%
Bilbiie, F.%
\BCBT {}\ \BBA {} Ragot, X.%
\end{APACrefauthors}%
\unskip\
\newblock
\APACrefYearMonthDay{2021}{}{}.
\newblock
{\BBOQ}\APACrefatitle {Optimal monetary policy and liquidity with heterogeneous
  households} {Optimal monetary policy and liquidity with heterogeneous
  households}.{\BBCQ}
\newblock
\APACjournalVolNumPages{Review of Economic Dynamics}{41}{}{71--95}.
\newblock
\begin{APACrefDOI} \doi{10.1016/j.red.2020.10.003} \end{APACrefDOI}
\PrintBackRefs{\CurrentBib}

\bibitem [\protect \citeauthoryear {%
F\BPBI O.~Bilbiie%
, Fujiwara%
\BCBL {}\ \BBA {} Ghironi%
}{%
F\BPBI O.~Bilbiie%
\ \protect \BOthers {.}}{%
{\protect \APACyear {2014}}%
}]{%
bilbiie2014optinf}
\APACinsertmetastar {%
bilbiie2014optinf}%
\begin{APACrefauthors}%
Bilbiie, F\BPBI O.%
, Fujiwara, I.%
\BCBL {}\ \BBA {} Ghironi, F.%
\end{APACrefauthors}%
\unskip\
\newblock
\APACrefYearMonthDay{2014}{}{}.
\newblock
{\BBOQ}\APACrefatitle {Optimal monetary policy with endogenous entry and
  product variety} {Optimal monetary policy with endogenous entry and product
  variety}.{\BBCQ}
\newblock
\APACjournalVolNumPages{Journal of Monetary Economics}{64}{}{1--20}.
\newblock
\begin{APACrefDOI} \doi{10.1016/j.jmoneco.2014.02.006} \end{APACrefDOI}
\PrintBackRefs{\CurrentBib}

\bibitem [\protect \citeauthoryear {%
Billi%
}{%
Billi%
}{%
{\protect \APACyear {2011}}%
}]{%
billi2008optinf}
\APACinsertmetastar {%
billi2008optinf}%
\begin{APACrefauthors}%
Billi, R\BPBI M.%
\end{APACrefauthors}%
\unskip\
\newblock
\APACrefYearMonthDay{2011}{}{}.
\newblock
{\BBOQ}\APACrefatitle {Optimal Inflation for the U.S. Economy} {Optimal
  inflation for the u.s. economy}.{\BBCQ}
\newblock
\APACjournalVolNumPages{American Economic Journal:
  Macroeconomics}{3}{3}{29--52}.
\newblock
\APACrefnote{Published version of FRB Kansas City Working Paper RWP 07-03
  (2008).}
\newblock
\begin{APACrefDOI} \doi{10.1257/mac.3.3.29} \end{APACrefDOI}
\PrintBackRefs{\CurrentBib}

\bibitem [\protect \citeauthoryear {%
Blanchard%
, Dell'Ariccia%
\BCBL {}\ \BBA {} Mauro%
}{%
Blanchard%
\ \protect \BOthers {.}}{%
{\protect \APACyear {2010}}%
}]{%
blanchard2010rethinking}
\APACinsertmetastar {%
blanchard2010rethinking}%
\begin{APACrefauthors}%
Blanchard, O.%
, Dell'Ariccia, G.%
\BCBL {}\ \BBA {} Mauro, P.%
\end{APACrefauthors}%
\unskip\
\newblock
\APACrefYearMonthDay{2010}{}{}.
\newblock
{\BBOQ}\APACrefatitle {Rethinking Macroeconomic Policy} {Rethinking
  macroeconomic policy}.{\BBCQ}
\newblock
\APACjournalVolNumPages{Journal of Money, Credit and
  Banking}{42}{s1}{199--215}.
\newblock
\begin{APACrefDOI} \doi{10.2139/ssrn.1555117} \end{APACrefDOI}
\PrintBackRefs{\CurrentBib}

\bibitem [\protect \citeauthoryear {%
Blanchard%
\ \BBA {} Gali%
}{%
Blanchard%
\ \BBA {} Gali%
}{%
{\protect \APACyear {2007}}%
}]{%
blanchard2007realwage}
\APACinsertmetastar {%
blanchard2007realwage}%
\begin{APACrefauthors}%
Blanchard, O.%
\BCBT {}\ \BBA {} Gali, J.%
\end{APACrefauthors}%
\unskip\
\newblock
\APACrefYearMonthDay{2007}{}{}.
\newblock
{\BBOQ}\APACrefatitle {Real Wage Rigidities and the {New Keynesian} Model}
  {Real wage rigidities and the {New Keynesian} model}.{\BBCQ}
\newblock
\APACjournalVolNumPages{Journal of Money, Credit and Banking}{39}{s1}{35--65}.
\newblock
\begin{APACrefDOI} \doi{10.1111/j.1538-4616.2007.00015.x} \end{APACrefDOI}
\PrintBackRefs{\CurrentBib}

\bibitem [\protect \citeauthoryear {%
Blanchard%
\ \BBA {} Gal\'{i}%
}{%
Blanchard%
\ \BBA {} Gal\'{i}%
}{%
{\protect \APACyear {2008}}%
}]{%
blanchard2008optinf}
\APACinsertmetastar {%
blanchard2008optinf}%
\begin{APACrefauthors}%
Blanchard, O.%
\BCBT {}\ \BBA {} Gal\'{i}, J.%
\end{APACrefauthors}%
\unskip\
\newblock
\APACrefYearMonthDay{2008}{}{}.
\newblock
\APACrefbtitle {Labor Markets and Monetary Policy: A {N}ew-{K}eynesian Model
  with Unemployment} {Labor markets and monetary policy: A {N}ew-{K}eynesian
  model with unemployment}\ \APACbVolEdTR {}{NBER Working Paper\ \BNUM\ 13897}.
\newblock
\APACaddressInstitution{}{National Bureau of Economic Research}.
\newblock
\begin{APACrefDOI} \doi{10.3386/w13897} \end{APACrefDOI}
\PrintBackRefs{\CurrentBib}

\bibitem [\protect \citeauthoryear {%
Blanco%
}{%
Blanco%
}{%
{\protect \APACyear {2021}}%
}]{%
blanco2015optinf}
\APACinsertmetastar {%
blanco2015optinf}%
\begin{APACrefauthors}%
Blanco, A.%
\end{APACrefauthors}%
\unskip\
\newblock
\APACrefYearMonthDay{2021}{}{}.
\newblock
{\BBOQ}\APACrefatitle {Optimal Inflation Target in an Economy with Menu Costs
  and an Occasionally Binding Zero Lower Bound} {Optimal inflation target in an
  economy with menu costs and an occasionally binding zero lower bound}.{\BBCQ}
\newblock
\APACjournalVolNumPages{American Economic Journal:
  Macroeconomics}{13}{2}{108--141}.
\newblock
\APACrefnote{Published version of the 2015 working paper.}
\newblock
\begin{APACrefDOI} \doi{10.1257/mac.20180198} \end{APACrefDOI}
\PrintBackRefs{\CurrentBib}

\bibitem [\protect \citeauthoryear {%
Boehm%
\ \BBA {} House%
}{%
Boehm%
\ \BBA {} House%
}{%
{\protect \APACyear {2014}}%
}]{%
boehm2014optinf}
\APACinsertmetastar {%
boehm2014optinf}%
\begin{APACrefauthors}%
Boehm, C\BPBI E.%
\BCBT {}\ \BBA {} House, C\BPBI L.%
\end{APACrefauthors}%
\unskip\
\newblock
\APACrefYearMonthDay{2014}{}{}.
\newblock
\APACrefbtitle {Optimal {T}aylor Rules in {New Keynesian} Models} {Optimal
  {T}aylor rules in {New Keynesian} models}\ \APACbVolEdTR{}{\BTR{}}.
\newblock
\APACaddressInstitution{}{NBER Working Paper}.
\newblock
\begin{APACrefDOI} \doi{10.3386/w20237} \end{APACrefDOI}
\PrintBackRefs{\CurrentBib}

\bibitem [\protect \citeauthoryear {%
Bonciani%
\ \BBA {} Oh%
}{%
Bonciani%
\ \BBA {} Oh%
}{%
{\protect \APACyear {2026}}%
}]{%
bonciani2026optinf}
\APACinsertmetastar {%
bonciani2026optinf}%
\begin{APACrefauthors}%
Bonciani, D.%
\BCBT {}\ \BBA {} Oh, J.%
\end{APACrefauthors}%
\unskip\
\newblock
\APACrefYearMonthDay{2026}{}{}.
\newblock
{\BBOQ}\APACrefatitle {Unemployment risk, liquidity traps, and monetary policy}
  {Unemployment risk, liquidity traps, and monetary policy}.{\BBCQ}
\newblock
\APACjournalVolNumPages{Journal of Economic Dynamics and
  Control}{182}{}{105238}.
\newblock
\begin{APACrefDOI} \doi{10.1016/j.jedc.2025.105238} \end{APACrefDOI}
\PrintBackRefs{\CurrentBib}

\bibitem [\protect \citeauthoryear {%
Boskin%
, Dulberger%
, Gordon%
, Griliches%
\BCBL {}\ \BBA {} Jorgenson%
}{%
Boskin%
\ \protect \BOthers {.}}{%
{\protect \APACyear {1996}}%
}]{%
boskin1996toward}
\APACinsertmetastar {%
boskin1996toward}%
\begin{APACrefauthors}%
Boskin, M\BPBI J.%
, Dulberger, E\BPBI R.%
, Gordon, R\BPBI J.%
, Griliches, Z.%
\BCBL {}\ \BBA {} Jorgenson, D\BPBI W.%
\end{APACrefauthors}%
\unskip\
\newblock
\APACrefYearMonthDay{1996}{}{}.
\newblock
\APACrefbtitle {Toward a More Accurate Measure of the Cost of Living} {Toward a
  more accurate measure of the cost of living}\ \APACbVolEdTR{}{\BTR{}}.
\newblock
\APACaddressInstitution{}{Final Report to the Senate Finance Committee from the
  Advisory Commission to Study the Consumer Price Index}.
\PrintBackRefs{\CurrentBib}

\bibitem [\protect \citeauthoryear {%
Brock%
}{%
Brock%
}{%
{\protect \APACyear {1974}}%
}]{%
brock1974money}
\APACinsertmetastar {%
brock1974money}%
\begin{APACrefauthors}%
Brock, W\BPBI A.%
\end{APACrefauthors}%
\unskip\
\newblock
\APACrefYearMonthDay{1974}{}{}.
\newblock
{\BBOQ}\APACrefatitle {Money and Growth: The Case of Long-Run Perfect
  Foresight} {Money and growth: The case of long-run perfect foresight}.{\BBCQ}
\newblock
\APACjournalVolNumPages{International Economic Review}{15}{3}{750--777}.
\newblock
\begin{APACrefDOI} \doi{10.2307/2525739} \end{APACrefDOI}
\PrintBackRefs{\CurrentBib}

\bibitem [\protect \citeauthoryear {%
Broda%
\ \BBA {} Weinstein%
}{%
Broda%
\ \BBA {} Weinstein%
}{%
{\protect \APACyear {2010}}%
}]{%
broda2010product}
\APACinsertmetastar {%
broda2010product}%
\begin{APACrefauthors}%
Broda, C.%
\BCBT {}\ \BBA {} Weinstein, D\BPBI E.%
\end{APACrefauthors}%
\unskip\
\newblock
\APACrefYearMonthDay{2010}{}{}.
\newblock
{\BBOQ}\APACrefatitle {Product Creation and Destruction: Evidence and Price
  Implications} {Product creation and destruction: Evidence and price
  implications}.{\BBCQ}
\newblock
\APACjournalVolNumPages{American Economic Review}{100}{3}{691--723}.
\newblock
\begin{APACrefDOI} \doi{10.1257/aer.100.3.691} \end{APACrefDOI}
\PrintBackRefs{\CurrentBib}

\bibitem [\protect \citeauthoryear {%
Bryan%
\ \BBA {} Cecchetti%
}{%
Bryan%
\ \BBA {} Cecchetti%
}{%
{\protect \APACyear {1993}}%
}]{%
bryan2002measured}
\APACinsertmetastar {%
bryan2002measured}%
\begin{APACrefauthors}%
Bryan, M\BPBI F.%
\BCBT {}\ \BBA {} Cecchetti, S\BPBI G.%
\end{APACrefauthors}%
\unskip\
\newblock
\APACrefYearMonthDay{1993}{}{}.
\newblock
\APACrefbtitle {The {C}onsumer {P}rice {I}ndex as a Measure of Inflation} {The
  {C}onsumer {P}rice {I}ndex as a measure of inflation}\ \APACbVolEdTR {}{NBER
  Working Paper\ \BNUM\ 4505}.
\newblock
\APACaddressInstitution{}{National Bureau of Economic Research}.
\newblock
\begin{APACrefURL} \url{http://www.nber.org/papers/w4505} \end{APACrefURL}
\newblock
\APACrefnote{Also published in {\em Federal Reserve Bank of Cleveland Economic
  Review} (1993, Q4) and discussed in subsequent measurement-error literature,
  including Bryan, Cecchetti and O'Sullivan (2002)}
\newblock
\begin{APACrefDOI} \doi{10.3386/w4505} \end{APACrefDOI}
\PrintBackRefs{\CurrentBib}

\bibitem [\protect \citeauthoryear {%
Burnham%
\ \BBA {} Anderson%
}{%
Burnham%
\ \BBA {} Anderson%
}{%
{\protect \APACyear {2002}}%
}]{%
burnham2002model}
\APACinsertmetastar {%
burnham2002model}%
\begin{APACrefauthors}%
Burnham, K\BPBI P.%
\BCBT {}\ \BBA {} Anderson, D\BPBI R.%
\end{APACrefauthors}%
\unskip\
\newblock
\APACrefYear{2002}.
\newblock
\APACrefbtitle {Model Selection and Multimodel Inference: A Practical
  Information-Theoretic Approach} {Model selection and multimodel inference: A
  practical information-theoretic approach}\ (\PrintOrdinal{2}\ \BEd).
\newblock
\APACaddressPublisher{New York}{Springer}.
\PrintBackRefs{\CurrentBib}

\bibitem [\protect \citeauthoryear {%
Calvo%
}{%
Calvo%
}{%
{\protect \APACyear {1983}}%
}]{%
calvo1983staggered}
\APACinsertmetastar {%
calvo1983staggered}%
\begin{APACrefauthors}%
Calvo, G\BPBI A.%
\end{APACrefauthors}%
\unskip\
\newblock
\APACrefYearMonthDay{1983}{}{}.
\newblock
{\BBOQ}\APACrefatitle {Staggered Prices in a Utility-Maximizing Framework}
  {Staggered prices in a utility-maximizing framework}.{\BBCQ}
\newblock
\APACjournalVolNumPages{Journal of Monetary Economics}{12}{3}{383--398}.
\newblock
\begin{APACrefDOI} \doi{10.1016/0304-3932(83)90060-0} \end{APACrefDOI}
\PrintBackRefs{\CurrentBib}

\bibitem [\protect \citeauthoryear {%
Carlsson%
\ \BBA {} Westermark%
}{%
Carlsson%
\ \BBA {} Westermark%
}{%
{\protect \APACyear {2016}}%
}]{%
carlsson2016optinf}
\APACinsertmetastar {%
carlsson2016optinf}%
\begin{APACrefauthors}%
Carlsson, M.%
\BCBT {}\ \BBA {} Westermark, A.%
\end{APACrefauthors}%
\unskip\
\newblock
\APACrefYearMonthDay{2016}{}{}.
\newblock
{\BBOQ}\APACrefatitle {Labor market frictions and optimal steady-state
  inflation} {Labor market frictions and optimal steady-state
  inflation}.{\BBCQ}
\newblock
\APACjournalVolNumPages{Journal of Monetary Economics}{78}{}{67--79}.
\newblock
\begin{APACrefDOI} \doi{10.1016/j.jmoneco.2016.01.002} \end{APACrefDOI}
\PrintBackRefs{\CurrentBib}

\bibitem [\protect \citeauthoryear {%
Choi%
\ \BBA {} Foerster%
}{%
Choi%
\ \BBA {} Foerster%
}{%
{\protect \APACyear {2021}}%
}]{%
choi2021optinf}
\APACinsertmetastar {%
choi2021optinf}%
\begin{APACrefauthors}%
Choi, J.%
\BCBT {}\ \BBA {} Foerster, A.%
\end{APACrefauthors}%
\unskip\
\newblock
\APACrefYearMonthDay{2021}{}{}.
\newblock
{\BBOQ}\APACrefatitle {Optimal monetary policy regime switches} {Optimal
  monetary policy regime switches}.{\BBCQ}
\newblock
\APACjournalVolNumPages{Review of Economic Dynamics}{42}{}{333--346}.
\newblock
\begin{APACrefDOI} \doi{10.1016/j.red.2020.11.007} \end{APACrefDOI}
\PrintBackRefs{\CurrentBib}

\bibitem [\protect \citeauthoryear {%
Chugh%
}{%
Chugh%
}{%
{\protect \APACyear {2006}}%
}]{%
chugh2006optinf}
\APACinsertmetastar {%
chugh2006optinf}%
\begin{APACrefauthors}%
Chugh, S\BPBI K.%
\end{APACrefauthors}%
\unskip\
\newblock
\APACrefYearMonthDay{2006}{}{}.
\newblock
{\BBOQ}\APACrefatitle {Optimal fiscal and monetary policy with sticky wages and
  sticky prices} {Optimal fiscal and monetary policy with sticky wages and
  sticky prices}.{\BBCQ}
\newblock
\APACjournalVolNumPages{Review of Economic Dynamics}{9}{4}{683--714}.
\newblock
\begin{APACrefDOI} \doi{10.1016/j.red.2006.07.001} \end{APACrefDOI}
\PrintBackRefs{\CurrentBib}

\bibitem [\protect \citeauthoryear {%
Chugh%
}{%
Chugh%
}{%
{\protect \APACyear {2007}}%
}]{%
chugh2007optinf}
\APACinsertmetastar {%
chugh2007optinf}%
\begin{APACrefauthors}%
Chugh, S\BPBI K.%
\end{APACrefauthors}%
\unskip\
\newblock
\APACrefYearMonthDay{2007}{}{}.
\newblock
{\BBOQ}\APACrefatitle {Optimal inflation persistence: {Ramsey} taxation with
  capital and habits} {Optimal inflation persistence: {Ramsey} taxation with
  capital and habits}.{\BBCQ}
\newblock
\APACjournalVolNumPages{Journal of Monetary Economics}{54}{6}{1809--1836}.
\newblock
\begin{APACrefDOI} \doi{10.1016/j.jmoneco.2006.07.005} \end{APACrefDOI}
\PrintBackRefs{\CurrentBib}

\bibitem [\protect \citeauthoryear {%
Chugh%
}{%
Chugh%
}{%
{\protect \APACyear {2009}}%
}]{%
chugh2009optinf}
\APACinsertmetastar {%
chugh2009optinf}%
\begin{APACrefauthors}%
Chugh, S\BPBI K.%
\end{APACrefauthors}%
\unskip\
\newblock
\APACrefYearMonthDay{2009}{}{}.
\newblock
{\BBOQ}\APACrefatitle {Does the Timing of the Cash-in-Advance Constraint Matter
  for Optimal Fiscal and Monetary Policy?} {Does the timing of the
  cash-in-advance constraint matter for optimal fiscal and monetary
  policy?}{\BBCQ}
\newblock
\APACjournalVolNumPages{Macroeconomic Dynamics}{13}{S1}{133--150}.
\newblock
\begin{APACrefDOI} \doi{10.1017/s1365100509080158} \end{APACrefDOI}
\PrintBackRefs{\CurrentBib}

\bibitem [\protect \citeauthoryear {%
Coibion%
, Gorodnichenko%
\BCBL {}\ \BBA {} Wieland%
}{%
Coibion%
\ \protect \BOthers {.}}{%
{\protect \APACyear {2012}}%
}]{%
coibion2012optimal}
\APACinsertmetastar {%
coibion2012optimal}%
\begin{APACrefauthors}%
Coibion, O.%
, Gorodnichenko, Y.%
\BCBL {}\ \BBA {} Wieland, J.%
\end{APACrefauthors}%
\unskip\
\newblock
\APACrefYearMonthDay{2012}{}{}.
\newblock
{\BBOQ}\APACrefatitle {The Optimal Inflation Rate in {New Keynesian} Models:
  Should Central Banks Raise Their Inflation Targets in Light of the Zero Lower
  Bound?} {The optimal inflation rate in {New Keynesian} models: Should central
  banks raise their inflation targets in light of the zero lower bound?}{\BBCQ}
\newblock
\APACjournalVolNumPages{Review of Economic Studies}{79}{4}{1371--1406}.
\newblock
\begin{APACrefDOI} \doi{10.1093/restud/rds013} \end{APACrefDOI}
\PrintBackRefs{\CurrentBib}

\bibitem [\protect \citeauthoryear {%
Cook%
\ \protect \BOthers {.}}{%
Cook%
\ \protect \BOthers {.}}{%
{\protect \APACyear {2026}}%
}]{%
cook2026reporting}
\APACinsertmetastar {%
cook2026reporting}%
\begin{APACrefauthors}%
Cook, D\BPBI O.%
\BCBT {}\ \BOthersPeriod {.}
\end{APACrefauthors}%
\unskip\
\newblock
\APACrefYearMonthDay{2026}{}{}.
\newblock
{\BBOQ}\APACrefatitle {Guidance for the Use of {AI} in the Meta-Analysis of
  Economics Research ({MAER-Net})} {Guidance for the use of {AI} in the
  meta-analysis of economics research ({MAER-Net})}.{\BBCQ}
\newblock
\APACjournalVolNumPages{Journal of Economic Surveys}{}{}{}.
\newblock
\APACrefnote{Forthcoming}
\newblock
\begin{APACrefDOI} \doi{10.1111/joes.70105} \end{APACrefDOI}
\PrintBackRefs{\CurrentBib}

\bibitem [\protect \citeauthoryear {%
Cooley%
\ \BBA {} Hansen%
}{%
Cooley%
\ \BBA {} Hansen%
}{%
{\protect \APACyear {1989}}%
}]{%
cooley1989optinf}
\APACinsertmetastar {%
cooley1989optinf}%
\begin{APACrefauthors}%
Cooley, T\BPBI F.%
\BCBT {}\ \BBA {} Hansen, G\BPBI D.%
\end{APACrefauthors}%
\unskip\
\newblock
\APACrefYearMonthDay{1989}{}{}.
\newblock
{\BBOQ}\APACrefatitle {The Inflation Tax in a Real Business Cycle Model} {The
  inflation tax in a real business cycle model}.{\BBCQ}
\newblock
\APACjournalVolNumPages{The American Economic Review}{79}{4}{733--748}.
\PrintBackRefs{\CurrentBib}

\bibitem [\protect \citeauthoryear {%
Correia%
\ \BBA {} Teles%
}{%
Correia%
\ \BBA {} Teles%
}{%
{\protect \APACyear {1999}}%
}]{%
correia1999optinf}
\APACinsertmetastar {%
correia1999optinf}%
\begin{APACrefauthors}%
Correia, I.%
\BCBT {}\ \BBA {} Teles, P.%
\end{APACrefauthors}%
\unskip\
\newblock
\APACrefYearMonthDay{1999}{}{}.
\newblock
{\BBOQ}\APACrefatitle {The Optimal Inflation Tax} {The optimal inflation
  tax}.{\BBCQ}
\newblock
\APACjournalVolNumPages{Review of Economic Dynamics}{2}{2}{325--346}.
\newblock
\begin{APACrefDOI} \doi{10.1006/redy.1998.0040} \end{APACrefDOI}
\PrintBackRefs{\CurrentBib}

\bibitem [\protect \citeauthoryear {%
da Costa%
\ \BBA {} Werning%
}{%
da Costa%
\ \BBA {} Werning%
}{%
{\protect \APACyear {2008}}%
}]{%
dacosta2008optinf}
\APACinsertmetastar {%
dacosta2008optinf}%
\begin{APACrefauthors}%
da Costa, C\BPBI E.%
\BCBT {}\ \BBA {} Werning, I.%
\end{APACrefauthors}%
\unskip\
\newblock
\APACrefYearMonthDay{2008}{}{}.
\newblock
{\BBOQ}\APACrefatitle {On the Optimality of the {Friedman} Rule with
  Heterogeneous Agents and Nonlinear Income Taxation} {On the optimality of the
  {Friedman} rule with heterogeneous agents and nonlinear income
  taxation}.{\BBCQ}
\newblock
\APACjournalVolNumPages{Journal of Political Economy}{116}{1}{82--112}.
\newblock
\begin{APACrefDOI} \doi{10.1086/529397} \end{APACrefDOI}
\PrintBackRefs{\CurrentBib}

\bibitem [\protect \citeauthoryear {%
Darracq~Pari\`{e}s%
\ \BBA {} Loublier%
}{%
Darracq~Pari\`{e}s%
\ \BBA {} Loublier%
}{%
{\protect \APACyear {2010}}%
}]{%
darracqparies2010optinf}
\APACinsertmetastar {%
darracqparies2010optinf}%
\begin{APACrefauthors}%
Darracq~Pari\`{e}s, M.%
\BCBT {}\ \BBA {} Loublier, A.%
\end{APACrefauthors}%
\unskip\
\newblock
\APACrefYearMonthDay{2010}{}{}.
\newblock
\APACrefbtitle {Epstein-Zin preferences and their use in macro-finance models:
  Implications for optimal monetary policy} {Epstein-zin preferences and their
  use in macro-finance models: Implications for optimal monetary policy}\
  \APACbVolEdTR{}{\BTR{}}.
\newblock
\APACaddressInstitution{}{ECB Working Paper Series}.
\newblock
\begin{APACrefDOI} \doi{10.2139/ssrn.1617249} \end{APACrefDOI}
\PrintBackRefs{\CurrentBib}

\bibitem [\protect \citeauthoryear {%
Daudignon%
\ \BBA {} Tristani%
}{%
Daudignon%
\ \BBA {} Tristani%
}{%
{\protect \APACyear {2025}}%
}]{%
daudignon2025optinf}
\APACinsertmetastar {%
daudignon2025optinf}%
\begin{APACrefauthors}%
Daudignon, S.%
\BCBT {}\ \BBA {} Tristani, O.%
\end{APACrefauthors}%
\unskip\
\newblock
\APACrefYearMonthDay{2025}{}{}.
\newblock
{\BBOQ}\APACrefatitle {Monetary Policy and the Drifting Natural Rate of
  Interest} {Monetary policy and the drifting natural rate of interest}.{\BBCQ}
\newblock
\APACjournalVolNumPages{Journal of Money, Credit and Banking}{}{}{}.
\newblock
\begin{APACrefDOI} \doi{10.1111/jmcb.13274} \end{APACrefDOI}
\PrintBackRefs{\CurrentBib}

\bibitem [\protect \citeauthoryear {%
De\'{a}k%
, Levine%
, Mirza%
\BCBL {}\ \BBA {} Pham%
}{%
De\'{a}k%
\ \protect \BOthers {.}}{%
{\protect \APACyear {2026}}%
}]{%
deak2024optinf}
\APACinsertmetastar {%
deak2024optinf}%
\begin{APACrefauthors}%
De\'{a}k, S.%
, Levine, P.%
, Mirza, A.%
\BCBL {}\ \BBA {} Pham, S\BPBI T.%
\end{APACrefauthors}%
\unskip\
\newblock
\APACrefYearMonthDay{2026}{}{}.
\newblock
{\BBOQ}\APACrefatitle {Negotiating the wilderness of bounded rationality
  through robust policy} {Negotiating the wilderness of bounded rationality
  through robust policy}.{\BBCQ}
\newblock
\APACjournalVolNumPages{Journal of Economic Behavior \&
  Organization}{245}{}{107476}.
\newblock
\begin{APACrefURL}
  \url{https://www.sciencedirect.com/science/article/pii/S0167268126000624}
  \end{APACrefURL}
\newblock
\begin{APACrefDOI} \doi{https://doi.org/10.1016/j.jebo.2026.107476}
  \end{APACrefDOI}
\PrintBackRefs{\CurrentBib}

\bibitem [\protect \citeauthoryear {%
Di~Bartolomeo%
, Tirelli%
\BCBL {}\ \BBA {} Acocella%
}{%
Di~Bartolomeo%
\ \protect \BOthers {.}}{%
{\protect \APACyear {2013}}%
}]{%
dibartolomeo2013optinf}
\APACinsertmetastar {%
dibartolomeo2013optinf}%
\begin{APACrefauthors}%
Di~Bartolomeo, G.%
, Tirelli, P.%
\BCBL {}\ \BBA {} Acocella, N.%
\end{APACrefauthors}%
\unskip\
\newblock
\APACrefYearMonthDay{2013}{}{}.
\newblock
\APACrefbtitle {The comeback of inflation as an optimal public finance tool}
  {The comeback of inflation as an optimal public finance tool}\ \APACbVolEdTR
  {}{SSRN Working Paper\ \BNUM\ 2369183}.
\newblock
\APACaddressInstitution{}{SSRN Electronic Journal}.
\newblock
\begin{APACrefDOI} \doi{10.2139/ssrn.2369183} \end{APACrefDOI}
\PrintBackRefs{\CurrentBib}

\bibitem [\protect \citeauthoryear {%
Diercks%
}{%
Diercks%
}{%
{\protect \APACyear {2019}}%
{\protect \APACexlab {{\protect \BCnt {1}}}}}]{%
diercks2019equity}
\APACinsertmetastar {%
diercks2019equity}%
\begin{APACrefauthors}%
Diercks, A\BPBI M.%
\end{APACrefauthors}%
\unskip\
\newblock
\APACrefYearMonthDay{2019{\protect \BCnt {1}}}{}{}.
\newblock
\APACrefbtitle {The equity premium, long-run risks to r-star, and asymmetric
  optimal monetary policy} {The equity premium, long-run risks to r-star, and
  asymmetric optimal monetary policy}\ \APACbVolEdTR{}{\BTR{}\ \BNUM\ 3435372}.
\newblock
\APACaddressInstitution{}{SSRN}.
\newblock
\begin{APACrefDOI} \doi{10.2139/ssrn.3435372} \end{APACrefDOI}
\PrintBackRefs{\CurrentBib}

\bibitem [\protect \citeauthoryear {%
Diercks%
}{%
Diercks%
}{%
{\protect \APACyear {2019}}%
{\protect \APACexlab {{\protect \BCnt {2}}}}}]{%
diercks2019reader}
\APACinsertmetastar {%
diercks2019reader}%
\begin{APACrefauthors}%
Diercks, A\BPBI M.%
\end{APACrefauthors}%
\unskip\
\newblock
\APACrefYearMonthDay{2019{\protect \BCnt {2}}}{}{}.
\newblock
\APACrefbtitle {The reader's guide to optimal monetary policy} {The reader's
  guide to optimal monetary policy}\ \APACbVolEdTR{}{\BTR{}\ \BNUM\ 2989237}.
\newblock
\APACaddressInstitution{}{SSRN}.
\PrintBackRefs{\CurrentBib}

\bibitem [\protect \citeauthoryear {%
Dordal~i Carreras%
, Coibion%
, Gorodnichenko%
\BCBL {}\ \BBA {} Wieland%
}{%
Dordal~i Carreras%
\ \protect \BOthers {.}}{%
{\protect \APACyear {2016}}%
}]{%
dordalicarreras2016optinf}
\APACinsertmetastar {%
dordalicarreras2016optinf}%
\begin{APACrefauthors}%
Dordal~i Carreras, M.%
, Coibion, O.%
, Gorodnichenko, Y.%
\BCBL {}\ \BBA {} Wieland, J.%
\end{APACrefauthors}%
\unskip\
\newblock
\APACrefYearMonthDay{2016}{}{}.
\newblock
\APACrefbtitle {Infrequent but Long-Lived Zero-Bound Episodes and the Optimal
  Rate of Inflation.} {Infrequent but long-lived zero-bound episodes and the
  optimal rate of inflation.}
\PrintBackRefs{\CurrentBib}

\bibitem [\protect \citeauthoryear {%
Doucouliagos%
\ \BBA {} Stanley%
}{%
Doucouliagos%
\ \BBA {} Stanley%
}{%
{\protect \APACyear {2013}}%
}]{%
doucouliagos2013there}
\APACinsertmetastar {%
doucouliagos2013there}%
\begin{APACrefauthors}%
Doucouliagos, C.%
\BCBT {}\ \BBA {} Stanley, T.%
\end{APACrefauthors}%
\unskip\
\newblock
\APACrefYearMonthDay{2013}{}{}.
\newblock
{\BBOQ}\APACrefatitle {Are All Economic Facts Greatly Exaggerated? Theory
  Competition and Selectivity} {Are all economic facts greatly exaggerated?
  theory competition and selectivity}.{\BBCQ}
\newblock
\APACjournalVolNumPages{Journal of Economic Surveys}{27}{2}{316-339}.
\newblock
\begin{APACrefURL}
  \url{https://onlinelibrary.wiley.com/doi/abs/10.1111/j.1467-6419.2011.00706.x}
  \end{APACrefURL}
\newblock
\begin{APACrefDOI} \doi{https://doi.org/10.1111/j.1467-6419.2011.00706.x}
  \end{APACrefDOI}
\PrintBackRefs{\CurrentBib}

\bibitem [\protect \citeauthoryear {%
Drazen%
}{%
Drazen%
}{%
{\protect \APACyear {1979}}%
}]{%
drazen1979optimal}
\APACinsertmetastar {%
drazen1979optimal}%
\begin{APACrefauthors}%
Drazen, A.%
\end{APACrefauthors}%
\unskip\
\newblock
\APACrefYearMonthDay{1979}{}{}.
\newblock
{\BBOQ}\APACrefatitle {The Optimal Rate of Inflation Revisited} {The optimal
  rate of inflation revisited}.{\BBCQ}
\newblock
\APACjournalVolNumPages{Journal of Monetary Economics}{5}{2}{231--248}.
\newblock
\begin{APACrefDOI} \doi{10.1016/0304-3932(79)90005-9} \end{APACrefDOI}
\PrintBackRefs{\CurrentBib}

\bibitem [\protect \citeauthoryear {%
Edge%
, Laubach%
\BCBL {}\ \BBA {} Williams%
}{%
Edge%
\ \protect \BOthers {.}}{%
{\protect \APACyear {2010}}%
}]{%
edge2010optinf}
\APACinsertmetastar {%
edge2010optinf}%
\begin{APACrefauthors}%
Edge, R\BPBI M.%
, Laubach, T.%
\BCBL {}\ \BBA {} Williams, J\BPBI C.%
\end{APACrefauthors}%
\unskip\
\newblock
\APACrefYearMonthDay{2010}{}{}.
\newblock
{\BBOQ}\APACrefatitle {Welfare-maximizing monetary policy under parameter
  uncertainty} {Welfare-maximizing monetary policy under parameter
  uncertainty}.{\BBCQ}
\newblock
\APACjournalVolNumPages{Journal of Applied Econometrics}{25}{1}{129--143}.
\newblock
\begin{APACrefDOI} \doi{10.1002/jae.1136} \end{APACrefDOI}
\PrintBackRefs{\CurrentBib}

\bibitem [\protect \citeauthoryear {%
Eggertsson%
\ \BBA {} Woodford%
}{%
Eggertsson%
\ \BBA {} Woodford%
}{%
{\protect \APACyear {2003}}%
}]{%
eggertsson2003zero}
\APACinsertmetastar {%
eggertsson2003zero}%
\begin{APACrefauthors}%
Eggertsson, G\BPBI B.%
\BCBT {}\ \BBA {} Woodford, M.%
\end{APACrefauthors}%
\unskip\
\newblock
\APACrefYearMonthDay{2003}{}{}.
\newblock
{\BBOQ}\APACrefatitle {The Zero Bound on Interest Rates and Optimal Monetary
  Policy} {The zero bound on interest rates and optimal monetary
  policy}.{\BBCQ}
\newblock
\APACjournalVolNumPages{Brookings Papers on Economic
  Activity}{2003}{1}{139--211}.
\newblock
\begin{APACrefDOI} \doi{10.1353/eca.2003.0010} \end{APACrefDOI}
\PrintBackRefs{\CurrentBib}

\bibitem [\protect \citeauthoryear {%
Eicher%
, Papageorgiou%
\BCBL {}\ \BBA {} Raftery%
}{%
Eicher%
\ \protect \BOthers {.}}{%
{\protect \APACyear {2011}}%
}]{%
eicher2011default}
\APACinsertmetastar {%
eicher2011default}%
\begin{APACrefauthors}%
Eicher, T\BPBI S.%
, Papageorgiou, C.%
\BCBL {}\ \BBA {} Raftery, A\BPBI E.%
\end{APACrefauthors}%
\unskip\
\newblock
\APACrefYearMonthDay{2011}{}{}.
\newblock
{\BBOQ}\APACrefatitle {Default priors and predictive performance in {B}ayesian
  model averaging, with application to growth determinants} {Default priors and
  predictive performance in {B}ayesian model averaging, with application to
  growth determinants}.{\BBCQ}
\newblock
\APACjournalVolNumPages{Journal of Applied Econometrics}{26}{1}{30--55}.
\newblock
\begin{APACrefDOI} \doi{10.1002/jae.1112} \end{APACrefDOI}
\PrintBackRefs{\CurrentBib}

\bibitem [\protect \citeauthoryear {%
Elminejad%
, Havranek%
\BCBL {}\ \BBA {} Horvath%
}{%
Elminejad%
\ \protect \BOthers {.}}{%
{\protect \APACyear {2023}}%
}]{%
havranek2018monetary}
\APACinsertmetastar {%
havranek2018monetary}%
\begin{APACrefauthors}%
Elminejad, A.%
, Havranek, T.%
\BCBL {}\ \BBA {} Horvath, R.%
\end{APACrefauthors}%
\unskip\
\newblock
\APACrefYearMonthDay{2023}{}{}.
\newblock
{\BBOQ}\APACrefatitle {Intertemporal Substitution in Labor Supply: A
  Meta-Analysis} {Intertemporal substitution in labor supply: A
  meta-analysis}.{\BBCQ}
\newblock
\APACjournalVolNumPages{Review of Economic Dynamics}{51}{}{1095--1113}.
\newblock
\begin{APACrefDOI} \doi{10.1016/j.red.2023.10.001} \end{APACrefDOI}
\PrintBackRefs{\CurrentBib}

\bibitem [\protect \citeauthoryear {%
Erceg%
, Henderson%
\BCBL {}\ \BBA {} Levin%
}{%
Erceg%
\ \protect \BOthers {.}}{%
{\protect \APACyear {2000}}%
}]{%
erceg2000optimal}
\APACinsertmetastar {%
erceg2000optimal}%
\begin{APACrefauthors}%
Erceg, C\BPBI J.%
, Henderson, D\BPBI W.%
\BCBL {}\ \BBA {} Levin, A\BPBI T.%
\end{APACrefauthors}%
\unskip\
\newblock
\APACrefYearMonthDay{2000}{}{}.
\newblock
{\BBOQ}\APACrefatitle {Optimal Monetary Policy with Staggered Wage and Price
  Contracts} {Optimal monetary policy with staggered wage and price
  contracts}.{\BBCQ}
\newblock
\APACjournalVolNumPages{Journal of Monetary Economics}{46}{2}{281--313}.
\newblock
\begin{APACrefDOI} \doi{10.1016/s0304-3932(00)00028-3} \end{APACrefDOI}
\PrintBackRefs{\CurrentBib}

\bibitem [\protect \citeauthoryear {%
Fagan%
\ \BBA {} Messina%
}{%
Fagan%
\ \BBA {} Messina%
}{%
{\protect \APACyear {2009}}%
}]{%
fagan2009optinf}
\APACinsertmetastar {%
fagan2009optinf}%
\begin{APACrefauthors}%
Fagan, G.%
\BCBT {}\ \BBA {} Messina, J.%
\end{APACrefauthors}%
\unskip\
\newblock
\APACrefYearMonthDay{2009}{}{}.
\newblock
\APACrefbtitle {Downward wage rigidity and optimal steady-state inflation}
  {Downward wage rigidity and optimal steady-state inflation}\
  \APACbVolEdTR{}{\BTR{}}.
\newblock
\APACaddressInstitution{}{ECB Working Paper Series}.
\newblock
\begin{APACrefDOI} \doi{10.2139/ssrn.1386926} \end{APACrefDOI}
\PrintBackRefs{\CurrentBib}

\bibitem [\protect \citeauthoryear {%
Faia%
}{%
Faia%
}{%
{\protect \APACyear {2008}}%
}]{%
faia2008optinf}
\APACinsertmetastar {%
faia2008optinf}%
\begin{APACrefauthors}%
Faia, E.%
\end{APACrefauthors}%
\unskip\
\newblock
\APACrefYearMonthDay{2008}{}{}.
\newblock
{\BBOQ}\APACrefatitle {{Ramsey} Monetary Policy with Capital Accumulation and
  Nominal Rigidities} {{Ramsey} monetary policy with capital accumulation and
  nominal rigidities}.{\BBCQ}
\newblock
\APACjournalVolNumPages{Macroeconomic Dynamics}{12}{S1}{90--99}.
\newblock
\begin{APACrefDOI} \doi{10.1017/s1365100507070083} \end{APACrefDOI}
\PrintBackRefs{\CurrentBib}

\bibitem [\protect \citeauthoryear {%
Faia%
}{%
Faia%
}{%
{\protect \APACyear {2009}}%
}]{%
faia2009optinf}
\APACinsertmetastar {%
faia2009optinf}%
\begin{APACrefauthors}%
Faia, E.%
\end{APACrefauthors}%
\unskip\
\newblock
\APACrefYearMonthDay{2009}{}{}.
\newblock
{\BBOQ}\APACrefatitle {{Ramsey} monetary policy with labor market frictions}
  {{Ramsey} monetary policy with labor market frictions}.{\BBCQ}
\newblock
\APACjournalVolNumPages{Journal of Monetary Economics}{56}{4}{570--581}.
\newblock
\begin{APACrefDOI} \doi{10.1016/j.jmoneco.2009.03.009} \end{APACrefDOI}
\PrintBackRefs{\CurrentBib}

\bibitem [\protect \citeauthoryear {%
Faia%
, Lechthaler%
\BCBL {}\ \BBA {} Merkl%
}{%
Faia%
\ \protect \BOthers {.}}{%
{\protect \APACyear {2014}}%
}]{%
faia2014optinf}
\APACinsertmetastar {%
faia2014optinf}%
\begin{APACrefauthors}%
Faia, E.%
, Lechthaler, W.%
\BCBL {}\ \BBA {} Merkl, C.%
\end{APACrefauthors}%
\unskip\
\newblock
\APACrefYearMonthDay{2014}{}{}.
\newblock
{\BBOQ}\APACrefatitle {Labor Selection, Turnover Costs, and Optimal Monetary
  Policy} {Labor selection, turnover costs, and optimal monetary
  policy}.{\BBCQ}
\newblock
\APACjournalVolNumPages{Journal of Money, Credit and Banking}{46}{1}{115--144}.
\newblock
\begin{APACrefDOI} \doi{10.1111/jmcb.12099} \end{APACrefDOI}
\PrintBackRefs{\CurrentBib}

\bibitem [\protect \citeauthoryear {%
Faia%
\ \BBA {} Monacelli%
}{%
Faia%
\ \BBA {} Monacelli%
}{%
{\protect \APACyear {2007}}%
}]{%
faia2007optinf}
\APACinsertmetastar {%
faia2007optinf}%
\begin{APACrefauthors}%
Faia, E.%
\BCBT {}\ \BBA {} Monacelli, T.%
\end{APACrefauthors}%
\unskip\
\newblock
\APACrefYearMonthDay{2007}{}{}.
\newblock
{\BBOQ}\APACrefatitle {Optimal interest rate rules, asset prices, and credit
  frictions} {Optimal interest rate rules, asset prices, and credit
  frictions}.{\BBCQ}
\newblock
\APACjournalVolNumPages{Journal of Economic Dynamics \&
  Control}{31}{10}{3228--3254}.
\newblock
\begin{APACrefDOI} \doi{10.1016/j.jedc.2006.11.006} \end{APACrefDOI}
\PrintBackRefs{\CurrentBib}

\bibitem [\protect \citeauthoryear {%
Fasolo%
}{%
Fasolo%
}{%
{\protect \APACyear {2014}}%
}]{%
fasolo2014optinf}
\APACinsertmetastar {%
fasolo2014optinf}%
\begin{APACrefauthors}%
Fasolo, A\BPBI M.%
\end{APACrefauthors}%
\unskip\
\newblock
\APACrefYearMonthDay{2014}{}{}.
\newblock
\APACrefbtitle {The {Ramsey} Steady State under Optimal Monetary and Fiscal
  Policy for Small Open Economies} {The {Ramsey} steady state under optimal
  monetary and fiscal policy for small open economies}\ \APACbVolEdTR
  {}{Working Paper Series\ \BNUM~357}.
\newblock
\APACaddressInstitution{}{Banco Central do Brasil}.
\PrintBackRefs{\CurrentBib}

\bibitem [\protect \citeauthoryear {%
Feldstein%
}{%
Feldstein%
}{%
{\protect \APACyear {1997}}%
}]{%
feldstein1997optinf}
\APACinsertmetastar {%
feldstein1997optinf}%
\begin{APACrefauthors}%
Feldstein, M\BPBI S.%
\end{APACrefauthors}%
\unskip\
\newblock
\APACrefYearMonthDay{1997}{}{}.
\newblock
{\BBOQ}\APACrefatitle {The Costs and Benefits of Going from Low Inflation to
  Price Stability} {The costs and benefits of going from low inflation to price
  stability}.{\BBCQ}
\newblock
\BIn{} \APACrefbtitle {Reducing Inflation: Motivation and Strategy} {Reducing
  inflation: Motivation and strategy}\ (\BPGS\ 123--166).
\newblock
\APACaddressPublisher{}{University of Chicago Press}.
\newblock
\begin{APACrefDOI} \doi{10.3386/w5469} \end{APACrefDOI}
\PrintBackRefs{\CurrentBib}

\bibitem [\protect \citeauthoryear {%
Fernandez%
, Ley%
\BCBL {}\ \BBA {} Steel%
}{%
Fernandez%
\ \protect \BOthers {.}}{%
{\protect \APACyear {2001}}%
}]{%
fernandez2001benchmark}
\APACinsertmetastar {%
fernandez2001benchmark}%
\begin{APACrefauthors}%
Fernandez, C.%
, Ley, E.%
\BCBL {}\ \BBA {} Steel, M\BPBI F.%
\end{APACrefauthors}%
\unskip\
\newblock
\APACrefYearMonthDay{2001}{}{}.
\newblock
{\BBOQ}\APACrefatitle {Benchmark priors for {B}ayesian model averaging}
  {Benchmark priors for {B}ayesian model averaging}.{\BBCQ}
\newblock
\APACjournalVolNumPages{Journal of Econometrics}{100}{2}{381--427}.
\newblock
\begin{APACrefDOI} \doi{10.1016/S0304-4076(00)00076-2} \end{APACrefDOI}
\PrintBackRefs{\CurrentBib}

\bibitem [\protect \citeauthoryear {%
Filiani%
}{%
Filiani%
}{%
{\protect \APACyear {2021}}%
}]{%
filiani2021optinf}
\APACinsertmetastar {%
filiani2021optinf}%
\begin{APACrefauthors}%
Filiani, P.%
\end{APACrefauthors}%
\unskip\
\newblock
\APACrefYearMonthDay{2021}{}{}.
\newblock
{\BBOQ}\APACrefatitle {Optimal monetary--fiscal policy in the euro area
  liquidity crisis} {Optimal monetary--fiscal policy in the euro area liquidity
  crisis}.{\BBCQ}
\newblock
\APACjournalVolNumPages{Journal of Macroeconomics}{70}{}{103364}.
\newblock
\begin{APACrefDOI} \doi{10.1016/j.jmacro.2021.103364} \end{APACrefDOI}
\PrintBackRefs{\CurrentBib}

\bibitem [\protect \citeauthoryear {%
Finocchiaro%
, Lombardo%
, Mendicino%
\BCBL {}\ \BBA {} Weil%
}{%
Finocchiaro%
\ \protect \BOthers {.}}{%
{\protect \APACyear {2018}}%
}]{%
finocchiaro2018optinf}
\APACinsertmetastar {%
finocchiaro2018optinf}%
\begin{APACrefauthors}%
Finocchiaro, D.%
, Lombardo, G.%
, Mendicino, C.%
\BCBL {}\ \BBA {} Weil, P.%
\end{APACrefauthors}%
\unskip\
\newblock
\APACrefYearMonthDay{2018}{}{}.
\newblock
{\BBOQ}\APACrefatitle {Optimal inflation with corporate taxation and financial
  constraints} {Optimal inflation with corporate taxation and financial
  constraints}.{\BBCQ}
\newblock
\APACjournalVolNumPages{Journal of Monetary Economics}{95}{}{18-31}.
\newblock
\begin{APACrefDOI} \doi{10.1016/j.jmoneco.2018.02.003} \end{APACrefDOI}
\PrintBackRefs{\CurrentBib}

\bibitem [\protect \citeauthoryear {%
Friedman%
}{%
Friedman%
}{%
{\protect \APACyear {1969}}%
}]{%
friedman1969}
\APACinsertmetastar {%
friedman1969}%
\begin{APACrefauthors}%
Friedman, M.%
\end{APACrefauthors}%
\unskip\
\newblock
\APACrefYearMonthDay{1969}{}{}.
\newblock
\APACrefbtitle {The Optimum Quantity of Money.} {The optimum quantity of
  money.}
\newblock
\APAChowpublished {in The Optimum Quantity of Money and Other Essays, Aldine
  Pub. Co, Chicago}.
\PrintBackRefs{\CurrentBib}

\bibitem [\protect \citeauthoryear {%
Garga%
\ \BBA {} Singh%
}{%
Garga%
\ \BBA {} Singh%
}{%
{\protect \APACyear {2021}}%
}]{%
garga2021optinf}
\APACinsertmetastar {%
garga2021optinf}%
\begin{APACrefauthors}%
Garga, V.%
\BCBT {}\ \BBA {} Singh, S.%
\end{APACrefauthors}%
\unskip\
\newblock
\APACrefYearMonthDay{2021}{}{}.
\newblock
{\BBOQ}\APACrefatitle {Output hysteresis and optimal monetary policy} {Output
  hysteresis and optimal monetary policy}.{\BBCQ}
\newblock
\APACjournalVolNumPages{Journal of Monetary Economics}{117}{}{871-886}.
\newblock
\begin{APACrefDOI} \doi{10.1016/j.jmoneco.2020.06.005} \end{APACrefDOI}
\PrintBackRefs{\CurrentBib}

\bibitem [\protect \citeauthoryear {%
George%
\ \protect \BOthers {.}}{%
George%
\ \protect \BOthers {.}}{%
{\protect \APACyear {2010}}%
}]{%
george2010dilution}
\APACinsertmetastar {%
george2010dilution}%
\begin{APACrefauthors}%
George, E\BPBI I.%
\BCBT {}\ \BOthersPeriod {.}
\end{APACrefauthors}%
\unskip\
\newblock
\APACrefYearMonthDay{2010}{}{}.
\newblock
{\BBOQ}\APACrefatitle {Dilution priors: Compensating for model space
  redundancy} {Dilution priors: Compensating for model space
  redundancy}.{\BBCQ}
\newblock
\APACjournalVolNumPages{Borrowing Strength: Theory Powering Applications--A
  Festschrift for Lawrence D. Brown}{6}{}{158--165}.
\newblock
\begin{APACrefDOI} \doi{10.1214/10-imscoll611} \end{APACrefDOI}
\PrintBackRefs{\CurrentBib}

\bibitem [\protect \citeauthoryear {%
Gerber%
, Malhotra%
\BCBL {}\ \protect \BOthers {.}}{%
Gerber%
\ \protect \BOthers {.}}{%
{\protect \APACyear {2008}}%
}]{%
gerber2008statistical}
\APACinsertmetastar {%
gerber2008statistical}%
\begin{APACrefauthors}%
Gerber, A.%
, Malhotra, N.%
\BCBL {}\ \BOthersPeriod {.}\end{APACrefauthors}%
\unskip\
\newblock
\APACrefYearMonthDay{2008}{}{}.
\newblock
{\BBOQ}\APACrefatitle {Do statistical reporting standards affect what is
  published? Publication bias in two leading political science journals} {Do
  statistical reporting standards affect what is published? publication bias in
  two leading political science journals}.{\BBCQ}
\newblock
\APACjournalVolNumPages{Quarterly Journal of Political
  Science}{3}{3}{313--326}.
\newblock
\begin{APACrefDOI} \doi{10.1561/100.00008024} \end{APACrefDOI}
\PrintBackRefs{\CurrentBib}

\bibitem [\protect \citeauthoryear {%
Golosov%
\ \BBA {} Lucas%
}{%
Golosov%
\ \BBA {} Lucas%
}{%
{\protect \APACyear {2007}}%
}]{%
golosov2007menu}
\APACinsertmetastar {%
golosov2007menu}%
\begin{APACrefauthors}%
Golosov, M.%
\BCBT {}\ \BBA {} Lucas, R\BPBI E., Jr.%
\end{APACrefauthors}%
\unskip\
\newblock
\APACrefYearMonthDay{2007}{}{}.
\newblock
{\BBOQ}\APACrefatitle {Menu Costs and {P}hillips Curves} {Menu costs and
  {P}hillips curves}.{\BBCQ}
\newblock
\APACjournalVolNumPages{Journal of Political Economy}{115}{2}{171--199}.
\newblock
\begin{APACrefDOI} \doi{10.1086/512625} \end{APACrefDOI}
\PrintBackRefs{\CurrentBib}

\bibitem [\protect \citeauthoryear {%
Goodfriend%
\ \BBA {} King%
}{%
Goodfriend%
\ \BBA {} King%
}{%
{\protect \APACyear {1997}}%
}]{%
goodfriend1997optinf}
\APACinsertmetastar {%
goodfriend1997optinf}%
\begin{APACrefauthors}%
Goodfriend, M.%
\BCBT {}\ \BBA {} King, R\BPBI G.%
\end{APACrefauthors}%
\unskip\
\newblock
\APACrefYearMonthDay{1997}{}{}.
\newblock
\APACrefbtitle {The New Neoclassical Synthesis and the Role of Monetary Policy}
  {The new neoclassical synthesis and the role of monetary policy}\
  \APACbVolEdTR{}{\BTR{}}.
\newblock
\APACaddressInstitution{}{NBER Macroeconomics Annual}.
\newblock
\begin{APACrefDOI} \doi{10.2139/ssrn.2123683} \end{APACrefDOI}
\PrintBackRefs{\CurrentBib}

\bibitem [\protect \citeauthoryear {%
Gordon%
}{%
Gordon%
}{%
{\protect \APACyear {2006}}%
}]{%
gordon2006boskin}
\APACinsertmetastar {%
gordon2006boskin}%
\begin{APACrefauthors}%
Gordon, R\BPBI J.%
\end{APACrefauthors}%
\unskip\
\newblock
\APACrefYearMonthDay{2006}{}{}.
\newblock
\APACrefbtitle {The Boskin Commission Report: A Retrospective One Decade Later}
  {The boskin commission report: A retrospective one decade later}\
  \APACbVolEdTR{}{\BTR{}}.
\newblock
\APACaddressInstitution{}{NBER Working Paper No. 12311}.
\newblock
\begin{APACrefDOI} \doi{10.3386/w12311} \end{APACrefDOI}
\PrintBackRefs{\CurrentBib}

\bibitem [\protect \citeauthoryear {%
Grandmont%
\ \BBA {} Younes%
}{%
Grandmont%
\ \BBA {} Younes%
}{%
{\protect \APACyear {1973}}%
}]{%
grandmont1973efficiency}
\APACinsertmetastar {%
grandmont1973efficiency}%
\begin{APACrefauthors}%
Grandmont, J\BHBI M.%
\BCBT {}\ \BBA {} Younes, Y.%
\end{APACrefauthors}%
\unskip\
\newblock
\APACrefYearMonthDay{1973}{04}{}.
\newblock
{\BBOQ}\APACrefatitle {On the Efficiency of a Monetary Equilibrium} {On the
  efficiency of a monetary equilibrium}.{\BBCQ}
\newblock
\APACjournalVolNumPages{The Review of Economic Studies}{40}{2}{149-165}.
\newblock
\begin{APACrefURL} \url{https://doi.org/10.2307/2296645} \end{APACrefURL}
\newblock
\begin{APACrefDOI} \doi{10.2307/2296645} \end{APACrefDOI}
\PrintBackRefs{\CurrentBib}

\bibitem [\protect \citeauthoryear {%
Hampl%
\ \BBA {} Havr\'{a}nek%
}{%
Hampl%
\ \BBA {} Havr\'{a}nek%
}{%
{\protect \APACyear {2017}}%
}]{%
hampl2017inflation}
\APACinsertmetastar {%
hampl2017inflation}%
\begin{APACrefauthors}%
Hampl, M.%
\BCBT {}\ \BBA {} Havr\'{a}nek, T.%
\end{APACrefauthors}%
\unskip\
\newblock
\APACrefYearMonthDay{2017}{December}{}.
\newblock
\APACrefbtitle {Should Inflation Measures Used by Central Banks Incorporate
  House Prices? The {C}zech {N}ational {B}ank's Approach} {Should inflation
  measures used by central banks incorporate house prices? the {C}zech
  {N}ational {B}ank's approach}\ \APACbVolEdTR{}{\BTR{}}.
\newblock
\APACaddressInstitution{}{Czech National Bank, Research and Policy Notes
  2017/01; IES Working Paper 2017/12}.
\newblock
\APACrefnote{Available at
  \href{https://www.cnb.cz/en/economic-research/research-publications/research-and-policy-notes/Should-Inflation-Measures-Used-by-Central-Banks-Incorporate-House-Prices-The-Czech-National-Banks-Approach-00002/}{cnb.cz}}
\PrintBackRefs{\CurrentBib}

\bibitem [\protect \citeauthoryear {%
Hausman%
}{%
Hausman%
}{%
{\protect \APACyear {2003}}%
}]{%
hausman2003sources}
\APACinsertmetastar {%
hausman2003sources}%
\begin{APACrefauthors}%
Hausman, J.%
\end{APACrefauthors}%
\unskip\
\newblock
\APACrefYearMonthDay{2003}{}{}.
\newblock
{\BBOQ}\APACrefatitle {Sources of Bias and Solutions to Bias in the Consumer
  Price Index} {Sources of bias and solutions to bias in the consumer price
  index}.{\BBCQ}
\newblock
\APACjournalVolNumPages{Journal of Economic Perspectives}{17}{1}{23--44}.
\newblock
\begin{APACrefDOI} \doi{10.1257/089533003321164930} \end{APACrefDOI}
\PrintBackRefs{\CurrentBib}

\bibitem [\protect \citeauthoryear {%
Havranek%
}{%
Havranek%
}{%
{\protect \APACyear {2026}}%
}]{%
havranek2026auditduel}
\APACinsertmetastar {%
havranek2026auditduel}%
\begin{APACrefauthors}%
Havranek, T.%
\end{APACrefauthors}%
\unskip\
\newblock
\APACrefYearMonthDay{2026}{}{}.
\newblock
\APACrefbtitle {Research Audit-Duel Protocol.} {Research audit-duel protocol.}
\newblock
\APAChowpublished {GitHub repository}.
\newblock
\APACrefnote{\url{https://github.com/tjhavranek/research-audit-duel-protocol}}
\PrintBackRefs{\CurrentBib}

\bibitem [\protect \citeauthoryear {%
Havranek%
, Horvath%
, Irsova%
\BCBL {}\ \BBA {} Rusnak%
}{%
Havranek%
\ \protect \BOthers {.}}{%
{\protect \APACyear {2015}}%
}]{%
havranek2015cross}
\APACinsertmetastar {%
havranek2015cross}%
\begin{APACrefauthors}%
Havranek, T.%
, Horvath, R.%
, Irsova, Z.%
\BCBL {}\ \BBA {} Rusnak, M.%
\end{APACrefauthors}%
\unskip\
\newblock
\APACrefYearMonthDay{2015}{}{}.
\newblock
{\BBOQ}\APACrefatitle {Cross-country heterogeneity in intertemporal
  substitution} {Cross-country heterogeneity in intertemporal
  substitution}.{\BBCQ}
\newblock
\APACjournalVolNumPages{Journal of International Economics}{96}{1}{100--118}.
\newblock
\begin{APACrefDOI} \doi{10.1016/j.jinteco.2015.01.012} \end{APACrefDOI}
\PrintBackRefs{\CurrentBib}

\bibitem [\protect \citeauthoryear {%
Havr\'{a}nek%
\ \BBA {} Ir\v{s}ov\'{a}%
}{%
Havr\'{a}nek%
\ \BBA {} Ir\v{s}ov\'{a}%
}{%
{\protect \APACyear {2026}}%
}]{%
havranek2026madresearch}
\APACinsertmetastar {%
havranek2026madresearch}%
\begin{APACrefauthors}%
Havr\'{a}nek, T.%
\BCBT {}\ \BBA {} Ir\v{s}ov\'{a}, Z.%
\end{APACrefauthors}%
\unskip\
\newblock
\APACrefYearMonthDay{2026}{}{}.
\newblock
\APACrefbtitle {mad-research: {Claude Code} skills for adversarial multi-agent
  debate on research documents.} {mad-research: {Claude Code} skills for
  adversarial multi-agent debate on research documents.}
\newblock
\APAChowpublished {GitHub repository, MIT License.
  \href{https://doi.org/10.5281/zenodo.19105954}{doi:10.5281/zenodo.19105954}}.
\newblock
\APACrefnote{Open-source successor to the audit-duel protocol; includes a
  meta-analytic data-collection module.
  \url{https://github.com/tjhavranek/mad-research}}
\PrintBackRefs{\CurrentBib}

\bibitem [\protect \citeauthoryear {%
Havranek%
\ \protect \BOthers {.}}{%
Havranek%
\ \protect \BOthers {.}}{%
{\protect \APACyear {2020}}%
}]{%
havranek2020reporting}
\APACinsertmetastar {%
havranek2020reporting}%
\begin{APACrefauthors}%
Havranek, T.%
, Stanley, T\BPBI D.%
, Doucouliagos, H.%
, Bom, P.%
, Geyer-Klingeberg, J.%
, Iwasaki, I.%
\BDBL {}van Aert, R\BPBI C\BPBI M.%
\end{APACrefauthors}%
\unskip\
\newblock
\APACrefYearMonthDay{2020}{}{}.
\newblock
{\BBOQ}\APACrefatitle {Reporting Guidelines for Meta-Analysis in Economics}
  {Reporting guidelines for meta-analysis in economics}.{\BBCQ}
\newblock
\APACjournalVolNumPages{Journal of Economic Surveys}{34}{3}{469--475}.
\newblock
\begin{APACrefDOI} \doi{10.1111/joes.12363} \end{APACrefDOI}
\PrintBackRefs{\CurrentBib}

\bibitem [\protect \citeauthoryear {%
Hendrickson%
\ \BBA {} Salter%
}{%
Hendrickson%
\ \BBA {} Salter%
}{%
{\protect \APACyear {2016}}%
}]{%
hendrickson2016optinf}
\APACinsertmetastar {%
hendrickson2016optinf}%
\begin{APACrefauthors}%
Hendrickson, J.%
\BCBT {}\ \BBA {} Salter, A.%
\end{APACrefauthors}%
\unskip\
\newblock
\APACrefYearMonthDay{2016}{}{}.
\newblock
{\BBOQ}\APACrefatitle {Money, liquidity, and the structure of production}
  {Money, liquidity, and the structure of production}.{\BBCQ}
\newblock
\APACjournalVolNumPages{Journal of Economic Dynamics \&
  Control}{73}{}{314-328}.
\newblock
\begin{APACrefDOI} \doi{10.1016/j.jedc.2016.10.001} \end{APACrefDOI}
\PrintBackRefs{\CurrentBib}

\bibitem [\protect \citeauthoryear {%
Hu%
\ \BBA {} Kam%
}{%
Hu%
\ \BBA {} Kam%
}{%
{\protect \APACyear {2009}}%
}]{%
hu2009optinf}
\APACinsertmetastar {%
hu2009optinf}%
\begin{APACrefauthors}%
Hu, Y.%
\BCBT {}\ \BBA {} Kam, T.%
\end{APACrefauthors}%
\unskip\
\newblock
\APACrefYearMonthDay{2009}{}{}.
\newblock
{\BBOQ}\APACrefatitle {Bonds with transactions service and optimal {Ramsey}
  policy} {Bonds with transactions service and optimal {Ramsey} policy}.{\BBCQ}
\newblock
\APACjournalVolNumPages{Journal of Macroeconomics}{31}{4}{633--653}.
\newblock
\begin{APACrefDOI} \doi{10.1016/j.jmacro.2009.01.006} \end{APACrefDOI}
\PrintBackRefs{\CurrentBib}

\bibitem [\protect \citeauthoryear {%
Imrohoroglu%
}{%
Imrohoroglu%
}{%
{\protect \APACyear {1992}}%
}]{%
imrohoroglu1992optinf}
\APACinsertmetastar {%
imrohoroglu1992optinf}%
\begin{APACrefauthors}%
Imrohoroglu, A.%
\end{APACrefauthors}%
\unskip\
\newblock
\APACrefYearMonthDay{1992}{}{}.
\newblock
{\BBOQ}\APACrefatitle {The welfare cost of inflation under imperfect insurance}
  {The welfare cost of inflation under imperfect insurance}.{\BBCQ}
\newblock
\APACjournalVolNumPages{Journal of Economic Dynamics and
  Control}{16}{1}{79--91}.
\newblock
\begin{APACrefDOI} \doi{10.1016/0165-1889(92)90006-Z} \end{APACrefDOI}
\PrintBackRefs{\CurrentBib}

\bibitem [\protect \citeauthoryear {%
Ioannidis%
, Stanley%
\BCBL {}\ \BBA {} Doucouliagos%
}{%
Ioannidis%
\ \protect \BOthers {.}}{%
{\protect \APACyear {2017}}%
}]{%
ioannidis2017power}
\APACinsertmetastar {%
ioannidis2017power}%
\begin{APACrefauthors}%
Ioannidis, J\BPBI P.%
, Stanley, T.%
\BCBL {}\ \BBA {} Doucouliagos, H.%
\end{APACrefauthors}%
\unskip\
\newblock
\APACrefYearMonthDay{2017}{}{}.
\newblock
{\BBOQ}\APACrefatitle {THE POWER OF BIAS IN ECONOMICS RESEARCH} {The power of
  bias in economics research}.{\BBCQ}
\newblock
\APACjournalVolNumPages{The Economic Journal}{127}{605}{F236--F265}.
\newblock
\begin{APACrefDOI} \doi{10.1111/ecoj.12461} \end{APACrefDOI}
\PrintBackRefs{\CurrentBib}

\bibitem [\protect \citeauthoryear {%
Irsova%
, Bom%
, Havr\'{a}nek%
\BCBL {}\ \BBA {} Rachinger%
}{%
Irsova%
\ \protect \BOthers {.}}{%
{\protect \APACyear {2025}}%
}]{%
irsova2025spurious}
\APACinsertmetastar {%
irsova2025spurious}%
\begin{APACrefauthors}%
Irsova, Z.%
, Bom, P\BPBI R\BPBI D.%
, Havr\'{a}nek, T.%
\BCBL {}\ \BBA {} Rachinger, H.%
\end{APACrefauthors}%
\unskip\
\newblock
\APACrefYearMonthDay{2025}{}{}.
\newblock
{\BBOQ}\APACrefatitle {Spurious Precision in Meta-Analysis of Observational
  Research} {Spurious precision in meta-analysis of observational
  research}.{\BBCQ}
\newblock
\APACjournalVolNumPages{Nature Communications}{}{}{}.
\newblock
\begin{APACrefDOI} \doi{10.1038/s41467-025-63261-0} \end{APACrefDOI}
\PrintBackRefs{\CurrentBib}

\bibitem [\protect \citeauthoryear {%
Irsova%
, Doucouliagos%
, Havr\'{a}nek%
\BCBL {}\ \BBA {} Stanley%
}{%
Irsova%
\ \protect \BOthers {.}}{%
{\protect \APACyear {2024}}%
}]{%
irsova2024practitioner}
\APACinsertmetastar {%
irsova2024practitioner}%
\begin{APACrefauthors}%
Irsova, Z.%
, Doucouliagos, H.%
, Havr\'{a}nek, T.%
\BCBL {}\ \BBA {} Stanley, T\BPBI D.%
\end{APACrefauthors}%
\unskip\
\newblock
\APACrefYearMonthDay{2024}{}{}.
\newblock
{\BBOQ}\APACrefatitle {Meta-Analysis of Social Science Research: A
  Practitioner's Guide} {Meta-analysis of social science research: A
  practitioner's guide}.{\BBCQ}
\newblock
\APACjournalVolNumPages{Journal of Economic Surveys}{38}{5}{1547--1566}.
\newblock
\begin{APACrefDOI} \doi{10.1111/joes.12595} \end{APACrefDOI}
\PrintBackRefs{\CurrentBib}

\bibitem [\protect \citeauthoryear {%
Jiang%
}{%
Jiang%
}{%
{\protect \APACyear {2022}}%
}]{%
jiang2022optinf}
\APACinsertmetastar {%
jiang2022optinf}%
\begin{APACrefauthors}%
Jiang, S.%
\end{APACrefauthors}%
\unskip\
\newblock
\APACrefYearMonthDay{2022}{}{}.
\newblock
{\BBOQ}\APACrefatitle {Optimal Credit, Monetary, and Fiscal Policy under
  Occasional Financial Frictions and the Zero Lower Bound} {Optimal credit,
  monetary, and fiscal policy under occasional financial frictions and the zero
  lower bound}.{\BBCQ}
\newblock
\APACjournalVolNumPages{International Journal of Central
  Banking}{2022}{72}{151-197}.
\newblock
\begin{APACrefDOI} \doi{10.2139/ssrn.3711147} \end{APACrefDOI}
\PrintBackRefs{\CurrentBib}

\bibitem [\protect \citeauthoryear {%
Jung%
}{%
Jung%
}{%
{\protect \APACyear {2025}}%
}]{%
jung2025optinf}
\APACinsertmetastar {%
jung2025optinf}%
\begin{APACrefauthors}%
Jung, Y.%
\end{APACrefauthors}%
\unskip\
\newblock
\APACrefYearMonthDay{2025}{}{}.
\newblock
{\BBOQ}\APACrefatitle {Optimal Monetary Policy in a Small Open Economy With
  Habit Persistence} {Optimal monetary policy in a small open economy with
  habit persistence}.{\BBCQ}
\newblock
\APACjournalVolNumPages{International Finance}{29}{1}{2--28}.
\newblock
\begin{APACrefDOI} \doi{10.1111/infi.70008} \end{APACrefDOI}
\PrintBackRefs{\CurrentBib}

\bibitem [\protect \citeauthoryear {%
Khan%
, King%
\BCBL {}\ \BBA {} Wolman%
}{%
Khan%
\ \protect \BOthers {.}}{%
{\protect \APACyear {2003}}%
}]{%
khan2003optimal}
\APACinsertmetastar {%
khan2003optimal}%
\begin{APACrefauthors}%
Khan, A.%
, King, R\BPBI G.%
\BCBL {}\ \BBA {} Wolman, A\BPBI L.%
\end{APACrefauthors}%
\unskip\
\newblock
\APACrefYearMonthDay{2003}{}{}.
\newblock
{\BBOQ}\APACrefatitle {Optimal Monetary Policy} {Optimal monetary
  policy}.{\BBCQ}
\newblock
\APACjournalVolNumPages{The Review of Economic Studies}{70}{4}{825--860}.
\newblock
\begin{APACrefDOI} \doi{10.1111/1467-937X.00269} \end{APACrefDOI}
\PrintBackRefs{\CurrentBib}

\bibitem [\protect \citeauthoryear {%
Kiarsi%
}{%
Kiarsi%
}{%
{\protect \APACyear {2021}}%
}]{%
kiarsi2021optinf}
\APACinsertmetastar {%
kiarsi2021optinf}%
\begin{APACrefauthors}%
Kiarsi, M.%
\end{APACrefauthors}%
\unskip\
\newblock
\APACrefYearMonthDay{2021}{}{}.
\newblock
{\BBOQ}\APACrefatitle {The rise of market power and {Ramsey}-optimal policy
  implications} {The rise of market power and {Ramsey}-optimal policy
  implications}.{\BBCQ}
\newblock
\APACjournalVolNumPages{Economic Notes}{50}{1}{}.
\newblock
\begin{APACrefDOI} \doi{10.1111/ecno.12175} \end{APACrefDOI}
\PrintBackRefs{\CurrentBib}

\bibitem [\protect \citeauthoryear {%
Kiley%
\ \BBA {} Roberts%
}{%
Kiley%
\ \BBA {} Roberts%
}{%
{\protect \APACyear {2017}}%
}]{%
kiley2017monetary}
\APACinsertmetastar {%
kiley2017monetary}%
\begin{APACrefauthors}%
Kiley, M\BPBI T.%
\BCBT {}\ \BBA {} Roberts, J\BPBI M.%
\end{APACrefauthors}%
\unskip\
\newblock
\APACrefYearMonthDay{2017}{}{}.
\newblock
{\BBOQ}\APACrefatitle {Monetary Policy in a Low Interest Rate World} {Monetary
  policy in a low interest rate world}.{\BBCQ}
\newblock
\APACjournalVolNumPages{Brookings Papers on Economic
  Activity}{2017}{1}{317--396}.
\newblock
\begin{APACrefDOI} \doi{10.1353/eca.2017.0004} \end{APACrefDOI}
\PrintBackRefs{\CurrentBib}

\bibitem [\protect \citeauthoryear {%
Kim%
\ \BBA {} Ruge-Murcia%
}{%
Kim%
\ \BBA {} Ruge-Murcia%
}{%
{\protect \APACyear {2019}}%
}]{%
kim2016optinf}
\APACinsertmetastar {%
kim2016optinf}%
\begin{APACrefauthors}%
Kim, J.%
\BCBT {}\ \BBA {} Ruge-Murcia, F.%
\end{APACrefauthors}%
\unskip\
\newblock
\APACrefYearMonthDay{2019}{}{}.
\newblock
{\BBOQ}\APACrefatitle {EXTREME EVENTS AND OPTIMAL MONETARY POLICY} {Extreme
  events and optimal monetary policy}.{\BBCQ}
\newblock
\APACjournalVolNumPages{International Economic Review}{60}{2}{939-963}.
\newblock
\begin{APACrefURL}
  \url{https://onlinelibrary.wiley.com/doi/abs/10.1111/iere.12372}
  \end{APACrefURL}
\newblock
\begin{APACrefDOI} \doi{https://doi.org/10.1111/iere.12372} \end{APACrefDOI}
\PrintBackRefs{\CurrentBib}

\bibitem [\protect \citeauthoryear {%
Kim%
\ \BBA {} Ruge-Murcia%
}{%
Kim%
\ \BBA {} Ruge-Murcia%
}{%
{\protect \APACyear {2009}}%
}]{%
kimokane2005optimal}
\APACinsertmetastar {%
kimokane2005optimal}%
\begin{APACrefauthors}%
Kim, J.%
\BCBT {}\ \BBA {} Ruge-Murcia, F\BPBI J.%
\end{APACrefauthors}%
\unskip\
\newblock
\APACrefYearMonthDay{2009}{}{}.
\newblock
{\BBOQ}\APACrefatitle {How Much Inflation is Necessary to Grease the Wheels?}
  {How much inflation is necessary to grease the wheels?}{\BBCQ}
\newblock
\APACjournalVolNumPages{Journal of Monetary Economics}{56}{3}{365--377}.
\newblock
\begin{APACrefDOI} \doi{10.1016/j.jmoneco.2009.03.004} \end{APACrefDOI}
\PrintBackRefs{\CurrentBib}

\bibitem [\protect \citeauthoryear {%
Kim%
\ \BBA {} Ruge-Murcia%
}{%
Kim%
\ \BBA {} Ruge-Murcia%
}{%
{\protect \APACyear {2011}}%
}]{%
kim2011optinf}
\APACinsertmetastar {%
kim2011optinf}%
\begin{APACrefauthors}%
Kim, J.%
\BCBT {}\ \BBA {} Ruge-Murcia, F\BPBI J.%
\end{APACrefauthors}%
\unskip\
\newblock
\APACrefYearMonthDay{2011}{}{}.
\newblock
{\BBOQ}\APACrefatitle {Monetary policy when wages are downwardly rigid:
  {Friedman} meets Tobin} {Monetary policy when wages are downwardly rigid:
  {Friedman} meets tobin}.{\BBCQ}
\newblock
\APACjournalVolNumPages{Journal of Economic Dynamics \&
  Control}{35}{12}{2064--2077}.
\newblock
\begin{APACrefDOI} \doi{10.1016/j.jedc.2011.08.002} \end{APACrefDOI}
\PrintBackRefs{\CurrentBib}

\bibitem [\protect \citeauthoryear {%
King%
}{%
King%
}{%
{\protect \APACyear {1996}}%
}]{%
king1996optinf}
\APACinsertmetastar {%
king1996optinf}%
\begin{APACrefauthors}%
King, M\BPBI A.%
\end{APACrefauthors}%
\unskip\
\newblock
\APACrefYearMonthDay{1996}{}{}.
\newblock
{\BBOQ}\APACrefatitle {How Should Central Banks Reduce Inflation? ---
  Conceptual Issues} {How should central banks reduce inflation? --- conceptual
  issues}.{\BBCQ}
\newblock
\APACjournalVolNumPages{Economic Review, Federal Reserve Bank of Kansas
  City}{81}{4}{25--52}.
\PrintBackRefs{\CurrentBib}

\bibitem [\protect \citeauthoryear {%
Kirsanova%
, Leith%
, Machado%
\BCBL {}\ \BBA {} Ribeiro%
}{%
Kirsanova%
\ \protect \BOthers {.}}{%
{\protect \APACyear {2025}}%
}]{%
kirsanova2025optinf}
\APACinsertmetastar {%
kirsanova2025optinf}%
\begin{APACrefauthors}%
Kirsanova, T.%
, Leith, C.%
, Machado, C.%
\BCBL {}\ \BBA {} Ribeiro, A.%
\end{APACrefauthors}%
\unskip\
\newblock
\APACrefYearMonthDay{2025}{}{}.
\newblock
{\BBOQ}\APACrefatitle {(Re)Evaluating recent macroeconomic policy in the {US}}
  {(re)evaluating recent macroeconomic policy in the {US}}.{\BBCQ}
\newblock
\APACjournalVolNumPages{European Economic Review}{178}{}{105091-105091}.
\newblock
\begin{APACrefDOI} \doi{10.1016/j.euroecorev.2025.105091} \end{APACrefDOI}
\PrintBackRefs{\CurrentBib}

\bibitem [\protect \citeauthoryear {%
Kohlbrecher%
}{%
Kohlbrecher%
}{%
{\protect \APACyear {2016}}%
}]{%
kohlbrecher2016optinf}
\APACinsertmetastar {%
kohlbrecher2016optinf}%
\begin{APACrefauthors}%
Kohlbrecher, B.%
\end{APACrefauthors}%
\unskip\
\newblock
\APACrefYearMonthDay{2016}{}{}.
\newblock
{\BBOQ}\APACrefatitle {Optimal Monetary Policy under Rigid Wages and Decreasing
  Returns} {Optimal monetary policy under rigid wages and decreasing
  returns}.{\BBCQ}
\newblock
\BIn{} \APACrefbtitle {Beitr\"{a}ge zur Jahrestagung des Vereins f\"{u}r
  Socialpolitik 2016: Demographischer Wandel -- Session: Transmission of
  Monetary Policy.} {Beitr\"{a}ge zur jahrestagung des vereins f\"{u}r
  socialpolitik 2016: Demographischer wandel -- session: Transmission of
  monetary policy.}
\newblock
\APACaddressPublisher{Kiel und Hamburg}{ZBW -- Deutsche Zentralbibliothek
  f\"{u}r Wirtschaftswissenschaften, Leibniz-Informationszentrum Wirtschaft}.
\newblock
\begin{APACrefURL} \url{https://hdl.handle.net/10419/145867} \end{APACrefURL}
\PrintBackRefs{\CurrentBib}

\bibitem [\protect \citeauthoryear {%
Kollmann%
}{%
Kollmann%
}{%
{\protect \APACyear {2002}}%
}]{%
kollmann2002optinf}
\APACinsertmetastar {%
kollmann2002optinf}%
\begin{APACrefauthors}%
Kollmann, R.%
\end{APACrefauthors}%
\unskip\
\newblock
\APACrefYearMonthDay{2002}{}{}.
\newblock
{\BBOQ}\APACrefatitle {Monetary policy rules in the open economy: effects on
  welfare and business cycles} {Monetary policy rules in the open economy:
  effects on welfare and business cycles}.{\BBCQ}
\newblock
\APACjournalVolNumPages{Journal of Monetary Economics}{49}{5}{989--1015}.
\newblock
\begin{APACrefDOI} \doi{10.1016/s0304-3932(02)00132-0} \end{APACrefDOI}
\PrintBackRefs{\CurrentBib}

\bibitem [\protect \citeauthoryear {%
Kollmann%
}{%
Kollmann%
}{%
{\protect \APACyear {2008}}%
}]{%
kollmann2008optinf}
\APACinsertmetastar {%
kollmann2008optinf}%
\begin{APACrefauthors}%
Kollmann, R.%
\end{APACrefauthors}%
\unskip\
\newblock
\APACrefYearMonthDay{2008}{}{}.
\newblock
{\BBOQ}\APACrefatitle {Welfare-Maximizing Operational Monetary and Tax Policy
  Rules} {Welfare-maximizing operational monetary and tax policy rules}.{\BBCQ}
\newblock
\APACjournalVolNumPages{Macroeconomic Dynamics}{12}{S1}{112--125}.
\newblock
\begin{APACrefDOI} \doi{10.1017/s1365100507060397} \end{APACrefDOI}
\PrintBackRefs{\CurrentBib}

\bibitem [\protect \citeauthoryear {%
Leach%
}{%
Leach%
}{%
{\protect \APACyear {1983}}%
}]{%
leach1983inflation}
\APACinsertmetastar {%
leach1983inflation}%
\begin{APACrefauthors}%
Leach, J.%
\end{APACrefauthors}%
\unskip\
\newblock
\APACrefYearMonthDay{1983}{}{}.
\newblock
{\BBOQ}\APACrefatitle {Inflation as a Commodity Tax} {Inflation as a commodity
  tax}.{\BBCQ}
\newblock
\APACjournalVolNumPages{Canadian Journal of Economics}{16}{3}{508--516}.
\newblock
\begin{APACrefDOI} \doi{10.2307/135161} \end{APACrefDOI}
\PrintBackRefs{\CurrentBib}

\bibitem [\protect \citeauthoryear {%
Lebow%
\ \BBA {} Rudd%
}{%
Lebow%
\ \BBA {} Rudd%
}{%
{\protect \APACyear {2003}}%
}]{%
lebow2003measurement}
\APACinsertmetastar {%
lebow2003measurement}%
\begin{APACrefauthors}%
Lebow, D\BPBI E.%
\BCBT {}\ \BBA {} Rudd, J\BPBI B.%
\end{APACrefauthors}%
\unskip\
\newblock
\APACrefYearMonthDay{2003}{}{}.
\newblock
{\BBOQ}\APACrefatitle {Measurement Error in the Consumer Price Index: Where Do
  We Stand?} {Measurement error in the consumer price index: Where do we
  stand?}{\BBCQ}
\newblock
\APACjournalVolNumPages{Journal of Economic Literature}{41}{1}{159--201}.
\newblock
\begin{APACrefDOI} \doi{10.1257/002205103321544729} \end{APACrefDOI}
\PrintBackRefs{\CurrentBib}

\bibitem [\protect \citeauthoryear {%
Leith%
, Moldovan%
\BCBL {}\ \BBA {} Rossi%
}{%
Leith%
\ \protect \BOthers {.}}{%
{\protect \APACyear {2012}}%
}]{%
leith2012optinf}
\APACinsertmetastar {%
leith2012optinf}%
\begin{APACrefauthors}%
Leith, C.%
, Moldovan, I.%
\BCBL {}\ \BBA {} Rossi, R.%
\end{APACrefauthors}%
\unskip\
\newblock
\APACrefYearMonthDay{2012}{}{}.
\newblock
{\BBOQ}\APACrefatitle {Optimal monetary policy in a {New Keynesian} model with
  habits in consumption} {Optimal monetary policy in a {New Keynesian} model
  with habits in consumption}.{\BBCQ}
\newblock
\APACjournalVolNumPages{Review of Economic Dynamics}{15}{3}{416--435}.
\newblock
\begin{APACrefDOI} \doi{10.1016/j.red.2012.03.001} \end{APACrefDOI}
\PrintBackRefs{\CurrentBib}

\bibitem [\protect \citeauthoryear {%
Levin%
, Lopez-Salido%
\BCBL {}\ \BBA {} Yun%
}{%
Levin%
\ \protect \BOthers {.}}{%
{\protect \APACyear {2007}}%
}]{%
levin2007optinf}
\APACinsertmetastar {%
levin2007optinf}%
\begin{APACrefauthors}%
Levin, A.%
, Lopez-Salido, D.%
\BCBL {}\ \BBA {} Yun, T.%
\end{APACrefauthors}%
\unskip\
\newblock
\APACrefYearMonthDay{2007}{August}{}.
\newblock
\APACrefbtitle {Strategic Complementarities and Optimal Monetary Policy}
  {Strategic complementarities and optimal monetary policy}\ \APACbVolEdTR
  {}{CEPR Discussion Paper\ \BNUM\ DP6423}.
\newblock
\APACaddressInstitution{}{Centre for Economic Policy Research (CEPR)}.
\newblock
\APACrefnote{Available at SSRN: https://ssrn.com/abstract=1138536}
\PrintBackRefs{\CurrentBib}

\bibitem [\protect \citeauthoryear {%
Lewis%
}{%
Lewis%
}{%
{\protect \APACyear {2013}}%
}]{%
lewis2013optinf}
\APACinsertmetastar {%
lewis2013optinf}%
\begin{APACrefauthors}%
Lewis, V.%
\end{APACrefauthors}%
\unskip\
\newblock
\APACrefYearMonthDay{2013}{}{}.
\newblock
{\BBOQ}\APACrefatitle {Optimal monetary policy and firm entry} {Optimal
  monetary policy and firm entry}.{\BBCQ}
\newblock
\APACjournalVolNumPages{Macroeconomic Dynamics}{17}{8}{1687--1710}.
\newblock
\begin{APACrefDOI} \doi{10.1017/s1365100512000272} \end{APACrefDOI}
\PrintBackRefs{\CurrentBib}

\bibitem [\protect \citeauthoryear {%
Ley%
\ \BBA {} Steel%
}{%
Ley%
\ \BBA {} Steel%
}{%
{\protect \APACyear {2009}}%
}]{%
ley2009effect}
\APACinsertmetastar {%
ley2009effect}%
\begin{APACrefauthors}%
Ley, E.%
\BCBT {}\ \BBA {} Steel, M\BPBI F.%
\end{APACrefauthors}%
\unskip\
\newblock
\APACrefYearMonthDay{2009}{}{}.
\newblock
{\BBOQ}\APACrefatitle {On the effect of prior assumptions in {B}ayesian model
  averaging with applications to growth regression} {On the effect of prior
  assumptions in {B}ayesian model averaging with applications to growth
  regression}.{\BBCQ}
\newblock
\APACjournalVolNumPages{Journal of applied econometrics}{24}{4}{651--674}.
\newblock
\begin{APACrefDOI} \doi{10.1002/jae.1057} \end{APACrefDOI}
\PrintBackRefs{\CurrentBib}

\bibitem [\protect \citeauthoryear {%
Lipi\'{n}ska%
}{%
Lipi\'{n}ska%
}{%
{\protect \APACyear {2015}}%
}]{%
lipinska2015optimal}
\APACinsertmetastar {%
lipinska2015optimal}%
\begin{APACrefauthors}%
Lipi\'{n}ska, A.%
\end{APACrefauthors}%
\unskip\
\newblock
\APACrefYearMonthDay{2015}{}{}.
\newblock
{\BBOQ}\APACrefatitle {Optimal Monetary Policy for the {EMU} Accession
  Countries: A New {K}eynesian Approach} {Optimal monetary policy for the {EMU}
  accession countries: A new {K}eynesian approach}.{\BBCQ}
\newblock
\APACjournalVolNumPages{Macroeconomic Dynamics}{19}{7}{1427--1475}.
\newblock
\begin{APACrefDOI} \doi{10.1017/S1365100513000898} \end{APACrefDOI}
\PrintBackRefs{\CurrentBib}

\bibitem [\protect \citeauthoryear {%
Lubik%
\ \BBA {} Teo%
}{%
Lubik%
\ \BBA {} Teo%
}{%
{\protect \APACyear {2010}}%
}]{%
lubik2010optinf}
\APACinsertmetastar {%
lubik2010optinf}%
\begin{APACrefauthors}%
Lubik, T\BPBI A.%
\BCBT {}\ \BBA {} Teo, W\BPBI L.%
\end{APACrefauthors}%
\unskip\
\newblock
\APACrefYearMonthDay{2010}{}{}.
\newblock
\APACrefbtitle {Inventories and Optimal Monetary Policy} {Inventories and
  optimal monetary policy}\ \APACbVolEdTR{}{\BTR{}}.
\newblock
\APACaddressInstitution{}{CAMA Working Paper}.
\newblock
\begin{APACrefDOI} \doi{10.2139/ssrn.1668128} \end{APACrefDOI}
\PrintBackRefs{\CurrentBib}

\bibitem [\protect \citeauthoryear {%
Lucas%
}{%
Lucas%
}{%
{\protect \APACyear {1980}}%
}]{%
lucas1980equilibrium}
\APACinsertmetastar {%
lucas1980equilibrium}%
\begin{APACrefauthors}%
Lucas, R\BPBI E., Jr.%
\end{APACrefauthors}%
\unskip\
\newblock
\APACrefYearMonthDay{1980}{}{}.
\newblock
{\BBOQ}\APACrefatitle {Equilibrium in a Pure Currency Economy} {Equilibrium in
  a pure currency economy}.{\BBCQ}
\newblock
\APACjournalVolNumPages{Economic Inquiry}{18}{2}{203--220}.
\newblock
\begin{APACrefDOI} \doi{10.1111/j.1465-7295.1980.tb00570.x} \end{APACrefDOI}
\PrintBackRefs{\CurrentBib}

\bibitem [\protect \citeauthoryear {%
Lucas%
}{%
Lucas%
}{%
{\protect \APACyear {2000}}%
}]{%
lucas2000optinf}
\APACinsertmetastar {%
lucas2000optinf}%
\begin{APACrefauthors}%
Lucas, R\BPBI E., Jr.%
\end{APACrefauthors}%
\unskip\
\newblock
\APACrefYearMonthDay{2000}{}{}.
\newblock
{\BBOQ}\APACrefatitle {Inflation and Welfare} {Inflation and welfare}.{\BBCQ}
\newblock
\APACjournalVolNumPages{Econometrica}{68}{2}{247--274}.
\newblock
\begin{APACrefDOI} \doi{10.1111/1468-0262.00109} \end{APACrefDOI}
\PrintBackRefs{\CurrentBib}

\bibitem [\protect \citeauthoryear {%
Mathur%
}{%
Mathur%
}{%
{\protect \APACyear {2024}}%
}]{%
mathur2024}
\APACinsertmetastar {%
mathur2024}%
\begin{APACrefauthors}%
Mathur, M\BPBI B.%
\end{APACrefauthors}%
\unskip\
\newblock
\APACrefYearMonthDay{2024}{}{}.
\newblock
{\BBOQ}\APACrefatitle {P-Hacking in Meta-Analyses: A Formalization and New
  Meta-Analytic Methods} {P-hacking in meta-analyses: A formalization and new
  meta-analytic methods}.{\BBCQ}
\newblock
\APACjournalVolNumPages{Research Synthesis Methods}{15}{3}{483-499}.
\newblock
\begin{APACrefDOI} \doi{10.1002/jrsm.1701} \end{APACrefDOI}
\PrintBackRefs{\CurrentBib}

\bibitem [\protect \citeauthoryear {%
Matveev%
}{%
Matveev%
}{%
{\protect \APACyear {2021}}%
}]{%
matveev2021optinf}
\APACinsertmetastar {%
matveev2021optinf}%
\begin{APACrefauthors}%
Matveev, D.%
\end{APACrefauthors}%
\unskip\
\newblock
\APACrefYearMonthDay{2021}{}{}.
\newblock
{\BBOQ}\APACrefatitle {Time-Consistent Management of a Liquidity Trap with
  Government Debt} {Time-consistent management of a liquidity trap with
  government debt}.{\BBCQ}
\newblock
\APACjournalVolNumPages{Journal of Money, Credit and
  Banking}{53}{8}{2129-2165}.
\newblock
\begin{APACrefDOI} \doi{10.1111/jmcb.12820} \end{APACrefDOI}
\PrintBackRefs{\CurrentBib}

\bibitem [\protect \citeauthoryear {%
Menna%
\ \BBA {} Tirelli%
}{%
Menna%
\ \BBA {} Tirelli%
}{%
{\protect \APACyear {2017}}%
}]{%
menna2017optinf}
\APACinsertmetastar {%
menna2017optinf}%
\begin{APACrefauthors}%
Menna, L.%
\BCBT {}\ \BBA {} Tirelli, P.%
\end{APACrefauthors}%
\unskip\
\newblock
\APACrefYearMonthDay{2017}{}{}.
\newblock
{\BBOQ}\APACrefatitle {Optimal inflation to reduce inequality} {Optimal
  inflation to reduce inequality}.{\BBCQ}
\newblock
\APACjournalVolNumPages{Review of Economic Dynamics}{24}{}{79--94}.
\newblock
\begin{APACrefDOI} \doi{10.1016/j.red.2017.01.004} \end{APACrefDOI}
\PrintBackRefs{\CurrentBib}

\bibitem [\protect \citeauthoryear {%
Mineyama%
}{%
Mineyama%
}{%
{\protect \APACyear {2022}}%
}]{%
mineyama2022optinf}
\APACinsertmetastar {%
mineyama2022optinf}%
\begin{APACrefauthors}%
Mineyama, T.%
\end{APACrefauthors}%
\unskip\
\newblock
\APACrefYearMonthDay{2022}{}{}.
\newblock
{\BBOQ}\APACrefatitle {Revisiting the optimal inflation rate with downward
  nominal wage rigidity: The role of heterogeneity} {Revisiting the optimal
  inflation rate with downward nominal wage rigidity: The role of
  heterogeneity}.{\BBCQ}
\newblock
\APACjournalVolNumPages{Journal of Economic Dynamics \&
  Control}{139}{}{104350}.
\newblock
\begin{APACrefDOI} \doi{10.1016/j.jedc.2022.104350} \end{APACrefDOI}
\PrintBackRefs{\CurrentBib}

\bibitem [\protect \citeauthoryear {%
Miura%
}{%
Miura%
}{%
{\protect \APACyear {2023}}%
}]{%
miura2023optinf}
\APACinsertmetastar {%
miura2023optinf}%
\begin{APACrefauthors}%
Miura, S.%
\end{APACrefauthors}%
\unskip\
\newblock
\APACrefYearMonthDay{2023}{}{}.
\newblock
{\BBOQ}\APACrefatitle {Optimal inflation rate and fair wage} {Optimal inflation
  rate and fair wage}.{\BBCQ}
\newblock
\APACjournalVolNumPages{The Quarterly Review of Economics and
  Finance}{88}{}{158--167}.
\newblock
\begin{APACrefDOI} \doi{10.1016/j.qref.2022.12.013} \end{APACrefDOI}
\PrintBackRefs{\CurrentBib}

\bibitem [\protect \citeauthoryear {%
Montoro%
}{%
Montoro%
}{%
{\protect \APACyear {2012}}%
}]{%
montoro2012optinf}
\APACinsertmetastar {%
montoro2012optinf}%
\begin{APACrefauthors}%
Montoro, C.%
\end{APACrefauthors}%
\unskip\
\newblock
\APACrefYearMonthDay{2012}{}{}.
\newblock
{\BBOQ}\APACrefatitle {Oil shocks and optimal monetary policy} {Oil shocks and
  optimal monetary policy}.{\BBCQ}
\newblock
\APACjournalVolNumPages{Macroeconomic Dynamics}{16}{2}{240--277}.
\newblock
\begin{APACrefDOI} \doi{10.1017/s1365100510000106} \end{APACrefDOI}
\PrintBackRefs{\CurrentBib}

\bibitem [\protect \citeauthoryear {%
Motta%
\ \BBA {} Tirelli%
}{%
Motta%
\ \BBA {} Tirelli%
}{%
{\protect \APACyear {2012}}%
}]{%
motta2012optinf}
\APACinsertmetastar {%
motta2012optinf}%
\begin{APACrefauthors}%
Motta, G.%
\BCBT {}\ \BBA {} Tirelli, P.%
\end{APACrefauthors}%
\unskip\
\newblock
\APACrefYearMonthDay{2012}{}{}.
\newblock
{\BBOQ}\APACrefatitle {Optimal Simple Monetary and Fiscal Rules under Limited
  Asset Market Participation} {Optimal simple monetary and fiscal rules under
  limited asset market participation}.{\BBCQ}
\newblock
\APACjournalVolNumPages{Journal of Money, Credit and
  Banking}{44}{7}{1351--1374}.
\newblock
\begin{APACrefDOI} \doi{10.1111/j.1538-4616.2012.00535.x} \end{APACrefDOI}
\PrintBackRefs{\CurrentBib}

\bibitem [\protect \citeauthoryear {%
Moulton%
}{%
Moulton%
}{%
{\protect \APACyear {2018}}%
}]{%
moulton2018measurement}
\APACinsertmetastar {%
moulton2018measurement}%
\begin{APACrefauthors}%
Moulton, B\BPBI R.%
\end{APACrefauthors}%
\unskip\
\newblock
\APACrefYearMonthDay{2018}{}{}.
\newblock
\APACrefbtitle {The Measurement of Output, Prices, and Productivity: What's
  Changed Since the Boskin Commission?} {The measurement of output, prices, and
  productivity: What's changed since the boskin commission?}\
  \APACbVolEdTR{}{\BTR{}}.
\newblock
\APACaddressInstitution{}{Hutchins Center on Fiscal and Monetary Policy at
  Brookings}.
\PrintBackRefs{\CurrentBib}

\bibitem [\protect \citeauthoryear {%
Mulligan%
}{%
Mulligan%
}{%
{\protect \APACyear {1997}}%
}]{%
mulligan1997optinf}
\APACinsertmetastar {%
mulligan1997optinf}%
\begin{APACrefauthors}%
Mulligan, C\BPBI B.%
\end{APACrefauthors}%
\unskip\
\newblock
\APACrefYearMonthDay{1997}{}{}.
\newblock
{\BBOQ}\APACrefatitle {Scale Economies, the Value of Time, and the Demand for
  Money: Longitudinal Evidence from Firms} {Scale economies, the value of time,
  and the demand for money: Longitudinal evidence from firms}.{\BBCQ}
\newblock
\APACjournalVolNumPages{Journal of Political Economy}{105}{5}{1061--1079}.
\newblock
\begin{APACrefDOI} \doi{10.1086/262105} \end{APACrefDOI}
\PrintBackRefs{\CurrentBib}

\bibitem [\protect \citeauthoryear {%
Nistic\`{o}%
}{%
Nistic\`{o}%
}{%
{\protect \APACyear {2016}}%
}]{%
nistico2016optinf}
\APACinsertmetastar {%
nistico2016optinf}%
\begin{APACrefauthors}%
Nistic\`{o}, S.%
\end{APACrefauthors}%
\unskip\
\newblock
\APACrefYearMonthDay{2016}{10}{}.
\newblock
{\BBOQ}\APACrefatitle {Optimal Monetary Policy and Financial Stability in a
  Non-Ricardian Economy} {Optimal monetary policy and financial stability in a
  non-ricardian economy}.{\BBCQ}
\newblock
\APACjournalVolNumPages{Journal of the European Economic
  Association}{14}{5}{1225-1252}.
\newblock
\begin{APACrefURL} \url{https://doi.org/10.1111/jeea.12182} \end{APACrefURL}
\newblock
\begin{APACrefDOI} \doi{10.1111/jeea.12182} \end{APACrefDOI}
\PrintBackRefs{\CurrentBib}

\bibitem [\protect \citeauthoryear {%
Nlemfu~Mukoko%
}{%
Nlemfu~Mukoko%
}{%
{\protect \APACyear {2016}}%
}]{%
mukoko2016optinf}
\APACinsertmetastar {%
mukoko2016optinf}%
\begin{APACrefauthors}%
Nlemfu~Mukoko, J\BPBI B.%
\end{APACrefauthors}%
\unskip\
\newblock
\APACrefYearMonthDay{2016}{}{}.
\newblock
\APACrefbtitle {On the Welfare Costs of Monetary Policy} {On the welfare costs
  of monetary policy}\ \APACbVolEdTR {}{MPRA Paper\ \BNUM\ 77382}.
\newblock
\APACaddressInstitution{}{Munich Personal RePEc Archive}.
\PrintBackRefs{\CurrentBib}

\bibitem [\protect \citeauthoryear {%
Nu\~{n}o%
\ \BBA {} Thomas%
}{%
Nu\~{n}o%
\ \BBA {} Thomas%
}{%
{\protect \APACyear {2022}}%
}]{%
nuno2022optinf}
\APACinsertmetastar {%
nuno2022optinf}%
\begin{APACrefauthors}%
Nu\~{n}o, G.%
\BCBT {}\ \BBA {} Thomas, C.%
\end{APACrefauthors}%
\unskip\
\newblock
\APACrefYearMonthDay{2022}{}{}.
\newblock
{\BBOQ}\APACrefatitle {Optimal Redistributive Inflation} {Optimal
  redistributive inflation}.{\BBCQ}
\newblock
\APACjournalVolNumPages{Annals of Economics and Statistics}{}{146}{}.
\newblock
\begin{APACrefDOI} \doi{10.2307/48674138} \end{APACrefDOI}
\PrintBackRefs{\CurrentBib}

\bibitem [\protect \citeauthoryear {%
Opatrny%
, Havranek%
, Irsova%
\BCBL {}\ \BBA {} Scasny%
}{%
Opatrny%
\ \protect \BOthers {.}}{%
{\protect \APACyear {2026}}%
}]{%
opatrny2024class}
\APACinsertmetastar {%
opatrny2024class}%
\begin{APACrefauthors}%
Opatrny, M.%
, Havranek, T.%
, Irsova, Z.%
\BCBL {}\ \BBA {} Scasny, M.%
\end{APACrefauthors}%
\unskip\
\newblock
\APACrefYearMonthDay{2026}{}{}.
\newblock
{\BBOQ}\APACrefatitle {Publication Bias and Model Uncertainty in Measuring the
  Effect of Class Size on Achievement} {Publication bias and model uncertainty
  in measuring the effect of class size on achievement}.{\BBCQ}
\newblock
\APACjournalVolNumPages{Journal of Labor Economics}{}{}{}.
\newblock
\begin{APACrefURL} \url{https://doi.org/10.1086/737989} \end{APACrefURL}
\newblock
\APACrefnote{Forthcoming}
\newblock
\begin{APACrefDOI} \doi{10.1086/737989} \end{APACrefDOI}
\PrintBackRefs{\CurrentBib}

\bibitem [\protect \citeauthoryear {%
Paciello%
\ \BBA {} Wiederholt%
}{%
Paciello%
\ \BBA {} Wiederholt%
}{%
{\protect \APACyear {2014}}%
}]{%
paciello2011optinf}
\APACinsertmetastar {%
paciello2011optinf}%
\begin{APACrefauthors}%
Paciello, L.%
\BCBT {}\ \BBA {} Wiederholt, M.%
\end{APACrefauthors}%
\unskip\
\newblock
\APACrefYearMonthDay{2014}{}{}.
\newblock
{\BBOQ}\APACrefatitle {Exogenous Information, Endogenous Information and
  Optimal Monetary Policy} {Exogenous information, endogenous information and
  optimal monetary policy}.{\BBCQ}
\newblock
\APACjournalVolNumPages{The Review of Economic Studies}{81}{1}{356--388}.
\newblock
\APACrefnote{Published version of EIEF Working Paper 11/01 (2011).}
\newblock
\begin{APACrefDOI} \doi{10.1093/restud/rdt024} \end{APACrefDOI}
\PrintBackRefs{\CurrentBib}

\bibitem [\protect \citeauthoryear {%
Paczos%
}{%
Paczos%
}{%
{\protect \APACyear {2020}}%
}]{%
paczos2020optinf}
\APACinsertmetastar {%
paczos2020optinf}%
\begin{APACrefauthors}%
Paczos, W.%
\end{APACrefauthors}%
\unskip\
\newblock
\APACrefYearMonthDay{2020}{}{}.
\newblock
{\BBOQ}\APACrefatitle {Optimal Inflation, Monetary Integration, and Asymmetric
  Sticky Prices} {Optimal inflation, monetary integration, and asymmetric
  sticky prices}.{\BBCQ}
\newblock
\APACjournalVolNumPages{Oxford Economic Papers}{72}{3}{710--730}.
\newblock
\begin{APACrefDOI} \doi{10.1093/oep/gpaa008} \end{APACrefDOI}
\PrintBackRefs{\CurrentBib}

\bibitem [\protect \citeauthoryear {%
Phelps%
}{%
Phelps%
}{%
{\protect \APACyear {1965}}%
}]{%
phelps1965anticipated}
\APACinsertmetastar {%
phelps1965anticipated}%
\begin{APACrefauthors}%
Phelps, E\BPBI S.%
\end{APACrefauthors}%
\unskip\
\newblock
\APACrefYearMonthDay{1965}{}{}.
\newblock
{\BBOQ}\APACrefatitle {Anticipated Inflation and Economic Welfare} {Anticipated
  inflation and economic welfare}.{\BBCQ}
\newblock
\APACjournalVolNumPages{Journal of Political Economy}{73}{1}{1--17}.
\newblock
\begin{APACrefDOI} \doi{10.1086/258988} \end{APACrefDOI}
\PrintBackRefs{\CurrentBib}

\bibitem [\protect \citeauthoryear {%
Pontiggia%
}{%
Pontiggia%
}{%
{\protect \APACyear {2012}}%
}]{%
pontiggia2012optinf}
\APACinsertmetastar {%
pontiggia2012optinf}%
\begin{APACrefauthors}%
Pontiggia, D.%
\end{APACrefauthors}%
\unskip\
\newblock
\APACrefYearMonthDay{2012}{}{}.
\newblock
{\BBOQ}\APACrefatitle {Optimal long-run inflation and the {New Keynesian}
  model} {Optimal long-run inflation and the {New Keynesian} model}.{\BBCQ}
\newblock
\APACjournalVolNumPages{Journal of Macroeconomics}{34}{4}{1077--1094}.
\newblock
\begin{APACrefDOI} \doi{10.1016/j.jmacro.2012.07.003} \end{APACrefDOI}
\PrintBackRefs{\CurrentBib}

\bibitem [\protect \citeauthoryear {%
Raissi%
}{%
Raissi%
}{%
{\protect \APACyear {2015}}%
}]{%
raissi2015optinf}
\APACinsertmetastar {%
raissi2015optinf}%
\begin{APACrefauthors}%
Raissi, M.%
\end{APACrefauthors}%
\unskip\
\newblock
\APACrefYearMonthDay{2015}{}{}.
\newblock
{\BBOQ}\APACrefatitle {Flexible inflation targeting and labor market
  inefficiencies} {Flexible inflation targeting and labor market
  inefficiencies}.{\BBCQ}
\newblock
\APACjournalVolNumPages{Economic Modelling}{46}{}{283--300}.
\newblock
\begin{APACrefDOI} \doi{10.1016/j.econmod.2014.12.025} \end{APACrefDOI}
\PrintBackRefs{\CurrentBib}

\bibitem [\protect \citeauthoryear {%
Ravenna%
\ \BBA {} Walsh%
}{%
Ravenna%
\ \BBA {} Walsh%
}{%
{\protect \APACyear {2011}}%
}]{%
ravenna2010optinf}
\APACinsertmetastar {%
ravenna2010optinf}%
\begin{APACrefauthors}%
Ravenna, F.%
\BCBT {}\ \BBA {} Walsh, C\BPBI E.%
\end{APACrefauthors}%
\unskip\
\newblock
\APACrefYearMonthDay{2011}{}{}.
\newblock
{\BBOQ}\APACrefatitle {Welfare-based optimal monetary policy with unemployment
  and sticky prices: A linear-quadratic framework} {Welfare-based optimal
  monetary policy with unemployment and sticky prices: A linear-quadratic
  framework}.{\BBCQ}
\newblock
\APACjournalVolNumPages{American Economic Journal:
  Macroeconomics}{3}{2}{130--162}.
\newblock
\begin{APACrefDOI} \doi{10.1257/mac.3.2.130} \end{APACrefDOI}
\PrintBackRefs{\CurrentBib}

\bibitem [\protect \citeauthoryear {%
Rotemberg%
}{%
Rotemberg%
}{%
{\protect \APACyear {1982}}%
}]{%
rotemberg1982sticky}
\APACinsertmetastar {%
rotemberg1982sticky}%
\begin{APACrefauthors}%
Rotemberg, J\BPBI J.%
\end{APACrefauthors}%
\unskip\
\newblock
\APACrefYearMonthDay{1982}{}{}.
\newblock
{\BBOQ}\APACrefatitle {Sticky Prices in the United States} {Sticky prices in
  the united states}.{\BBCQ}
\newblock
\APACjournalVolNumPages{Journal of Political Economy}{90}{6}{1187--1211}.
\newblock
\begin{APACrefDOI} \doi{10.1086/261117} \end{APACrefDOI}
\PrintBackRefs{\CurrentBib}

\bibitem [\protect \citeauthoryear {%
Rotemberg%
\ \BBA {} Woodford%
}{%
Rotemberg%
\ \BBA {} Woodford%
}{%
{\protect \APACyear {1997}}%
}]{%
rotemberg1997optinf}
\APACinsertmetastar {%
rotemberg1997optinf}%
\begin{APACrefauthors}%
Rotemberg, J\BPBI J.%
\BCBT {}\ \BBA {} Woodford, M.%
\end{APACrefauthors}%
\unskip\
\newblock
\APACrefYearMonthDay{1997}{}{}.
\newblock
\APACrefbtitle {An Optimization-Based Econometric Framework for the Evaluation
  of Monetary Policy} {An optimization-based econometric framework for the
  evaluation of monetary policy}\ \APACbVolEdTR{\BVOL~12}{\BTR{}}.
\newblock
\APACaddressInstitution{}{NBER Macroeconomics Annual}.
\newblock
\begin{APACrefDOI} \doi{10.1086/654340} \end{APACrefDOI}
\PrintBackRefs{\CurrentBib}

\bibitem [\protect \citeauthoryear {%
Schmitt-Groh\'{e}%
\ \BBA {} Uribe%
}{%
Schmitt-Groh\'{e}%
\ \BBA {} Uribe%
}{%
{\protect \APACyear {2004}}%
{\protect \APACexlab {{\protect \BCnt {1}}}}}]{%
schmittgrohe2004optinf}
\APACinsertmetastar {%
schmittgrohe2004optinf}%
\begin{APACrefauthors}%
Schmitt-Groh\'{e}, S.%
\BCBT {}\ \BBA {} Uribe, M.%
\end{APACrefauthors}%
\unskip\
\newblock
\APACrefYearMonthDay{2004{\protect \BCnt {1}}}{}{}.
\newblock
{\BBOQ}\APACrefatitle {Optimal fiscal and monetary policy under imperfect
  competition} {Optimal fiscal and monetary policy under imperfect
  competition}.{\BBCQ}
\newblock
\APACjournalVolNumPages{Journal of Macroeconomics}{26}{2}{183-209}.
\newblock
\begin{APACrefURL}
  \url{https://www.sciencedirect.com/science/article/pii/S0164070403000752}
  \end{APACrefURL}
\newblock
\APACrefnote{Monetary Policy}
\newblock
\begin{APACrefDOI} \doi{https://doi.org/10.1016/j.jmacro.2003.11.002}
  \end{APACrefDOI}
\PrintBackRefs{\CurrentBib}

\bibitem [\protect \citeauthoryear {%
Schmitt-Groh\'{e}%
\ \BBA {} Uribe%
}{%
Schmitt-Groh\'{e}%
\ \BBA {} Uribe%
}{%
{\protect \APACyear {2004}}%
{\protect \APACexlab {{\protect \BCnt {2}}}}}]{%
schmittgrohe2004optinfa}
\APACinsertmetastar {%
schmittgrohe2004optinfa}%
\begin{APACrefauthors}%
Schmitt-Groh\'{e}, S.%
\BCBT {}\ \BBA {} Uribe, M.%
\end{APACrefauthors}%
\unskip\
\newblock
\APACrefYearMonthDay{2004{\protect \BCnt {2}}}{}{}.
\newblock
{\BBOQ}\APACrefatitle {Optimal fiscal and monetary policy under sticky prices}
  {Optimal fiscal and monetary policy under sticky prices}.{\BBCQ}
\newblock
\APACjournalVolNumPages{Journal of Economic Theory}{114}{2}{198-230}.
\newblock
\begin{APACrefURL}
  \url{https://www.sciencedirect.com/science/article/pii/S002205310300111X}
  \end{APACrefURL}
\newblock
\begin{APACrefDOI} \doi{https://doi.org/10.1016/S0022-0531(03)00111-X}
  \end{APACrefDOI}
\PrintBackRefs{\CurrentBib}

\bibitem [\protect \citeauthoryear {%
Schmitt-Groh{\'e}%
\ \BBA {} Uribe%
}{%
Schmitt-Groh{\'e}%
\ \BBA {} Uribe%
}{%
{\protect \APACyear {2005}}%
{\protect \APACexlab {{\protect \BCnt {1}}}}}]{%
schmittgrohe2005optinf}
\APACinsertmetastar {%
schmittgrohe2005optinf}%
\begin{APACrefauthors}%
Schmitt-Groh{\'e}, S.%
\BCBT {}\ \BBA {} Uribe, M.%
\end{APACrefauthors}%
\unskip\
\newblock
\APACrefYearMonthDay{2005{\protect \BCnt {1}}}{}{}.
\newblock
{\BBOQ}\APACrefatitle {Optimal Fiscal and Monetary Policy in a Medium-Scale
  Macroeconomic Model} {Optimal fiscal and monetary policy in a medium-scale
  macroeconomic model}.{\BBCQ}
\newblock
\APACjournalVolNumPages{NBER Macroeconomics Annual}{20}{}{383--425}.
\PrintBackRefs{\CurrentBib}

\bibitem [\protect \citeauthoryear {%
Schmitt-Groh{\'e}%
\ \BBA {} Uribe%
}{%
Schmitt-Groh{\'e}%
\ \BBA {} Uribe%
}{%
{\protect \APACyear {2005}}%
{\protect \APACexlab {{\protect \BCnt {2}}}}}]{%
schmittgrohe2005optinfa}
\APACinsertmetastar {%
schmittgrohe2005optinfa}%
\begin{APACrefauthors}%
Schmitt-Groh{\'e}, S.%
\BCBT {}\ \BBA {} Uribe, M.%
\end{APACrefauthors}%
\unskip\
\newblock
\APACrefYearMonthDay{2005{\protect \BCnt {2}}}{December}{}.
\newblock
\APACrefbtitle {Optimal Inflation Stabilization in a Medium-Scale Macroeconomic
  Model} {Optimal inflation stabilization in a medium-scale macroeconomic
  model}\ \APACbVolEdTR {}{Working Paper\ \BNUM\ 11854}.
\newblock
\APACaddressInstitution{}{National Bureau of Economic Research}.
\newblock
\begin{APACrefURL} \url{http://www.nber.org/papers/w11854} \end{APACrefURL}
\newblock
\begin{APACrefDOI} \doi{10.3386/w11854} \end{APACrefDOI}
\PrintBackRefs{\CurrentBib}

\bibitem [\protect \citeauthoryear {%
Schmitt-Groh\'{e}%
\ \BBA {} Uribe%
}{%
Schmitt-Groh\'{e}%
\ \BBA {} Uribe%
}{%
{\protect \APACyear {2007}}%
}]{%
schmittgrohe2007optinf}
\APACinsertmetastar {%
schmittgrohe2007optinf}%
\begin{APACrefauthors}%
Schmitt-Groh\'{e}, S.%
\BCBT {}\ \BBA {} Uribe, M.%
\end{APACrefauthors}%
\unskip\
\newblock
\APACrefYearMonthDay{2007}{}{}.
\newblock
{\BBOQ}\APACrefatitle {Optimal simple and implementable monetary and fiscal
  rules} {Optimal simple and implementable monetary and fiscal rules}.{\BBCQ}
\newblock
\APACjournalVolNumPages{Journal of Monetary Economics}{54}{6}{1702--1725}.
\newblock
\begin{APACrefDOI} \doi{10.1016/j.jmoneco.2006.07.002} \end{APACrefDOI}
\PrintBackRefs{\CurrentBib}

\bibitem [\protect \citeauthoryear {%
Schmitt-Groh{\'e}%
\ \BBA {} Uribe%
}{%
Schmitt-Groh{\'e}%
\ \BBA {} Uribe%
}{%
{\protect \APACyear {2010}}%
}]{%
schmitt2010optimal}
\APACinsertmetastar {%
schmitt2010optimal}%
\begin{APACrefauthors}%
Schmitt-Groh{\'e}, S.%
\BCBT {}\ \BBA {} Uribe, M.%
\end{APACrefauthors}%
\unskip\
\newblock
\APACrefYearMonthDay{2010}{}{}.
\newblock
{\BBOQ}\APACrefatitle {The optimal rate of inflation} {The optimal rate of
  inflation}.{\BBCQ}
\newblock
\BIn{} \APACrefbtitle {Handbook of monetary economics} {Handbook of monetary
  economics}\ (\BVOL~3, \BPGS\ 653--722).
\newblock
\APACaddressPublisher{}{Elsevier}.
\newblock
\begin{APACrefDOI} \doi{10.1016/B978-0-444-53454-5.00001-3} \end{APACrefDOI}
\PrintBackRefs{\CurrentBib}

\bibitem [\protect \citeauthoryear {%
Schmitt-Groh\'{e}%
\ \BBA {} Uribe%
}{%
Schmitt-Groh\'{e}%
\ \BBA {} Uribe%
}{%
{\protect \APACyear {2012}}%
{\protect \APACexlab {{\protect \BCnt {1}}}}}]{%
schmittgrohe2012optinfa}
\APACinsertmetastar {%
schmittgrohe2012optinfa}%
\begin{APACrefauthors}%
Schmitt-Groh\'{e}, S.%
\BCBT {}\ \BBA {} Uribe, M.%
\end{APACrefauthors}%
\unskip\
\newblock
\APACrefYearMonthDay{2012{\protect \BCnt {1}}}{}{}.
\newblock
{\BBOQ}\APACrefatitle {Foreign Demand for Domestic Currency and the Optimal
  Rate of Inflation} {Foreign demand for domestic currency and the optimal rate
  of inflation}.{\BBCQ}
\newblock
\APACjournalVolNumPages{Journal of Money, Credit and
  Banking}{44}{6}{1207-1224}.
\newblock
\begin{APACrefURL}
  \url{https://onlinelibrary.wiley.com/doi/abs/10.1111/j.1538-4616.2012.00528.x}
  \end{APACrefURL}
\newblock
\begin{APACrefDOI} \doi{https://doi.org/10.1111/j.1538-4616.2012.00528.x}
  \end{APACrefDOI}
\PrintBackRefs{\CurrentBib}

\bibitem [\protect \citeauthoryear {%
Schmitt-Groh\'{e}%
\ \BBA {} Uribe%
}{%
Schmitt-Groh\'{e}%
\ \BBA {} Uribe%
}{%
{\protect \APACyear {2012}}%
{\protect \APACexlab {{\protect \BCnt {2}}}}}]{%
schmittgrohe2012optinf}
\APACinsertmetastar {%
schmittgrohe2012optinf}%
\begin{APACrefauthors}%
Schmitt-Groh\'{e}, S.%
\BCBT {}\ \BBA {} Uribe, M.%
\end{APACrefauthors}%
\unskip\
\newblock
\APACrefYearMonthDay{2012{\protect \BCnt {2}}}{}{}.
\newblock
{\BBOQ}\APACrefatitle {On quality bias and inflation targets} {On quality bias
  and inflation targets}.{\BBCQ}
\newblock
\APACjournalVolNumPages{Journal of Monetary Economics}{59}{4}{393-400}.
\newblock
\begin{APACrefURL}
  \url{https://www.sciencedirect.com/science/article/pii/S0304393212000220}
  \end{APACrefURL}
\newblock
\begin{APACrefDOI} \doi{https://doi.org/10.1016/j.jmoneco.2012.02.002}
  \end{APACrefDOI}
\PrintBackRefs{\CurrentBib}

\bibitem [\protect \citeauthoryear {%
Sidrauski%
}{%
Sidrauski%
}{%
{\protect \APACyear {1967}}%
}]{%
sidrauski1967inflation}
\APACinsertmetastar {%
sidrauski1967inflation}%
\begin{APACrefauthors}%
Sidrauski, M.%
\end{APACrefauthors}%
\unskip\
\newblock
\APACrefYearMonthDay{1967}{}{}.
\newblock
{\BBOQ}\APACrefatitle {Rational Choice and Patterns of Growth in a Monetary
  Economy} {Rational choice and patterns of growth in a monetary
  economy}.{\BBCQ}
\newblock
\APACjournalVolNumPages{American Economic Review}{57}{2}{534--544}.
\PrintBackRefs{\CurrentBib}

\bibitem [\protect \citeauthoryear {%
Sims%
}{%
Sims%
}{%
{\protect \APACyear {2013}}%
}]{%
sims2013optinf}
\APACinsertmetastar {%
sims2013optinf}%
\begin{APACrefauthors}%
Sims, E.%
\end{APACrefauthors}%
\unskip\
\newblock
\APACrefYearMonthDay{2013}{}{}.
\newblock
\APACrefbtitle {Growth or the Gap? Which Measure of Economic Activity Should be
  Targeted in Interest Rate Rules?} {Growth or the gap? which measure of
  economic activity should be targeted in interest rate rules?}\ \APACbVolEdTR
  {}{Working Paper}.
\newblock
\APACaddressInstitution{}{University of Notre Dame, NBER, and ifo}.
\PrintBackRefs{\CurrentBib}

\bibitem [\protect \citeauthoryear {%
Siu%
}{%
Siu%
}{%
{\protect \APACyear {2004}}%
}]{%
siu2004optinf}
\APACinsertmetastar {%
siu2004optinf}%
\begin{APACrefauthors}%
Siu, H\BPBI E.%
\end{APACrefauthors}%
\unskip\
\newblock
\APACrefYearMonthDay{2004}{}{}.
\newblock
{\BBOQ}\APACrefatitle {Optimal fiscal and monetary policy with sticky prices}
  {Optimal fiscal and monetary policy with sticky prices}.{\BBCQ}
\newblock
\APACjournalVolNumPages{Journal of Monetary Economics}{51}{3}{575--607}.
\newblock
\begin{APACrefDOI} \doi{10.1016/j.jmoneco.2003.07.008} \end{APACrefDOI}
\PrintBackRefs{\CurrentBib}

\bibitem [\protect \citeauthoryear {%
Stanley%
}{%
Stanley%
}{%
{\protect \APACyear {2005}}%
}]{%
stanley2005beyond}
\APACinsertmetastar {%
stanley2005beyond}%
\begin{APACrefauthors}%
Stanley, T\BPBI D.%
\end{APACrefauthors}%
\unskip\
\newblock
\APACrefYearMonthDay{2005}{}{}.
\newblock
{\BBOQ}\APACrefatitle {Beyond publication bias} {Beyond publication
  bias}.{\BBCQ}
\newblock
\APACjournalVolNumPages{Journal of economic surveys}{19}{3}{309--345}.
\newblock
\begin{APACrefDOI} \doi{10.1111/j.0950-0804.2005.00250.x} \end{APACrefDOI}
\PrintBackRefs{\CurrentBib}

\bibitem [\protect \citeauthoryear {%
Stanley%
\ \BBA {} Doucouliagos%
}{%
Stanley%
\ \BBA {} Doucouliagos%
}{%
{\protect \APACyear {2012}}%
}]{%
stanley2012meta}
\APACinsertmetastar {%
stanley2012meta}%
\begin{APACrefauthors}%
Stanley, T\BPBI D.%
\BCBT {}\ \BBA {} Doucouliagos, H.%
\end{APACrefauthors}%
\unskip\
\newblock
\APACrefYear{2012}.
\newblock
\APACrefbtitle {Meta-Regression Analysis in Economics and Business}
  {Meta-regression analysis in economics and business}.
\newblock
\APACaddressPublisher{Abingdon}{Routledge}.
\newblock
\begin{APACrefDOI} \doi{10.4324/9780203111710} \end{APACrefDOI}
\PrintBackRefs{\CurrentBib}

\bibitem [\protect \citeauthoryear {%
Stanley%
\ \BBA {} Doucouliagos%
}{%
Stanley%
\ \BBA {} Doucouliagos%
}{%
{\protect \APACyear {2014}}%
}]{%
stanley2014meta}
\APACinsertmetastar {%
stanley2014meta}%
\begin{APACrefauthors}%
Stanley, T\BPBI D.%
\BCBT {}\ \BBA {} Doucouliagos, H.%
\end{APACrefauthors}%
\unskip\
\newblock
\APACrefYearMonthDay{2014}{}{}.
\newblock
{\BBOQ}\APACrefatitle {Meta-regression approximations to reduce publication
  selection bias} {Meta-regression approximations to reduce publication
  selection bias}.{\BBCQ}
\newblock
\APACjournalVolNumPages{Research Synthesis Methods}{5}{1}{60--78}.
\newblock
\begin{APACrefDOI} \doi{10.1002/jrsm.1095} \end{APACrefDOI}
\PrintBackRefs{\CurrentBib}

\bibitem [\protect \citeauthoryear {%
Steel%
}{%
Steel%
}{%
{\protect \APACyear {2020}}%
}]{%
steel2020model}
\APACinsertmetastar {%
steel2020model}%
\begin{APACrefauthors}%
Steel, M\BPBI F.%
\end{APACrefauthors}%
\unskip\
\newblock
\APACrefYearMonthDay{2020}{}{}.
\newblock
{\BBOQ}\APACrefatitle {Model averaging and its use in economics} {Model
  averaging and its use in economics}.{\BBCQ}
\newblock
\APACjournalVolNumPages{Journal of Economic Literature}{58}{3}{644--719}.
\newblock
\begin{APACrefDOI} \doi{10.1257/jel.20191385} \end{APACrefDOI}
\PrintBackRefs{\CurrentBib}

\bibitem [\protect \citeauthoryear {%
Stockman%
}{%
Stockman%
}{%
{\protect \APACyear {1981}}%
}]{%
stockman1981anticipated}
\APACinsertmetastar {%
stockman1981anticipated}%
\begin{APACrefauthors}%
Stockman, A\BPBI C.%
\end{APACrefauthors}%
\unskip\
\newblock
\APACrefYearMonthDay{1981}{}{}.
\newblock
{\BBOQ}\APACrefatitle {Anticipated Inflation and the Capital Stock in a
  Cash-in-Advance Economy} {Anticipated inflation and the capital stock in a
  cash-in-advance economy}.{\BBCQ}
\newblock
\APACjournalVolNumPages{Journal of Monetary Economics}{8}{3}{387--393}.
\newblock
\begin{APACrefDOI} \doi{10.1016/0304-3932(81)90018-0} \end{APACrefDOI}
\PrintBackRefs{\CurrentBib}

\bibitem [\protect \citeauthoryear {%
B.~Talukdar%
}{%
B.~Talukdar%
}{%
{\protect \APACyear {2015}}%
}]{%
talukdar2015optinf}
\APACinsertmetastar {%
talukdar2015optinf}%
\begin{APACrefauthors}%
Talukdar, B.%
\end{APACrefauthors}%
\unskip\
\newblock
\APACrefYearMonthDay{2015}{}{}.
\newblock
{\BBOQ}\APACrefatitle {Learning-by-doing and the optimal {Ramsey} interest
  rate} {Learning-by-doing and the optimal {Ramsey} interest rate}.{\BBCQ}
\newblock
\APACjournalVolNumPages{Economics Bulletin}{35}{1}{53--60}.
\PrintBackRefs{\CurrentBib}

\bibitem [\protect \citeauthoryear {%
B\BPBI K.~Talukdar%
}{%
B\BPBI K.~Talukdar%
}{%
{\protect \APACyear {2014}}%
}]{%
talukdar2011optinf}
\APACinsertmetastar {%
talukdar2011optinf}%
\begin{APACrefauthors}%
Talukdar, B\BPBI K.%
\end{APACrefauthors}%
\unskip\
\newblock
\APACrefYearMonthDay{2014}{}{}.
\newblock
{\BBOQ}\APACrefatitle {Organizational Learning and Optimal Fiscal and Monetary
  Policy} {Organizational learning and optimal fiscal and monetary
  policy}.{\BBCQ}
\newblock
\APACjournalVolNumPages{The B.E. Journal of Macroeconomics}{14}{1}{445--475}.
\newblock
\APACrefnote{Published version of the 2011 working paper.}
\newblock
\begin{APACrefDOI} \doi{10.1515/bejm-2012-0002} \end{APACrefDOI}
\PrintBackRefs{\CurrentBib}

\bibitem [\protect \citeauthoryear {%
Tang%
}{%
Tang%
}{%
{\protect \APACyear {2010}}%
}]{%
tang2010optinf}
\APACinsertmetastar {%
tang2010optinf}%
\begin{APACrefauthors}%
Tang, J\BHBI H.%
\end{APACrefauthors}%
\unskip\
\newblock
\APACrefYearMonthDay{2010}{}{}.
\newblock
{\BBOQ}\APACrefatitle {Optimal monetary policy in a {New Keynesian} model with
  job search} {Optimal monetary policy in a {New Keynesian} model with job
  search}.{\BBCQ}
\newblock
\APACjournalVolNumPages{Journal of Economic Dynamics \&
  Control}{34}{3}{330--353}.
\newblock
\begin{APACrefDOI} \doi{10.1016/j.jedc.2009.09.009} \end{APACrefDOI}
\PrintBackRefs{\CurrentBib}

\bibitem [\protect \citeauthoryear {%
Taylor%
}{%
Taylor%
}{%
{\protect \APACyear {1980}}%
}]{%
taylor1980aggregate}
\APACinsertmetastar {%
taylor1980aggregate}%
\begin{APACrefauthors}%
Taylor, J\BPBI B.%
\end{APACrefauthors}%
\unskip\
\newblock
\APACrefYearMonthDay{1980}{}{}.
\newblock
{\BBOQ}\APACrefatitle {Aggregate Dynamics and Staggered Contracts} {Aggregate
  dynamics and staggered contracts}.{\BBCQ}
\newblock
\APACjournalVolNumPages{Journal of Political Economy}{88}{1}{1--23}.
\newblock
\begin{APACrefDOI} \doi{10.1086/260845} \end{APACrefDOI}
\PrintBackRefs{\CurrentBib}

\bibitem [\protect \citeauthoryear {%
Tulip%
}{%
Tulip%
}{%
{\protect \APACyear {2011}}%
}]{%
tulip2011optinf}
\APACinsertmetastar {%
tulip2011optinf}%
\begin{APACrefauthors}%
Tulip, P.%
\end{APACrefauthors}%
\unskip\
\newblock
\APACrefYearMonthDay{2011}{}{}.
\newblock
\APACrefbtitle {Fiscal Policy and the Inflation Target} {Fiscal policy and the
  inflation target}\ \APACbVolEdTR {}{SSRN Working Paper\ \BNUM\ 1928299}.
\newblock
\APACaddressInstitution{}{SSRN Electronic Journal}.
\newblock
\begin{APACrefDOI} \doi{10.2139/ssrn.1928299} \end{APACrefDOI}
\PrintBackRefs{\CurrentBib}

\bibitem [\protect \citeauthoryear {%
van Aert%
\ \BBA {} van Assen%
}{%
van Aert%
\ \BBA {} van Assen%
}{%
{\protect \APACyear {2026}}%
}]{%
vanaert2023correcting}
\APACinsertmetastar {%
vanaert2023correcting}%
\begin{APACrefauthors}%
van Aert, R\BPBI C\BPBI M.%
\BCBT {}\ \BBA {} van Assen, M\BPBI A\BPBI L\BPBI M.%
\end{APACrefauthors}%
\unskip\
\newblock
\APACrefYearMonthDay{2026}{}{}.
\newblock
{\BBOQ}\APACrefatitle {Correcting for publication bias in a meta-analysis with
  the p-uniform* method} {Correcting for publication bias in a meta-analysis
  with the p-uniform* method}.{\BBCQ}
\newblock
\APACjournalVolNumPages{Psychonomic Bulletin \& Review}{33}{3}{102}.
\newblock
\begin{APACrefDOI} \doi{10.3758/s13423-025-02812-4} \end{APACrefDOI}
\PrintBackRefs{\CurrentBib}

\bibitem [\protect \citeauthoryear {%
van~de Schoot%
\ \protect \BOthers {.}}{%
van~de Schoot%
\ \protect \BOthers {.}}{%
{\protect \APACyear {2021}}%
}]{%
vandeschoot2021asreview}
\APACinsertmetastar {%
vandeschoot2021asreview}%
\begin{APACrefauthors}%
van~de Schoot, R.%
, de Bruin, J.%
, Schram, R.%
, Zahedi, P.%
, de Boer, J.%
, Weijdema, F.%
\BDBL {}Oberski, D\BPBI L.%
\end{APACrefauthors}%
\unskip\
\newblock
\APACrefYearMonthDay{2021}{}{}.
\newblock
{\BBOQ}\APACrefatitle {An Open Source Machine Learning Framework for Efficient
  and Transparent Systematic Reviews} {An open source machine learning
  framework for efficient and transparent systematic reviews}.{\BBCQ}
\newblock
\APACjournalVolNumPages{Nature Machine Intelligence}{3}{2}{125--133}.
\newblock
\begin{APACrefDOI} \doi{10.1038/s42256-020-00287-7} \end{APACrefDOI}
\PrintBackRefs{\CurrentBib}

\bibitem [\protect \citeauthoryear {%
Venkateswaran%
\ \BBA {} Wright%
}{%
Venkateswaran%
\ \BBA {} Wright%
}{%
{\protect \APACyear {2013}}%
}]{%
venkateswaran2013optinf}
\APACinsertmetastar {%
venkateswaran2013optinf}%
\begin{APACrefauthors}%
Venkateswaran, V.%
\BCBT {}\ \BBA {} Wright, R.%
\end{APACrefauthors}%
\unskip\
\newblock
\APACrefYearMonthDay{2013}{}{}.
\newblock
\APACrefbtitle {Pledgability and Liquidity: A New Monetarist Model of Financial
  and Macroeconomic Activity} {Pledgability and liquidity: A new monetarist
  model of financial and macroeconomic activity}\
  \APACbVolEdTR{\BVOL~28}{\BTR{}}.
\newblock
\APACaddressInstitution{}{NBER}.
\newblock
\begin{APACrefDOI} \doi{10.1086/674600} \end{APACrefDOI}
\PrintBackRefs{\CurrentBib}

\bibitem [\protect \citeauthoryear {%
Weber%
}{%
Weber%
}{%
{\protect \APACyear {2012}}%
}]{%
weber2012optinf}
\APACinsertmetastar {%
weber2012optinf}%
\begin{APACrefauthors}%
Weber, H.%
\end{APACrefauthors}%
\unskip\
\newblock
\APACrefYearMonthDay{2012}{}{}.
\newblock
\APACrefbtitle {The optimal inflation rate and firm-level productivity growth}
  {The optimal inflation rate and firm-level productivity growth}\
  \APACbVolEdTR{}{\BTR{}}.
\newblock
\APACaddressInstitution{}{Kiel Working Paper}.
\PrintBackRefs{\CurrentBib}

\bibitem [\protect \citeauthoryear {%
Wolman%
}{%
Wolman%
}{%
{\protect \APACyear {2001}}%
}]{%
wolman2001optinf}
\APACinsertmetastar {%
wolman2001optinf}%
\begin{APACrefauthors}%
Wolman, A\BPBI L.%
\end{APACrefauthors}%
\unskip\
\newblock
\APACrefYearMonthDay{2001}{}{}.
\newblock
{\BBOQ}\APACrefatitle {A Primer on Optimal Monetary Policy with Staggered
  Price-Setting} {A primer on optimal monetary policy with staggered
  price-setting}.{\BBCQ}
\newblock
\APACjournalVolNumPages{Federal Reserve Bank of Richmond Economic
  Quarterly}{87}{4}{27--52}.
\PrintBackRefs{\CurrentBib}

\bibitem [\protect \citeauthoryear {%
Wolman%
}{%
Wolman%
}{%
{\protect \APACyear {2011}}%
}]{%
wolman2011optinf}
\APACinsertmetastar {%
wolman2011optinf}%
\begin{APACrefauthors}%
Wolman, A\BPBI L.%
\end{APACrefauthors}%
\unskip\
\newblock
\APACrefYearMonthDay{2011}{}{}.
\newblock
{\BBOQ}\APACrefatitle {The Optimal Rate of Inflation with Trending Relative
  Prices} {The optimal rate of inflation with trending relative prices}.{\BBCQ}
\newblock
\APACjournalVolNumPages{Journal of Money, Credit and
  Banking}{43}{2-3}{355--384}.
\newblock
\begin{APACrefDOI} \doi{10.1111/j.1538-4616.2010.00377.x} \end{APACrefDOI}
\PrintBackRefs{\CurrentBib}

\bibitem [\protect \citeauthoryear {%
Woodford%
}{%
Woodford%
}{%
{\protect \APACyear {2003}}%
}]{%
woodford2003interest}
\APACinsertmetastar {%
woodford2003interest}%
\begin{APACrefauthors}%
Woodford, M.%
\end{APACrefauthors}%
\unskip\
\newblock
\APACrefYear{2003}.
\newblock
\APACrefbtitle {Interest and Prices: Foundations of a Theory of Monetary
  Policy} {Interest and prices: Foundations of a theory of monetary policy}.
\newblock
\APACaddressPublisher{Princeton, NJ}{Princeton University Press}.
\newblock
\begin{APACrefDOI} \doi{10.1515/9781400830169} \end{APACrefDOI}
\PrintBackRefs{\CurrentBib}

\bibitem [\protect \citeauthoryear {%
Yun%
}{%
Yun%
}{%
{\protect \APACyear {2005}}%
}]{%
yun2005optinf}
\APACinsertmetastar {%
yun2005optinf}%
\begin{APACrefauthors}%
Yun, T.%
\end{APACrefauthors}%
\unskip\
\newblock
\APACrefYearMonthDay{2005}{}{}.
\newblock
{\BBOQ}\APACrefatitle {Optimal Monetary Policy with Relative Price Distortions}
  {Optimal monetary policy with relative price distortions}.{\BBCQ}
\newblock
\APACjournalVolNumPages{The American Economic Review}{95}{1}{89--109}.
\newblock
\begin{APACrefDOI} \doi{10.1257/0002828053828653} \end{APACrefDOI}
\PrintBackRefs{\CurrentBib}

\bibitem [\protect \citeauthoryear {%
Zeugner%
\ \BBA {} Feldkircher%
}{%
Zeugner%
\ \BBA {} Feldkircher%
}{%
{\protect \APACyear {2015}}%
}]{%
zeugner2015bayesian}
\APACinsertmetastar {%
zeugner2015bayesian}%
\begin{APACrefauthors}%
Zeugner, S.%
\BCBT {}\ \BBA {} Feldkircher, M.%
\end{APACrefauthors}%
\unskip\
\newblock
\APACrefYearMonthDay{2015}{}{}.
\newblock
{\BBOQ}\APACrefatitle {{B}ayesian model averaging employing fixed and flexible
  priors: The {BMS} package for {R}} {{B}ayesian model averaging employing
  fixed and flexible priors: The {BMS} package for {R}}.{\BBCQ}
\newblock
\APACjournalVolNumPages{Journal of Statistical Software}{68}{4}{1--37}.
\newblock
\begin{APACrefDOI} \doi{10.18637/jss.v068.i04} \end{APACrefDOI}
\PrintBackRefs{\CurrentBib}

\bibitem [\protect \citeauthoryear {%
Zigraiova%
\ \BBA {} Havranek%
}{%
Zigraiova%
\ \BBA {} Havranek%
}{%
{\protect \APACyear {2016}}%
}]{%
zigraiova2015bank}
\APACinsertmetastar {%
zigraiova2015bank}%
\begin{APACrefauthors}%
Zigraiova, D.%
\BCBT {}\ \BBA {} Havranek, T.%
\end{APACrefauthors}%
\unskip\
\newblock
\APACrefYearMonthDay{2016}{}{}.
\newblock
{\BBOQ}\APACrefatitle {Bank Competition and Financial Stability: Much Ado About
  Nothing?} {Bank competition and financial stability: Much ado about
  nothing?}{\BBCQ}
\newblock
\APACjournalVolNumPages{Journal of Economic Surveys}{30}{5}{944--981}.
\newblock
\begin{APACrefDOI} \doi{10.1111/joes.12131} \end{APACrefDOI}
\PrintBackRefs{\CurrentBib}

\end{thebibliography}

%
%

\appendix

\section{PRISMA flow diagram}
\label{app:prisma}

\begin{figure}[H]
\centering
\caption{PRISMA flow diagram for the optimal-inflation meta-analysis.}
\label{fig:prisma_v34}
\begin{tikzpicture}[
  rect/.style   = {rectangle, draw=black!60, minimum height=2.2em,
                   font=\scriptsize, text width=9em, text centered,
                   inner sep=3pt},
  mylabel/.style = {draw, rectangle, align=center, rounded corners,
                   font=\footnotesize\bf, inner sep=2ex,
                   minimum height=3em, minimum size=4em},
  every node/.append style={node distance=1.1cm and 1.2cm}
]

\node[rect]                 (step1)  {Records identified through broad Scopus query (n = 12{,}904)};
\node[rect, right=of step1] (step1b) {Additional records from living bibliography of \citet{diercks2019reader} (n = 425)};

\node[rect, below=of step1] (step2)  {Records after de-duplication and ASReview title / abstract screening (n = 709)};
\node[rect, right=of step2] (step3)  {Records excluded at title / abstract stage (n = 533)};

\node[rect, below=of step2] (step3b) {Reports sought for retrieval (n = 176); reports not retrieved because the full text could not be obtained (n = 9)};

\node[rect, below=of step3b] (step4)  {Full-text articles assessed for eligibility (n = 167)};
\node[rect, right=of step4] (step5)  {Full texts excluded as out of scope (n = 37)};

\node[rect, below=of step4] (step6)  {Studies included in the meta-analysis (n = 116 with quantitative optima; 777 estimates; scope set n = 130 incl.\ 14 narrative-only)};

\draw[->] (step1)  -- (step2);
\draw[->] (step1b) -- (step2);
\draw[->] (step2)  -- (step3);
\draw[->] (step2)  -- (step3b);
\draw[->] (step3b) -- (step4);
\draw[->] (step4)  -- (step5);
\draw[->] (step4)  -- (step6);

\begin{scope}[start chain=going below, xshift=-5.5cm, node distance=0.4cm]
  \node[mylabel, on chain] (id) {Identification};
  \node[mylabel, on chain] (sc) {Screening};
  \node[mylabel, on chain] (el) {Eligibility};
  \node[mylabel, on chain] (in) {Included};
\end{scope}

\end{tikzpicture}
\end{figure}
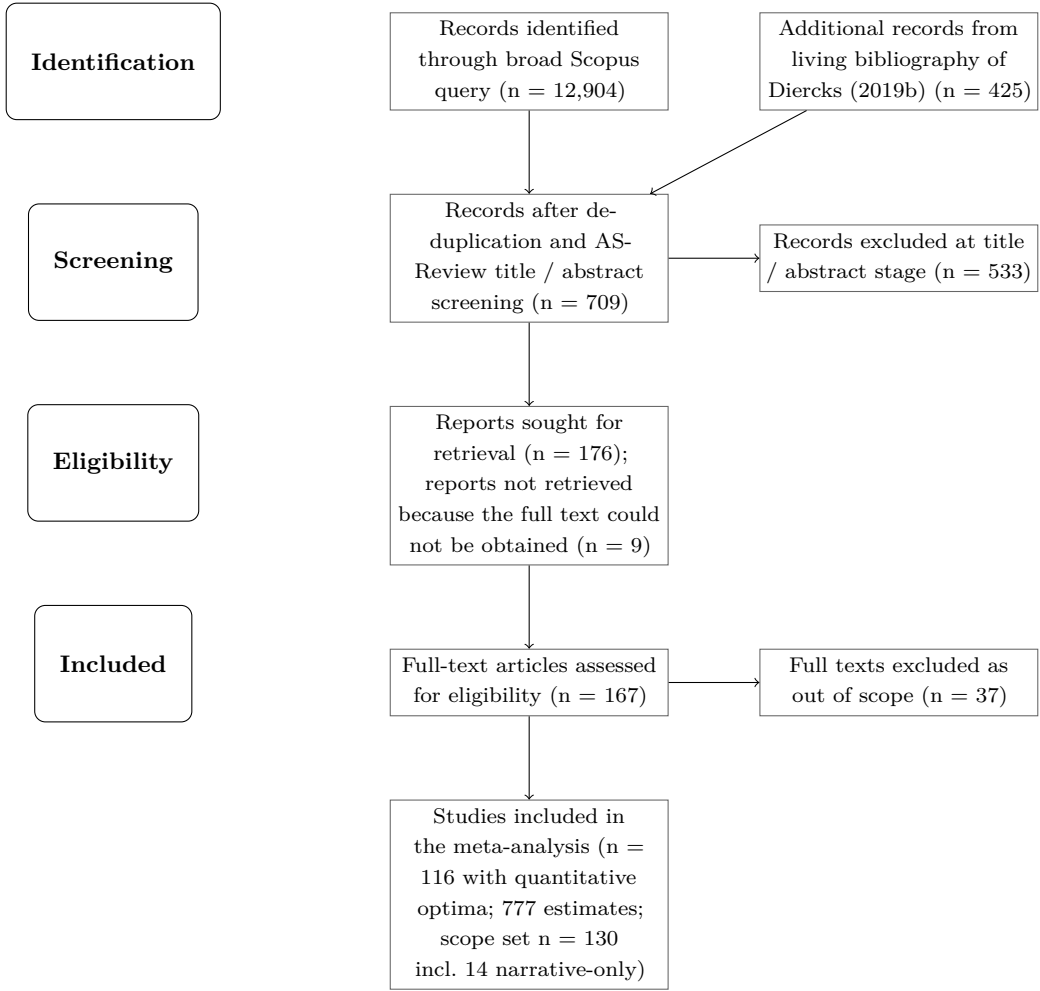

Figure~\ref{fig:prisma_v34} reproduces the PRISMA flow of records
through the identification, screening, eligibility, and inclusion
stages of our systematic review. The identification count combines
the $12{,}904$ records returned by the broad Scopus query
\texttt{TITLE-ABS-KEY("inflation") AND TITLE-ABS-KEY("monetary")} in
Economics and Business (April 2026, verified through the Scopus REST
API) with the $425$ records of the living bibliography of
\citet{diercks2019reader}. The narrower composite query described in
Section~\ref{sec:data_search} returns $380$ records and is contained
in the broad pool, so it contributes no additional records at the
identification stage. Identification also included targeted Scopus
queries on ``optimal inflation target'', ``optimal trend
inflation'', ``natural rate of interest optimal inflation'',
``quality bias inflation target'', ``EMU accession optimal
monetary policy'', and ``emerging market optimal inflation'';
records were de-duplicated against the main Scopus pool. Two policy
benchmarks that the editor or referees might expect to find were
checked explicitly: \citet{andrade2019optimal} (NBER WP 24328, 2018,
Idstudy~$9200$) and \citet{andrade2021should} (JEDC 2021,
Idstudy~$9201$) report calibrated optimal-inflation targets at the
estimated effective lower bound and are included in the analysis
set, contributing $19$ and $14$ winsorised estimates respectively;
\citet{lipinska2015optimal} contributes the only post-communist
structural calibration in the analysis set
($6$~estimates, Idstudy $9158$). Fully verified versions of all
primary studies are archived in the replication package.

\section{AI-extraction pipeline}
\label{app:ai-pipeline}

This appendix documents the large-language-model extraction pipeline
used to build the analysis dataset, the audit layers that gate its
output, and the self-consistency evidence that the rerun reported in
Section~\ref{sec:method} produced. The pipeline source is
\texttt{python/INFLATION\_2\_0.py}; the prompts are versioned under
\texttt{python/prompts/v3\_3/Super\_v3\_3.md}; the audit reports are
under \texttt{ai\_dataset/validation/}.

\subsection{Stage architecture and model assignment}
\label{app:ai-pipeline:arch}

The pipeline decomposes paper-to-row extraction into four stages,
each calling a single Anthropic Claude endpoint with a stage-specific
prompt and a stage-specific model. Cheaper Sonnet calls handle the
gating and metadata stages; more expensive Opus calls handle the two
stages whose errors propagate directly into the analysis dataset.

\begin{table}[H]
\centering
\small
\caption{AI-extraction pipeline stages.}
\label{tab:ai_stages}
\begin{tabular}{p{0.14\hsize}p{0.40\hsize}p{0.34\hsize}}
\hline
Stage & Task & Model \\ \hline
\texttt{pre\_scan} & Decide whether the paper reports any
quantitative optimal-inflation result; output count and source list. &
\texttt{claude-sonnet-4-5-20250929} \\
\texttt{metadata} & Extract bibliographic and study-level metadata
(authors, year, journal, country, model class). &
\texttt{claude-sonnet-4-5-20250929} \\
\texttt{structure} & Identify the model paradigm, frictions, agents,
calibration frequency, and policy regime. &
\texttt{claude-opus-4-5-20251001} \\
\texttt{results} & Read the result tables and emit one row per
optimal-inflation estimate, with the full closed-vocabulary moderator
set. & \texttt{claude-opus-4-5-20251001} \\ \hline
\end{tabular}
\end{table}

The Sonnet/Opus split is the result of an A/B test on the seven
hand-coded training papers cited in
Section~\ref{sec:method}. With Sonnet on all four stages, the
pipeline missed augmentation flags in Abo-Zaid (2013) and applied
the Ramsey-rule flag inconsistently in \citet{schmittgrohe2004optinf}
(2007); promoting \texttt{structure} and \texttt{results} to Opus
4.5 eliminated both errors. The promotion is recorded in
\texttt{python/config.py}.

\subsection{Audit layers}
\label{app:ai-pipeline:audit}

Three rule-based audit layers run after the LLM stages and gate the
release of the dataset. None of them is itself an LLM. The layers
are procedural checks that were executed during preparation of the
dataset; they leave their trace in the manuscript and the released
\texttt{.dta} rather than as separate row-level CSVs in the
replication package.

\begin{enumerate}
\item Citation audit: every in-text citation is matched against the
bibliography keys exported from \texttt{Bibl.bib}; orphan keys,
dangling references, and duplicated entries are flagged and
resolved before compile.
\item Claim checks: every numerical claim that the manuscript makes
about the dataset (sample size, share of US studies, share of
post-communist studies, etc.) is recomputed from the underlying
\texttt{.dta} and compared row by row; surviving claims are the
ones reported in the text.
\item PDF verification: every extracted estimate is paired back to
a page-anchor in the original PDF; rows that cannot be
page-anchored are flagged and either re-extracted or dropped, so
that the released \texttt{.dta} contains only page-anchored
estimates.
\end{enumerate}

\subsection{Training corpus, cross-check, and Round-2 rerun}
\label{app:ai-pipeline:audit_evidence}

The pipeline is grounded on three layers of human-vs-LLM comparison.

\paragraph{Seven hand-coded training papers.} The schema and prompt
were developed and frozen against seven papers coded by hand by the
authors before any LLM was invoked: \citet{cooley1989optinf},
\citet{amato2004optinf}, \citet{adam2006optinf},
\citet{amano2007optinf}, \citet{amano2009optinf},
\citet{abozaid2013optinf}, and \citet{abozaid2015optinf}. The
hand-coded set used a slightly older schema; the final schema
released to production added $15$ moderator columns, all of which
were re-checked by hand on the training papers before the production
run.

\paragraph{Thirty-one-paper cross-check.} Before the production run,
the pipeline was executed end-to-end on a $31$-paper full-text
cross-check sample (the list of papers and their hand-checked
reference values are documented in
\texttt{ai\_dataset/validation/audit\_round1\_report.md}),
deliberately biased toward the earliest, longest, and structurally
most complex papers in the corpus. The $31$ cross-check papers
amount to $26.7\%$ of the $116$-study analysis sample (well above
the conventional ten-percent audit benchmark) and to $23.8\%$ of
the $130$-study scope set, so the audit covers a substantial
fraction of the literature on which the headline rests. The
cross-check surfaced one definitional inconsistency in the Ramsey
flag (described in Section~\ref{sec:method}) which was fixed by
extending the policy-regime vocabulary and re-running the
structure stage. Within the $31$-paper full-text cross-check the audit
found no transcription errors in the numerical extraction of
optimal-inflation values: every reported estimate, every annualisation,
and every confidence-interval transcription matched the hand-checked
reference, as did the classification of the underlying estimator
(Ramsey planner, optimised Taylor rule, Friedman rule, etc.). This
should be read together with the smaller seven-paper hand-coded training
set reported in Section~\ref{app:ai-pipeline:irr}, where the paper-mean
MAE on $\pi^{\star}$ is $2.36$ pp/year and three categorical fields
($\texttt{Ramsey\_Rule}$, $\texttt{Government\_Included}$,
$\texttt{Other\_Agent\_Included}$) show low Cohen's $\kappa$. We do not
claim error-free extraction; we claim that the audit pool we hand-checked
row-by-row contained no transcription errors, while the smaller training
paper set exposes definitional ambiguities that motivated the schema
revisions documented in Section~\ref{sec:method}. After the
Ramsey-flag fix, the cross-check sample matched the hand-coded reference
at the moderator-class level on every paper.

\paragraph{Round-2 self-consistency rerun.} As a post-production
self-consistency check, the full pipeline was re-run with the same
prompts, the same model versions, and a fresh inference seed on a
stratified random sample of five studies drawn from the analysis
dataset (\texttt{random\_state=20260426}). Sample composition,
outcomes, and field-level comparison are reported in
Table~\ref{tab:round2}; the raw output is archived under
\texttt{ai\_dataset/validation/} and the audit
narrative under
\texttt{ai\_dataset/validation/audit\_round2\_report.md}.

\begin{table}[H]
\centering
\small
\caption{Round-2 self-consistency rerun (seed $=20{,}260{,}426$).}
\label{tab:round2}
\begin{threeparttable}
\begin{tabular}{p{0.32\hsize}cccp{0.28\hsize}}
\hline
Study & Rows & R2 rows & Stage & Outcome \\ \hline
Levin, Lopez-Salido \& Yun (2007)        & 7 & 0 & pre\_scan & Filtered out at inclusion gate \\
Kiley (2001)                             & 2 & 0 & pre\_scan & Filtered out at inclusion gate \\
Boehm \& House (2014)                    & 1 & 0 & pre\_scan & Filtered out at inclusion gate \\
Blanco (2015)                            & 4 & 4 & results   & 4/4 numerical fields exact match \\
\citet{schmittgrohe2004optinf}            & 3 & 4 & results   & Same Ramsey/fiscal/calibration class; extra $\lambda=1.35$ panel and within-panel statistic differences \\ \hline
\end{tabular}
\begin{tablenotes}\footnotesize
\item Notes: ``Rows'' is the number of preferred-specification rows for the
study in \texttt{stata/inflation\_v34.dta}. ``R2 rows'' is the number of
rows the Round-2 rerun emitted. ``Stage'' is the pipeline stage at which the
Round-2 outcome was determined: \texttt{pre\_scan} indicates that the
inclusion gate filtered the paper out, \texttt{results} indicates that the
paper passed the gate and was extracted to row level. Total Round-2
inference cost: USD $2.68$.
\end{tablenotes}
\end{threeparttable}
\end{table}

The pattern in Table~\ref{tab:round2} is informative for two
reasons. First, conditional on inclusion, the structured-extraction
stages reproduce the production numerical content closely: the four Blanco
(2015) optimal-inflation estimates ($5.0, 1.0, 0.5, 0.5$ percent),
the discount factor ($0.9967$), the country, the year, the
calibration frequency, the model paradigm, and the Ramsey flag are
all reproduced exactly, and the qualitative pattern in
\citet{schmittgrohe2004optinf} (mildly negative optimal inflation
under low markups, less negative under sticky prices, positive at
high markups) is preserved. Second, the dominant source of
run-to-run variation is the inclusion gate at \texttt{pre\_scan},
not the numerical-extraction layer: three of the five sampled
papers report calibrated steady-state inflation rates as
by-products of optimal-rule derivations rather than as headline
quantitative estimates, and Round-2 classified those by-products as
falling outside the inclusion criterion that Round-1 had applied
inclusively. This is the same inter-coder ambiguity that human
meta-analysts disclose under inclusion-criterion judgement; the
LLM-pipeline equivalent is documented here rather than left
implicit.

\paragraph{Implications for the headline.}
The Round-2 disagreements are confined to the inclusion margin and
to one within-paper panel-mapping disagreement in
\citet{schmittgrohe2004optinf}. All Round-2 numerical disagreements remain inside the
moderator-class envelope and inside the $116$-cluster precision of
the headline statistics. The headline numbers reported in
Section~\ref{sec:results} therefore stand without revision; the
appropriate qualification is that the inclusion margin is the
binding source of LLM-pipeline noise, which is the limitation we
flag.

\subsection{Reproducibility artefacts}
\label{app:ai-pipeline:repro}

\begin{itemize}
\item Pipeline: \texttt{python/INFLATION\_2\_0.py}.
\item Prompts (frozen production): \texttt{python/prompts/v3\_3/Super\_v3\_3.md}.
\item Configuration: \texttt{python/config.py} (model assignments, A/B-test
note, temperature settings).
\item Hand-coded reference set: the seven training papers cited
above (\texttt{ai\_dataset/\_study\_list\_v34.csv} flags them).
\item Cross-check pool: $31$-paper sample documented in
\texttt{ai\_dataset/validation/audit\_round1\_report.md}.
\item Production output (study list and enrichment):
\texttt{ai\_dataset/\_study\_list\_v34\_enriched.csv}.
\item Round-2 rerun audit:
\texttt{ai\_dataset/validation/audit\_round2\_report.md}
(deterministic seed $20{,}260{,}426$).
\item Stamped model versions: \texttt{claude-sonnet-4-5-20250929},
\texttt{claude-opus-4-5-20251001}.
\end{itemize}

\subsection{LLM-assisted adversarial checklist}
\label{app:ai-pipeline:adversarial}

Beyond the deterministic-rerun audit reported above, the manuscript
was also subjected to an adversarial AI stress test in which
independent LLM instances were prompted to challenge numerical
claims, factual claims, interpretation, and literature coverage.
The procedure operationalises the audit-duel idea of
\citet{havranek2026auditduel}, since generalised and re-released
as the open-source \texttt{mad-research} multi-agent debate
framework \citep{havranek2026madresearch}: independent
``challenger'' agents flag candidate inconsistencies in the
manuscript, ``defender'' agents reply with the corresponding
source or computation, and the authors review the residual flags
manually against the primary sources, the replication files, and
the manuscript text. The
procedure was used only to generate a checklist of possible issues
for human review; it did not produce estimates, coding decisions,
or substantive conclusions, and the headline numbers, the BMA
posterior, and the publication-bias diagnostics reported in the
main body are unchanged by it. The audit duel did not produce a
shippable artefact in this run; what survives is reflected in the
resolved manuscript text and in the open-source
\texttt{mad-research} framework cited above.

\subsection{Inter-rater agreement on the seven hand-coded training papers}
\label{app:ai-pipeline:irr}

For the categorical fields that are present in both the hand-coded
training schema (early, free-text-rich) and the released final schema
(closed-vocab-rich), we compute paper-level observed agreement,
Cohen's $\kappa$, and an approximate standard error. For the headline
numerical field we report the mean absolute error of the paper-mean
optimal inflation rate.

\begin{table}[H]
\centering
\caption{Inter-rater agreement, hand-coded training set vs.\ AI extraction.}
\label{tab:irr_v34}
\footnotesize
\begin{tabular*}{\linewidth}{@{\extracolsep{\fill}}lcccc@{}}
\toprule
Field & $n$ & Agreement & Cohen's $\kappa$ & SE($\kappa$) \\
\midrule
\texttt{Augmented\_base\_model}                  & 7 & 1.00 & $1.00$  & ---          \\
\texttt{Ramsey\_Rule}                            & 7 & 0.57 & $0.09$  & $0.40$       \\
\texttt{HH\_Included}                            & 7 & 1.00 & $1.00$  & ---          \\
\texttt{Firms\_Included}                         & 7 & 1.00 & $1.00$  & ---          \\
\texttt{Banks\_Included}                         & 7 & 0.86 & $0.00$  & $0.93$       \\
\texttt{Government\_Included}                    & 7 & 0.43 & $-0.40$ & $0.46$       \\
\texttt{Other\_Agent\_Included}                  & 7 & 0.43 & $-0.17$ & $0.38$       \\
\texttt{Empirical\_Research}                     & 7 & 0.86 & $0.00$  & $0.93$       \\
\midrule
Numerical (paper-mean $\pi^{\star}$, MAE pp/year)& 7 & ---  & ---     & MAE $=2.36$ \\
\bottomrule
\end{tabular*}
\end{table}

\begin{sloppypar}
Two honest observations follow. First, on the bookkeeping fields
(\texttt{Augmented\_base\_model},
\texttt{HH\_Included},
\texttt{Firms\_Included})
agreement is one. Second, on \texttt{Ramsey\_Rule},
\texttt{Government\_Included}, and
\texttt{Other\_Agent\_Included}
the agreement is materially below one.
\end{sloppypar}
The \texttt{Ramsey\_Rule} disagreement is the same definitional
inconsistency surfaced by the $31$-paper cross-check
(Section~\ref{app:ai-pipeline:audit_evidence}) and motivated the
introduction of the row-level \texttt{Policy\_Regime} closed vocabulary
in the production schema, which supersedes the paper-level \texttt{Ramsey\_Rule} flag in
all downstream analysis. The two disagreements in variables ending
with \texttt{\_Included} reflect different conventions for whether
modelled-but-passive sectors count as ``included''. With $n=7$ training papers the standard errors
on $\kappa$ are large and we present these statistics as diagnostic
rather than confirmatory; a larger inter-rater audit is on the
post-publication research agenda.

\subsection{Mapping to MAER-Net AI-extraction guidance}
\label{app:ai-pipeline:cook}

We map our pipeline against the MAER-Net AI-extraction guidance of
\citet{cook2026reporting}, marking each item as satisfied,
partially satisfied, or not satisfied. Partial compliance is
disclosed as such.

\begin{table}[H]
\centering
\caption{MAER-Net AI-extraction guidance: pipeline self-assessment.}
\label{tab:cook_compliance}
\footnotesize
\begin{tabular}{p{0.42\hsize}p{0.18\hsize}p{0.32\hsize}}
\toprule
Recommendation & Status & Evidence / caveat \\ \midrule
Prompts frozen before production run &
Satisfied & \texttt{python/prompts/v3\_3/Super\_v3\_3.md}, immutable since 22 April 2026. \\
Hand-coded training set used to develop and validate prompt &
Satisfied & Seven papers, $79$ estimates, hand-coded by one author
and frozen as the reference set against which the prompt was iterated. \\
Stamped model versions and deterministic decoding &
Satisfied & \texttt{claude-sonnet-4-5-20250929}, \texttt{claude-opus-4-5-20251001}; temperature $0$; deterministic order; round-2 rerun seed $20{,}260{,}426$. \\
Archived intermediate outputs of every stage &
Satisfied & Per-paper partial Excel files plus \texttt{checkpoint.jsonl}. \\
Inter-rater agreement (Cohen's $\kappa$) reported on the training set &
Partially satisfied & Computed (Table~\ref{tab:irr_v34}); $n=7$ papers limits inferential power. \\
Mean absolute error reported for numerical extractions &
Satisfied & MAE on paper-mean $\pi^{\star}$ in Table~\ref{tab:irr_v34}. \\
Round-2 self-consistency rerun & Satisfied &
Five-study stratified rerun, Section~\ref{app:ai-pipeline:audit_evidence}. \\
Independent human spot-check of production output & Satisfied &
$31$-paper full-text cross-check ($26.7\%$ of the $116$-study
analysis sample); the cross-check found no transcription errors
in the row-level optimal-inflation values or in the classification
of the underlying estimator (this is reported jointly with the
$7$-paper training-set MAE of $2.36$ pp/year on paper-mean
$\pi^{\star}$; see Table~\ref{tab:irr_v34}). \\
Public replication package with data, code, prompts, and audit reports &
Satisfied & Publicly available at \url{https://meta-analysis.cz/inflation}; structure documented in Section~\ref{app:ai-pipeline:repro}. \\
\bottomrule
\end{tabular}
\end{table}

\subsection{Forest plot by primary study (full sample)}
\label{app:ai-pipeline:forest}

For completeness we reproduce the horizontal box plot of estimates by
primary study, sorted by publication year. The figure is large by
construction ($116$ rows) and is provided as supplementary material
rather than as a primary in-text exhibit.

\begin{figure}[H]
\centering
\includegraphics[height=0.85\textheight, keepaspectratio]{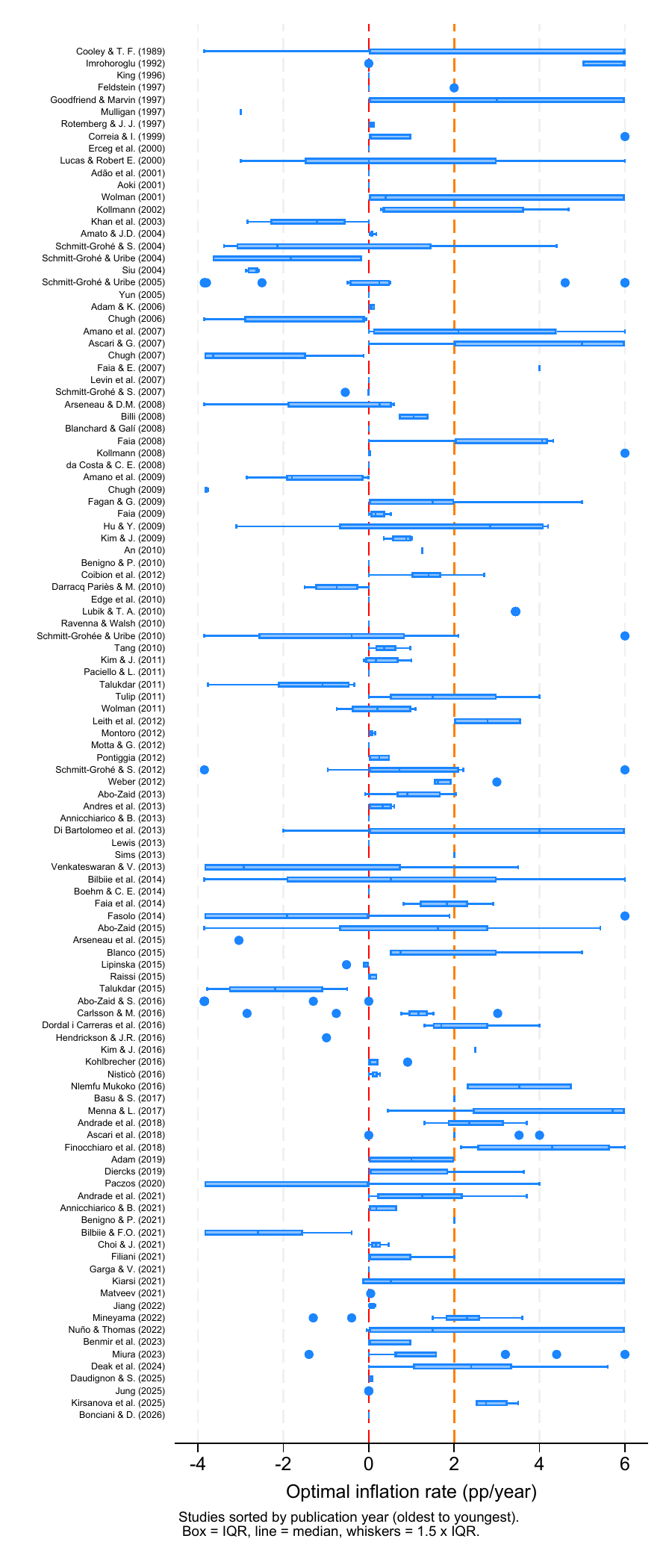}
\caption{Horizontal box plot of reported optimal inflation by primary study.}
\label{fig:bystudy_forest}
\begin{minipage}{\textwidth}
\footnotesize
\emph{Notes:} Horizontal box plot of the winsorised estimates of the
optimal long-run inflation rate (percent per year) for each of the
$116$ primary studies, sorted by publication year. Boxes show the
interquartile range, the line inside the box the median, and whiskers
extend to $1.5$ times the interquartile range. The vertical red line
marks zero measured inflation (price stability); the dashed orange line
marks the two-percent central-bank target.
\end{minipage}
\end{figure}


\end{document}